\documentclass[11pt,a4paper]{article}
\pdfoutput=1
\usepackage{jcappub}
\usepackage{xcolor}
\usepackage{graphicx}
\usepackage{float}
\usepackage{wrapfig}
\usepackage{bm}
\usepackage{ulem}

\usepackage{amsmath,amssymb}
\usepackage{subcaption}
\usepackage{verbatim}
\usepackage{makeidx}
\usepackage{braket}
\usepackage{multirow}
\usepackage[utf8]{inputenc}

\definecolor{red}{rgb}{1,0,0}
\definecolor{darkred}{rgb}{0.6,0,0}
\definecolor{darkgreen}{rgb}{0.992447,0.623778,0.034597}
\definecolor{ppink}{rgb}{1,0.4,0.4}
\definecolor{bblue}{rgb}{0.284602,0.317763,0.963947}
\definecolor{purple}{rgb}{0.5 ,0, 0.7}

\newcommand{\vev}[1]{ \left< {#1} \right> }

\newcommand{\dd}{\text{d}}

\newcommand{\Pl}{\text{Pl} }

\newcommand{\Si}{\text{Si}}
\newcommand{\Ci}{\text{Ci}}
\newcommand{\zo}{{(0)}}
\newcommand{\fo}{{(1)}}
\newcommand{\so}{{(2)}}
\newcommand{\rr}{{\text{r}}}
\newcommand{\mm}{{\text{m}}}
\newcommand{\sss}{{\text{s}}}

\makeatletter
\newcommand\footnoteref[1]{\protected@xdef\@thefnmark{\ref{#1}}\@footnotemark}
\gdef\@fpheader{}
\makeatother

\allowdisplaybreaks[1]

\title{
Analytic solutions of scalar perturbations induced by scalar perturbations
}
\author{Keisuke Inomata}
\affiliation{Research Center for the Early Universe (RESCEU), Graduate School of Science,
The University of Tokyo, Hongo 7-3-1, Bunkyo-ku, Tokyo 113-0033, Japan}

\abstract
{We study scalar perturbations induced by scalar perturbations through the non-linear interaction appearing at second order in perturbations.
We derive analytic solutions of the induced scalar perturbations in a perfect fluid. 
In particular, we consider the perturbations in a radiation-dominated era and a matter-dominated era.
With the analytic solutions, we also discuss how the induced scalar perturbations evolve from outside to inside horizon.
}

\begin{document}
\begin{flushright}
RESCEU-19/20
\end{flushright}
\maketitle

\section{Introduction}
\label{sec:intro}

The inflation theory predicts that the cosmological perturbations are generated from the quantum fluctuations of the fields, which means that valuable information about the early Universe is printed on the perturbations.
To reveal the history of the Universe, many authors have studied the perturbations for several decades.
Based on the behavior under the spatial coordinate transformation, the cosmological perturbations can be divided into three types: scalar, vector, and tensor perturbations.
In particular, scalar perturbations have been playing important roles in the study of the Universe.
For example, on large scales ($\gtrsim 1$Mpc), the amplitude and tilt of power spectrum of scalar perturbations have been determined through the observations of the large scale structure and the anisotropies of cosmic microwave background (CMB)~\cite{Aghanim:2018eyx}.
These observational results put constraints on parameters of inflation models~\cite{Akrami:2018odb}.
Also, scalar perturbations on small scales ($\lesssim 1$Mpc) have been attracting a lot of interest because of their rich phenomenology. 
If the amplitude of the small-scale perturbations are large enough, some unique compact objects, primordial black holes (PBHs)~\cite{1967SvA....10..602Z,Hawking:1971ei,Carr:1974nx,Carr:1975qj} and ultracompact minihalos (UCMHs)~\cite{Ricotti:2009bs,Scott:2009tu,Bringmann:2011ut}, are produced.
In particular, PBHs are one of the hottest topics in cosmology because they are good candidates of dark matter and the heavy BHs detected by LIGO-Virgo collaborations~\cite{Bird:2016dcv,Clesse:2016vqa,Sasaki:2016jop} (see also Refs.~\cite{Sasaki:2018dmp,Carr:2020gox,Carr:2020xqk,Green:2020jor} for recent reviews).

In the context of the small-scale scalar perturbations, many authors have recently been interested in the gravitational waves (GWs) induced by scalar perturbations because the induced GWs could be observed by the future observations depending on the amplitude of the small-scale perturbations~\cite{Ananda:2006af,Baumann:2007zm,Saito:2008jc,Saito:2009jt,Inomata:2016rbd,Ando:2017veq,Espinosa:2018eve,Kohri:2018awv,Cai:2018dig,Bartolo:2018evs,Bartolo:2018rku,Unal:2018yaa,Byrnes:2018txb,Inomata:2018epa,Clesse:2018ogk,Cai:2019amo,Cai:2019jah,Wang:2019kaf,Ben-Dayan:2019gll,Tada:2019amh,Inomata:2019zqy,Inomata:2019ivs,Yuan:2019udt,Xu:2019bdp,Cai:2019elf,Lu:2019sti, Yuan:2019wwo, Chen:2019xse, Hajkarim:2019nbx,Ozsoy:2019lyy,Domenech:2019quo,Fu:2019vqc,Ota:2020vfn,Lin:2020goi,Ballesteros:2020qam,Giovannini:2020qta,Pi:2020otn,Domenech:2020kqm,Lu:2020diy}.
Here, we draw attention to the fact that scalar perturbations also induce scalar perturbations on different scales through the non-linear interaction between scalar perturbations.\footnote{Scalar perturbations induced by tensor perturbations are discussed from the viewpoint of PBH production in Refs.~\cite{Nakama:2015nea,Nakama:2016enz}.}
Since the non-linear interaction appears at the second or higher order in perturbations, the induced scalar perturbations are smaller than the source (or first-order) scalar perturbations. 
However, the production of scalar perturbations on different scales can be important when we discuss the sharp power spectrum of the curvature perturbations, which are often considered in PBH scenarios~\cite{Kawasaki:2006zv,Kawaguchi:2007fz, Cai:2018tuh, Martin:2019nuw, Cai:2019bmk, Chen:2020uhe}, because the induced scalar perturbations can be dominant on scales where the source scalar perturbations are very small.
As concrete examples, we will see this important case in Fig.~\ref{fig:p_delta_rr} (Sec.~\ref{sec:solution_for_rd}) and Fig.~\ref{fig:p_delta_mm} (Sec.~\ref{sec:solution_for_md}) with the delta-function power spectrum of the source perturbations and in Fig.~\ref{fig:induced_scalar_real} (Appendix~\ref{app:power_realistic}) with the power spectrum for PBH scenarios.

Motivated by the above discussion, we shed light on the scalar perturbations induced by scalar perturbations in this paper.
Phenomenologically, the induced scalar perturbations have been studied in the context of the CMB anisotropies~\cite{Hu:1993tc,Dodelson:1993xz,Pyne:1995bs,Mollerach:1997up,Matarrese:1997ay,Goldberg:1999xm,Creminelli:2004pv,Bartolo:2004ty,Bartolo:2005xa,Tomita:2005et,Bartolo:2005fp,Bartolo:2005kv,Bartolo:2006cu,Bartolo:2006fj,Pitrou:2008ak,Khatri:2008kb} 
and the large scale structures~\cite{1981MNRAS.197..931J,1983MNRAS.203..345V,1984MNRAS.209..139J,Suto:1990wf,Makino:1991rp}.
Also, there are works focusing on the formalisms of the second-order perturbations, including scalar, vector, and tensor perturbations~\cite{tomita1967non,tomita1971non,tomita1972non,Matarrese:1993zf,Matarrese:1994wa,Mollerach:2003nq,Nakamura:2004rm,Nakamura:2004wr,Noh:2004bc,Nakamura:2006rk,Nakamura:2010yg,Domenech:2017ems,Uggla:2018cct,Nakamura:2019zbe}.
The main aim of this paper is to derive the analytic solutions of the induced scalar perturbations in two eras, a radiation-dominated (RD) era and a matter-dominated (MD) era.
Although Ref.~\cite{Bartolo:2006fj} derived the analytic solutions with simplified evolution of perturbations, we take into account the contributions that were neglected in the reference for the first time. 
On top of the reference, Refs.~\cite{Suto:1990wf,Makino:1991rp} derived the analytic solutions in a MD era with the Newtonian approximation, valid in the subhorizon limit~\cite{1983MNRAS.203..345V}. 
By contrast, in this paper, we derive the solutions that can describe the induced perturbations even on superhorizon scales.\footnote{Ref.~\cite{Hwang:2005hd} showed that if the $C$-gauge and the comoving condition are taken, the equations in the Newtonian approximation can be applied to the perturbations even on superhorizon scales. In this paper,  we derive the analytic solutions of the induced perturbations in the conformal Newtonian gauge to make a comprehensive analysis in both the eras. }

This paper is organized as follows.
In Sec.~\ref{sec:eom_so_scalar}, we summarize the basic equations for the scalar perturbations.
In Sec.~\ref{sec:solution_for_rd}, we derive the analytic solutions of the second-order perturbations and discuss their evolution in a RD.
In Sec.~\ref{sec:solution_for_md}, we discuss the second-order perturbations in a MD era, similarly to Sec.~\ref{sec:solution_for_rd}.
Finally, we devote Sec.~\ref{sec:conclusion} to the conclusion.

\section{Basic equations for scalar perturbations}
\label{sec:eom_so_scalar}

In this section, we summarize the basic equations for first- and second-order scalar perturbations.
Throughout this paper, we assume a flat FRW universe, follow the notation in Ref.~\cite{Malik:2008im}, and take the conformal Newtonian (longitudinal) gauge. 
Then, we can express the metric perturbations up to the second order as 
\begin{align}
 \dd s^2 &= g_{\mu\nu} \dd x^\mu \dd x^\nu \nonumber \\
& = a^2 \left\{- (1 + 2 \Phi^{(1)} + \Phi^{(2)})\dd \eta^2  + {V_i}^\so \dd \eta \dd x^i + \left[ (1-2 \Psi^{(1)} - \Psi^{(2)}) \delta_{ij} + \frac{1}{2} h^\so_{ij} \right] \dd x^i \dd x^j \right\},
\label{eq:def_metric_pertb}
\end{align}
where the superscript denotes the order in perturbations and we have neglected the first-order vector and tensor perturbations, $V^\fo_i$ and $h^\fo_{ij}$, because the main focus of this paper is on scalar perturbations induced by the first-order scalar perturbations.
In this paper, we assume a perfect fluid for simplicity. 
Then, we can express the energy-momentum tensor as
\begin{align}
	T^\mu_{\ \nu} &=  (\rho + P) u^\mu u_\nu + P \delta^\mu_{\ \nu},
\end{align}
where $\rho$ is the energy density, $P$ is the pressure, and $u_\mu$ is the 4-velocity.
We take the following notation for their perturbations up to the second order:
\begin{align}
	\label{eq:rho_pertb_def}
	\rho &= \rho^\zo + \delta \rho^\fo + \frac{1}{2} \delta \rho^\so, \\
	P &= P^\zo + \delta P^\fo + \frac{1}{2} \delta P^\so, \\
	\label{eq:u_i_pertb}
	u^i &= \frac{1}{a} \left( \delta v^{\fo,i} + \frac{1}{2} \delta v^{\so,i} + \frac{1}{2} \delta {v^i_V}^\so \right),
\end{align}
where the superscript ``$\zo$'' means the background value, $\delta {v^i_V}^\so$ is the second-order vector perturbation of the velocity perturbation, and the comma denotes the spatial derivative as $A^{,i} \equiv \delta^{ij} A_{,j} \equiv \delta^{ij}\, \partial A/\partial x^j$.
Note that the transformation between the superscript and the subscript in the 3-dimensional space is performed by $\delta_{ij}$ because we consider a flat universe.
Besides, we can express $u^0$ and $u_\mu (\equiv g_{\mu\nu} u^\nu)$ from their definitions as
\begin{align}
	u^0 
	&\equiv \frac{\dd x^0}{\sqrt{-g_{\mu\nu} \dd x^\mu \dd x^\nu}} \nonumber \\
	&= \frac{1}{a} \left( 1 - \Phi^\fo - \frac{1}{2} \Phi^\so + \frac{3}{2} \left(\Phi^\fo \right)^2 + \frac{1}{2} \delta v^{\fo,i} \delta v^\fo_{\quad ,i} \right), \\
	u_0 &\equiv g_{0 \mu}\, u^\mu \nonumber \\
	&= -a \left( 1 + \Phi^\fo + \frac{1}{2} \Phi^\so - \frac{1}{2} \left(\Phi^\fo \right)^2 + \frac{1}{2} \delta v^{\fo,i} \delta v^\fo_{\quad ,i}  \right), \\
	u_i &\equiv g_{i\mu}\, u^\mu \nonumber \\
	&= a \left( \delta v^{\fo}_{\quad ,i} + \frac{1}{2} \delta v^{\so}_{\quad ,i} + \frac{1}{2} \delta {v_{V\,i}}^\so  -2 \Psi^\fo \delta v^{\fo}_{\quad ,i} + \frac{1}{2}V_i^\so \right), \label{eq:u_i_pertb_def}
\end{align}
where we have neglected the terms of the third order or higher in perturbations.

In what follows, we consider the adiabatic perturbations for simplicity.
Then, the perturbations in a perfect fluid satisfy the following equations: (see Appendix~\ref{app:deriv_eq_for_pertb} for derivation)
\begin{align}
	&{\Phi^\fo}'' + 3 (1 + c_\sss^2) \mathcal H {\Phi^\fo}' + \left[2\mathcal H' + (3c_\sss^2 + 1) \mathcal H^2 \right] \Phi^\fo - c_\sss^2 \Phi^{\fo,i}_{\quad \  ,i} = 0, \label{eq:phi_fo_eom_re}\\
	&\delta^\fo \equiv \frac{\delta \rho^\fo}{\rho^\zo} = - \left( 2 \Phi^\fo + \frac{2}{\mathcal H} {\Phi^\fo}' - \frac{2}{3\mathcal H^2} \Phi^{\fo\, ,i}_{\qquad ,i} \right), \label{eq:delta_fo_eom_re}
\end{align}
where the prime denotes the derivative with respect to $\eta$, $\mathcal H (\equiv a'/a)$ is the conformal Hubble parameter, and $c_\sss^2 (\equiv \delta P^\fo/\delta \rho^\fo)$ is the square of the adiabatic sound speed. 
Note that we assume $c_\sss$ is constant for simplicity throughout this paper.
Solving these equations, we obtain the following solutions of the Fourier modes of the perturbations in a RD and a MD era:
\begin{align}	
	\label{eq:phi_zeta_rel}
	\Phi^\fo (\bm k, \eta) &= \begin{cases}
	 \displaystyle -\frac{2}{3} \zeta^\fo(\bm k) T_{\Phi,\rr}(x) & (\text{RD era}) \vspace{5pt} \\
	 \displaystyle -\frac{3}{5} \zeta^\fo(\bm k) T_{\Phi,\mm}(x) & (\text{MD era})
	 \end{cases},	 \\
	\delta^\fo (\bm k, \eta) &= \begin{cases}
	 \displaystyle -\frac{2}{3} \zeta^\fo(\bm k) T_{\delta,\rr}(x) & (\text{RD era}) \vspace{5pt} \\
	 \displaystyle -\frac{3}{5} \zeta^\fo(\bm k) T_{\delta,\mm}(x) & (\text{MD era})	 	 
	 \end{cases},
\end{align}
where $x\equiv k \eta$ with $k \equiv |\bm k|$ and the first-order curvature perturbation $\zeta^\fo$ and the transfer functions are defined as
\begin{align}
	\zeta^\fo &\equiv -\Psi^\fo + \frac{\delta \rho^\fo}{3(\rho^\zo + P^\zo)}, \\
	T_{\Phi,\rr}(x) &\equiv 3\sqrt{3} \frac{j_1(x/\sqrt{3})}{x} \\
	& = 
	\frac{3 [\sin(x/\sqrt{3}) - (x/\sqrt{3}) \cos (x/\sqrt{3})]}{(x/\sqrt{3})^3}, \\
	T_{\Phi,\mm}(x) &\equiv 1, \\
	T_{\delta,\rr}(x) &\equiv \frac{6x(-6 + x^2) \cos(x/\sqrt{3}) - 12 \sqrt{3} (-3 + x^2) \sin(x/\sqrt{3})}{x^3},\\
	T_{\delta,\mm}(x) &\equiv -2 - \frac{x^2}{6}. \label{eq:trans_delta_mm}
\end{align}
Note that $j_1$ is the spherical Bessel function of the first kind.

For the second-order perturbations, we can express the equation of motion for $\Psi^\so$ in a perfect fluid as~\cite{Bartolo:2006fj}\footnote{
Note that the notation of Ref.~\cite{Bartolo:2006fj} is different from that in this paper. They are related to each other as, $\Phi^\so_{\text{ref}} = \Phi^\so_{\text{this}} - 2 (\Phi^\fo_{\text{this}})^2$ and $\Psi^\so_{\text{ref}} = \Psi^\so_{\text{this}} + 2 (\Psi^\fo_{\text{this}})^2$, where the subscripts, ``ref'' and ``this'', denote the variables in the reference and this paper, respectively.} (see also Appendix~\ref{app:deriv_eq_for_pertb} for derivation)
\begin{align}
{\Psi^\so}'' + 3 (1 + c_\sss^2) \mathcal H {\Psi^\so}' + \left[2\mathcal H' + (3c_\sss^2 + 1) \mathcal H^2 \right] \Psi^\so - c_\sss^2 \Psi^{\so,i}_{\quad \  ,i} = S^\so, 
	\label{eq:psi_so_eom_re}
	\end{align}
where we have used the relation $\delta P^\so/ \delta \rho^\so = c_s^2 (\equiv \delta P^\fo/ \delta \rho^\fo)$, which are valid in the case the perturbations are adiabatic and the sound speed is constant~\cite{Uggla:2018cct}.
The source term $S^\so$ is defined as 
\begin{align}
	\label{eq:psi_so_source_re}
	S^\so \equiv& \left(3c_\sss^2 -\frac{1}{3} \right) \Phi^{\fo ,i} \Phi^\fo_{\quad ,i} + 8c_\sss^2 \Phi^\fo \Phi^{\fo ,i}_{\quad \ ,i} + (3c_\sss^2 + 1) \left({\Phi^\fo}' \right)^2 + \left[ (3c_\sss^2 + 1)\mathcal H^2 + 2 \mathcal H' \right]  N^j_{\ i} {B^i_{\ j}}^\so \nonumber \\
	& \ 
	 + \mathcal H  N^j_{\ i} \left({B^i_{\ j}}^\so \right)' + \frac{1}{3}  N^j_{\ i} \left({B^i_{\ j}}^\so \right)^{,k}_{\ ,k} + \left(\frac{1}{3} - c_\sss^2 \right) \frac{4}{3(1+w) \mathcal H^2}  \left( \mathcal H \Phi^{\fo,i} +  {\Phi^{\fo,i}}' \right)  \left(  \mathcal H \Phi^\fo_{,i} + {\Phi^\fo}'_{, i} \right),  
\end{align}
where $w (\equiv P^\zo/\rho^\zo)$ is the equation-of-state parameter and we assume $w$ is constant, similarly to $c_\sss$.
The ${B^i_{\ j}}^\so$ and $N^j_{\ i}$ are defined as 
\begin{align}
 {B^i_{\ j}}^\so \equiv \left[\frac{4(5+3w)}{3(1+w)} \Phi^{\fo ,i} \Phi^\fo_{\quad ,j} 
+ \frac{8}{3(1+w) \mathcal H} \left(\Phi^{\fo ,i} {\Phi^\fo_{\quad ,j}}\right)' 
 + \frac{8}{3(1+w) \mathcal H^2} {\Phi^{\fo ,i}}' {\Phi^\fo_{\quad ,j}}'\right],
 \label{eq:b_i_j_def}
\end{align}
\begin{align}
	N^j_{\ i} A^i_{\ j}(\bm x) &\equiv \frac{3}{2} \nabla^{-2} \left( \frac{\partial^j \partial_i}{\nabla^{2}} - \frac{1}{3} \delta^j_{\ i} \right) A^i_{\ j} (\bm x) \\
	&= \int \frac{\dd^3 k}{(2\pi)^3} \left( -\frac{3}{2k^2} \right) \left( \frac{k^j k_i}{k^2} - \frac{1}{3} \delta^j_{\ i} \right) A^i_{\ j} (\bm k), \label{eq:n_ji_def}
\end{align}
where $A^k_{\ j}$ is an arbitrary tensor.

Besides, we can express the second-order energy density perturbation with $\Psi^\so$ and $\Phi^\fo$ as 
\begin{align}
	\delta^\so \equiv& \frac{\delta \rho^\so}{\rho^\zo} \nonumber \\
	=& -2 \Psi^\so +2 N^j_{\ i} {B^i_{\ j}}^\so -\frac{2}{\mathcal H} {\Psi^\so}' + \frac{2}{3\mathcal H^2} \Psi^{\so,i}_{\quad \  ,i} 
	+ \frac{2}{\mathcal H^2} \left({\Phi^\fo}' \right)^2 + \frac{16}{3\mathcal H^2} \Phi^\fo \Phi^{\fo ,i}_{\quad \ ,i}  \nonumber \\
	& + \frac{1}{\mathcal H^2} \left( 2 - \frac{8}{9(1+w)} \right) \Phi^{\fo ,i} \Phi^\fo_{\quad ,i} - \frac{8}{9(1+w) \mathcal H^3} (\Phi^{\fo ,i} \Phi^\fo_{\quad ,i} )'  - \frac{8}{9(1+w) \mathcal H^4} {\Phi^\fo}'_{, i}  {\Phi^{\fo,i}}',\label{eq:delta_so_exp}
\end{align}
where its derivation is given in Appendix~\ref{app:deriv_eq_for_pertb}.
In the superhorizon limit ($k\eta \ll 1$), the perturbation becomes 
\begin{align}
	\delta^\so \simeq -2 \Psi^\so +2 N^j_{\ i} \left(\frac{4(5+3w)}{3(1+w)} \Phi^{\fo ,i} \Phi^\fo_{\quad ,j} \right),
	\label{eq:delta_so_ini}
\end{align}
where we have used ${\Phi^\fo}' = {\Psi^\so}' =0$ in the superhorizon limit. 

As a gauge-invariant variable, we introduce the second-order curvature perturbation in the uniform density gauge,\footnote{Strictly speaking, we need one more gauge condition on top of the uniform density one to define the second-order curvature perturbations uniquely. In this paper, we take the same gauge condition as that in Eq.~(7.71) of Ref.~\cite{Malik:2008im} (see also Appendix~\ref{app:deriv_eq_for_pertb}).}  which are expressed in the Newtonian gauge as~\cite{Malik:2008im}\footnote{We have fixed some typos in Eq.~(7.71) of Ref.~\cite{Malik:2008im}.}
\begin{align}
	\label{eq:zeta_so_def}
	\zeta^\so 
	=&
	-\Psi^\so + \frac{1}{3(1+w)} \delta^\so + \frac{1}{9 \mathcal H (1+w)^2} \left[\left(  \delta^\fo \right)^2\right]' - \frac{1+3w}{9 (1+w)^2} \left( \delta^\fo \right)^2  \nonumber \\
	& 
	- \frac{2}{3 \mathcal H (1+w)} \delta^\fo \left( {\Psi^\fo}' + 2 \mathcal H \Psi^\fo \right) - \frac{1}{18 \mathcal H^2 (1+w)^2} \left[ \delta^{\fo\, ,k} \delta^\fo_{,k} - \nabla^{-2} \left(  \delta^\fo_{,i }\delta^\fo_{,j} \right)^{,ij} \right],
\end{align}
where the expression in a general gauge is given in Appendix~\ref{app:deriv_eq_for_pertb}.
In the superhorizon limit, the curvature perturbation in a perfect fluid with $\Phi^\fo = \Psi^\fo$ can be expressed as
\begin{align}
	\zeta^\so \simeq - \frac{5+3w}{3(1+w)} \Psi^\so + \frac{2}{3(1+w)} N^j_{\ i} \left(\frac{4(5+3w)}{3(1+w)} \Phi^{\fo ,i} \Phi^\fo_{\quad ,j} \right) - \frac{1+3w}{9 (1+w)^2} \left( \delta^\fo \right)^2 - \frac{4}{3(1+w)}  \delta^\fo \Phi^\fo,
\end{align}
where we have used Eq.~(\ref{eq:delta_so_ini}). 
The initial condition of the second-order curvature perturbation depends on the primordial non-Gaussianity.
Following Ref.~\cite{Bartolo:2004if}, we assume the local-type non-Gaussianity throughout this paper, which is parametrized as $\zeta^\so = 2 a_{\text{NL}} (\zeta^\fo)^2$ in the superhorizon limit.
The value of $a_\text{NL}$ depends on the scenario and $a_\text{NL}=1$ corresponds to the Gaussian perturbation~\cite{Bartolo:2004ty,Bartolo:2005xa}.
With the parametrization of the second-order curvature perturbation, $\Psi^\so$ in the superhorizon limit can be expressed as 
\begin{align}
	\Psi^\so \simeq -\frac{6(1+w)}{5+3w} a_\text{NL} (\zeta^\fo)^2 + \frac{8}{3(1+w)} N^j_{\ i} \left(\Phi^{\fo ,i} \Phi^\fo_{\quad ,j} \right) - \frac{1+3w}{3 (1+w) (5+3w)} \left( \delta^\fo \right)^2 - \frac{4}{5+3w}  \delta^\fo \Phi^\fo.
	\label{eq:psi_so_ini_con}
\end{align}

\section{Induced scalar perturbations in a radiation-dominated era}
\label{sec:solution_for_rd}

In this section, we discuss the induced scalar perturbations in a RD era using the basic equations in the previous section.

\subsection{Analytic solutions}
First, we derive the analytic solutions for $\Psi^\so$, $\delta^\so$, and $\zeta^\so$.
Taking $c_\sss^2 = w = 1/3$, we can express Eq.~(\ref{eq:psi_so_eom_re}) in Fourier space as
\begin{align}
{\Psi^\so}''(\bm k,\eta) + \frac{4}{\eta} {\Psi^\so}'(\bm k,\eta) + \frac{k^2}{3} \Psi^\so (\bm k,\eta) = S^\so_\rr (\bm k,\eta),
\label{eq:psi_so_eom_re_rd}
\end{align}
where we have substituted $\mathcal H = 1/\eta$.
The source term $S^\so_\rr$ is given by
\begin{align}
	S^\so_\rr (\bm k,\eta) =& \int \frac{\dd^3 \tilde k}{(2\pi)^3} \left\{ -\left( \frac{2}{3} \tilde {\bm k} \cdot (\bm k - \tilde {\bm k}) + \frac{4}{3} ({\tilde k}^2 + {|\bm k - \bm{\tilde k}|}^2 ) \right) \Phi(\tilde {\bm k}) \Phi(\bm k - \tilde {\bm k}) + 2 \Phi'(\tilde {\bm k}) \Phi'(\bm k - \tilde {\bm k})  \right. \nonumber \\
	& \qquad \qquad \  
	+ \frac{3}{2k^2} \left( \frac{1}{k^2} ( \bm k \cdot \tilde {\bm k} ) ( \bm k \cdot (\bm k - \tilde {\bm k} ) ) - \frac{1}{3} \tilde {\bm k} \cdot (\bm k - \tilde {\bm k}) \right) \nonumber \\
	& \qquad \qquad \quad \times \left[
	\frac{1}{\eta} \left( 6 \Phi(\tilde {\bm k}) \Phi(\bm k - \tilde {\bm k}) + 4 \eta \Phi'(\tilde {\bm k}) \Phi(\bm k - \tilde {\bm k}) + 2 \eta^2 \Phi'(\tilde {\bm k}) \Phi'(\bm k - \tilde {\bm k}) \right)' \right. \nonumber \\
	& \qquad \qquad \left. \left. \phantom{\left( \frac{2}{3} \right)}
	- \frac{k^2}{3}  \left( 6 \Phi(\tilde {\bm k}) \Phi(\bm k - \tilde {\bm k}) + 4 \eta \Phi'(\tilde {\bm k}) \Phi(\bm k - \tilde {\bm k}) + 2 \eta^2 \Phi'(\tilde {\bm k}) \Phi'(\bm k - \tilde {\bm k}) \right) \right] \right\},
\end{align}
where we have omitted the superscript for the first-order perturbations. 
Using Eqs.~(\ref{eq:phi_fo_eom_re}) and (\ref{eq:phi_zeta_rel}), we can rewrite the source term as 
\begin{align}
	S^\so_\rr (\bm k,\eta)
	=&
	\int \frac{\dd^3 \tilde k}{(2\pi)^3} \left\{ -\left( \frac{2}{3} \tilde {\bm k} \cdot (\bm k - \tilde {\bm k}) + \frac{4}{3} ({\tilde k}^2 + |\bm k - \tilde {\bm k}|^2) \right) T_{\Phi,\rr}(\tilde k \eta) T_{\Phi,\rr}(|\bm k - \tilde {\bm k}| \eta)  \right. \nonumber \\
	& \qquad \qquad \ \,  
	+ 2 T_{\Phi,\rr}'(\tilde k \eta) T_{\Phi,\rr}'(|\bm k - \tilde {\bm k}| \eta) - \frac{3}{2k^2} \left( \frac{1}{k^2} ( \bm k \cdot \tilde {\bm k} ) ( \bm k \cdot (\bm k - \tilde {\bm k} ) ) - \frac{1}{3} \tilde {\bm k} \cdot (\bm k - \tilde {\bm k}) \right) \nonumber \\
	& \qquad \quad \left. \phantom{\left( \frac{2}{3} \right)} \times \left[ 
	\left( 2k^2 + \frac{2}{3}({\tilde k}^2 + |\bm k - \tilde {\bm k}|^2) \right) T_{\Phi,\rr}(\tilde k \eta) T_{\Phi,\rr}(|\bm k - \tilde {\bm k}| \eta) \right. \right. \nonumber \\
	& \qquad \quad\qquad \quad\ \ \, 
	+ \left( \frac{2k^2}{3} +  \frac{2{|\bm k - \tilde {\bm k}|}^2}{3} \right) \eta T_{\Phi,\rr}'(\tilde k \eta) T_{\Phi,\rr}(|\bm k - \tilde {\bm k}| \eta)  \nonumber \\ 
	& \qquad \quad\qquad \quad\ \ \, 
	 + \left( \frac{2k^2}{3} + \frac{2{\tilde k}^2}{3} \right) \eta T_{\Phi,\rr}(\tilde k \eta) T_{\Phi,\rr}'(|\bm k - \tilde {\bm k}| \eta) \nonumber \\
	& \qquad \quad  \ \ 
	\left.\left. \phantom{\left( \frac{{\tilde k}^2}{3} \right)} 
	 + \left( \frac{2 k^2\eta^2}{3} +8 \right) T_{\Phi,\rr}'(\tilde k \eta) T_{\Phi,\rr}'(|\bm k - \tilde {\bm k}| \eta) \right] \right\} 
	\left( \frac{2}{3} \right)^2 \zeta(\tilde {\bm k}) \zeta(\bm k - \tilde {\bm k}) \nonumber\\
	 =&
	\int \frac{\dd^3 \tilde k}{(2\pi)^3} k^2 uv f_\rr(u,v,x) \left( \frac{2}{3} \right)^2 \zeta(\tilde {\bm k}) \zeta(\bm k - \tilde {\bm k}),
	\label{eq:s_k_so_rad}
\end{align}
where $u \equiv |\bm k - \tilde{\bm k}|/k$, $v \equiv \tilde k/k$, and $f_\rr(u,v,x)$ is defined as 
\begin{align}
	f_\rr(u,v,x) \equiv& \left[ - \left( \frac{3u^2 + 3 v^2 +1}{3 u v} \right) - \left( \frac{2(u^2+v^2) - 3(u^2 - v^2)^2 +1}{8uv} \right) \left( 2 + \frac{2}{3}\left( u^2 + v^2 \right) \right) \right] 
	T_{\Phi,\rr}(v x) T_{\Phi,\rr}(u x) \nonumber \\
	&  
	- \left( \frac{2(u^2+v^2) - 3(u^2 - v^2)^2 +1}{8uv} \right) \left( \frac{2}{3} + \frac{2u^2}{3} \right) v x\, \frac{\dd T_{\Phi,\rr}(v x)}{\dd (vx)} T_{\Phi,\rr}( u x) \nonumber \\
	&  
	- \left( \frac{2(u^2+v^2) - 3(u^2 - v^2)^2 +1}{8uv} \right) \left( \frac{2}{3} + \frac{2v^2}{3} \right) u x\, T_{\Phi,\rr}( v x) \frac{\dd T_{\Phi,\rr}(ux)}{\dd (ux)} \nonumber \\
	&  
	+ \left[ \frac{2}{uv} - \left(  \frac{2(u^2+v^2) - 3(u^2 - v^2)^2 +1}{8uv} \right) \left( \frac{2x^2}{3} + 8 \right) \right] u v \, \frac{\dd T_{\Phi,\rr}(vx)}{\dd (vx)} \frac{\dd T_{\Phi,\rr}(ux)}{\dd (ux)}.
\end{align}

We solve Eq.~(\ref{eq:psi_so_eom_re}) using the Green function method.
Here, we define $z \equiv a^2 \Psi^\so$ and rewrite Eq.~(\ref{eq:psi_so_eom_re_rd}) as
\begin{align}
	z''(\bm k,\eta) + \left( \frac{k^2}{3} - \frac{2}{\eta^2} \right) z(\bm k,\eta) = a^2 S^\so_\rr(\bm k,\eta).
\end{align}
The Green function is defined as the solution of the following equation:
\begin{align}
	G''_\rr + \left( \frac{k^2}{3} - \frac{2}{\eta^2} \right) G_\rr = \delta(\eta - \bar \eta),
\end{align}
where the prime denotes the derivative with respect to $\eta$, not $\bar \eta$.
Then, we obtain the Green function as 
\begin{align}
	k G_\rr (k, \eta;\bar \eta) = -\Theta(\eta - \bar \eta) \frac{x \bar x}{\sqrt{3}} \left[ j_1(x/\sqrt{3}) y_1(\bar x/\sqrt{3}) - j_1(\bar x/\sqrt{3}) y_1(x/\sqrt{3}) \right],
\end{align}
where $y_1$ is the spherical Bessel function of the second kind.
Using this Green function, we can solve Eq.~(\ref{eq:psi_so_eom_re}) as 
\begin{align}
	\Psi^\so (\bm k,\eta) =& \Psi^\so(\bm k, 0) T_{\Phi,\rr}(x) + \int^\eta_0 \dd \bar\eta \left(\frac{a(\bar\eta)}{a(\eta)}\right)^2 G_\rr(k,\eta;\bar \eta) S^\so_\rr (\bm k,\eta) \\
	=&  \Psi^\so(\bm k, 0) T_{\Phi,\rr}(x) + \int \frac{\dd^3 \tilde k}{(2\pi)^3} uv I_{\Psi,\rr,\sss}(u,v,x) \left( \frac{2}{3} \right)^2 \zeta(\tilde {\bm k}) \zeta(\bm k - \tilde {\bm k}),
\end{align}	
where $I_{\Psi,\rr,\sss}$ is defined as 
\begin{align}
	I_{\Psi,\rr,\sss}(u,v,x) \equiv \int^x_0 \dd \bar x \left(\frac{\bar x}{x}\right)^2 G_\rr(k,\eta;\bar \eta) f_\rr(u,v,x). \label{eq:i_psi_rr_sss}
\end{align}
We can rewrite $I_{\Psi,\rr,\sss}$ as 
\begin{align}
	I_{\Psi,\rr,\sss}(u,v,x) 
	= & \int^x_0 \dd \bar x \, \left(\frac{\bar x}{x} \right)^2 \left\{ -\frac{x \bar x}{\sqrt{3}} \left[ j_1(x/\sqrt{3}) y_1(\bar x/\sqrt{3}) - j_1(\bar x/\sqrt{3}) y_1(x/\sqrt{3}) \right] \right\} f(u,v, \bar x) \nonumber \\
	= &\, \mathcal{J}(u,v,x) 3\sqrt{3} \frac{j_1(x/\sqrt{3})}{x} + \mathcal{Y}(u,v,x) 3\sqrt{3} \frac{y_1(x/\sqrt{3})}{x},	
	\label{eq:i_psi_r_s_re}
\end{align}
where $\mathcal J$ and $\mathcal Y$ are defined as 
\begin{align}
	\mathcal J(u,v,x) \equiv& \int^x_0 \dd \bar x \, \mathcal I_j(u,v,x), \\
	\mathcal Y(u,v,x) \equiv& \int^x_0 \dd \bar x \, \mathcal I_y(u,v,x), \\
	\mathcal I_j(u,v,\bar x) \equiv& \left( -\frac{{\bar x}^3}{9} \right) y_1(\bar x/\sqrt{3}) f_\rr(u,v,\bar x), \\
	\mathcal I_y(u,v,\bar x) \equiv& \frac{{\bar x}^3}{9}  j_1(\bar x/\sqrt{3}) f_\rr(u,v,\bar x).
\end{align}
Here, we expand the integrands $\mathcal I_j$ and $\mathcal I_y$ as 
\begin{align}
	\mathcal I_a(u,v, \bar x) =  \sum_{m=0}^7 \sum_{n=1}^8 \sin(\alpha_n \bar x  + \phi_n) \frac{M^a_{nm}}{\bar x^m},
	\label{eq:i_a_expand}
\end{align}
where the sub/superscript $a$ denotes $j$ or $y$ and $\alpha_n$, $\phi_n$, and $M^a_{nm}$ are independent of $\bar x$.
Specifically, $\alpha_n$ and $\phi_n$ are given as 
\begin{align}
	\alpha_n &= \begin{cases}
		\frac{1 - u - v}{\sqrt{3}} \quad & n=1,5 \\
		\frac{1 + u - v}{\sqrt{3}} \quad & n=2,6 \\
		\frac{1 - u + v}{\sqrt{3}} \quad & n=3,7 \\		
		\frac{1 + u + v}{\sqrt{3}} \quad & n=4,8 
		\end{cases}, \\
	\phi_n &= \begin{cases}
		\pi/2 \quad & n \leq 4 \\	
		0 \quad & n > 4
		\end{cases}.
\end{align}
The explicit expressions of $M^a_{nm}$ are given in Appendix~\ref{app:ana_sol_rd}.
For convenience, we divide $\mathcal J$ into two parts depending on the power of $\bar x$ of its integrand as
\begin{align}
	\mathcal J(u,v,x) =& \mathcal J_0(u,v,x) + \mathcal J_\text{h}(u,v,x), \\
	 \mathcal J_0(u,v,x) \equiv&  \int^x_0 \dd \bar x \, \sum_{n=1}^8 \sin(\alpha_n \bar x  + \phi_n) M^j_{n0}, \\
	 \mathcal J_\text{h}(u,v,x) \equiv&  \int^x_0 \dd \bar x \,  \sum_{m=1}^7 \sum_{n=1}^8 \sin(\alpha_n \bar x  + \phi_n) \frac{M^j_{nm}}{\bar x^m},\label{eq:mathcal_j_h_def}
\end{align}
and similarly for $\mathcal Y$. 
Note that the pioneering work, Ref.~\cite{Bartolo:2006fj}, focused only on the contributions from $\mathcal J_0$ and $\mathcal Y_0$.
Performing the integrals of $\mathcal J_0$ and $\mathcal Y_0$, we obtain the following expressions:
\begin{align}
	\mathcal J_{0} (u,v,x) =& \frac{3(-1 + 3(u^2-v^2)^2 -2(u^2 + v^2))}{16 u^2 v^2} \nonumber \\
	& \quad \times 
	\left( \frac{\cos\left( \frac{-1+u-v}{\sqrt{3}} x \right)}{-1+u-v} - \frac{\cos\left( \frac{1+u-v}{\sqrt{3}} x \right)}{1+u-v} - \frac{\cos\left( \frac{-1+u+v}{\sqrt{3}} x \right)}{-1+u+v} + \frac{\cos\left( \frac{1+u+v}{\sqrt{3}} x \right)}{1+u+v} \right)  \nonumber \\
	&- \frac{3(-1 + 3(u^2-v^2)^2 -2(u^2 + v^2))}{2uv ((u+v)^2-1)((u-v)^2 -1)} , \\
\mathcal Y_{0}(u,v,x) =& -\frac{3(-1 + 3(u^2-v^2)^2 -2(u^2 + v^2))}{16 u^2 v^2} \nonumber \\
	& \qquad \times 
	\left( \frac{\sin\left( \frac{-1+u-v}{\sqrt{3}} x \right)}{-1+u-v} + \frac{\sin\left( \frac{1+u-v}{\sqrt{3}} x \right)}{1+u-v} - \frac{\sin\left( \frac{-1+u+v}{\sqrt{3}} x \right)}{-1+u+v} - \frac{\sin\left( \frac{1+u+v}{\sqrt{3}} x \right)}{1+u+v} \right).
\end{align}
Note that $\mathcal J_0$ and $\mathcal Y_0$ do not converge even in the late-time limit ($x \gg 1$).
On the other hand, $\mathcal J_\text{h}$ and $\mathcal Y_\text{h}$ converge to $\mathcal J_\text{h,late}$ and $\mathcal Y_\text{h,late}$ in the late-time limit, which are given as 
\begin{align}
	\mathcal J_\text{h,late}(u,v) =&  \frac{1}{8 u^3 v^3}
\left[-9(u^6 + u^4 -u^2) -9(v^6 + v^4 - v^2) + 9  \right. \nonumber \\
& \qquad \qquad \left. + 6 u^2 v^2 (u^2 - v^2)^2 + 5 u^2 v^2(u^2 + v^2) + 8 u^2 v^2 \right] \nonumber \\
&  -\frac{3}{32 u^4 v^4} \left( 3(u^4-v^4)^2 - 6(u^4 +v^4) + 3 - 8u^2v^2(1 + u^2 + v^2) \right) \text{log}\left( \frac{1-(u-v)^2 }{(u+v)^2 -1} \right), \label{eq:j_h_late_def} \\
\mathcal Y_\text{h,late}(u,v) = & \frac{3 \pi }{32 u^4 v^4} \left( 3u^4(u^4 - 2) + 3v^4(v^4 -2) + 3 - 8u^2v^2(1 + u^2 + v^2) -6 u^4 v^4 \right). \label{eq:y_h_late_def}
\end{align}
The full expressions of $\mathcal J_\text{h}$ and $\mathcal Y_\text{h}$ are given in Appendix~\ref{app:ana_sol_rd}.
From Eq.~(\ref{eq:psi_so_ini_con}), we can express the contribution from the initial second-order perturbations as 
\begin{align}
	\Psi^\so (\bm k, 0) T_{\Phi,\rr}(x) = \int \frac{\dd^3 \tilde k}{(2\pi)^3} uv I_{\Psi,\rr,\text{i}}(u,v,x) \left( \frac{2}{3} \right)^2 \zeta(\tilde {\bm k}) \zeta(\bm k - \tilde {\bm k}),
\end{align}
where $I_{\Psi,\rr,\text{i}}$ is defined as
\begin{align}
	I_{\Psi,\rr,\text{i}}(u,v,x) &\equiv \mathcal J_\text{i}(u,v)T_{\Phi,\rr}(x), \\
	\mathcal J_\text{i}(u,v) &\equiv \frac{2(u^2+v^2) - 3(u^2 - v^2)^2 +5 -12 a_\text{NL}}{4uv}.
\end{align}
Combining all the contributions, we finally obtain the following expression:
\begin{align}
	\Psi^\so (\bm k,\eta) =& \int \frac{\dd^3 \tilde k}{(2\pi)^3} uv I_{\Psi,\rr}(u,v,x) \left( \frac{2}{3} \right)^2 \zeta(\tilde {\bm k}) \zeta(\bm k - \tilde {\bm k}),
	\label{eq:i_psi_rr_def}
\end{align}
where $I_{\Psi,\rr} \equiv I_{\Psi,\rr,\text{i}} + I_{\Psi,\rr,\sss}$.
This is the analytic solution of $\Psi^\so$.
Here, we mention the asymptotic behavior in the large-scale limit, $u \gg 1$ and $v \gg 1$\footnote{
The word ``large-scale limit'' here indicates the scales much larger than those of the source perturbations. On the other hand, the word ``late-time limit'' indicates the case where the induced perturbations (not the source perturbation) are on deep subhorizon scales. }, in which $\mathcal J_\text{i}$ can be approximated as
\begin{align}
	\mathcal J_{\text{i}}(u,v) \simeq \frac{2(u^2+v^2) - 3(u^2 - v^2)^2}{4uv} \sim \mathcal O(1) \qquad (u \gg 1, v \gg 1).
	\label{eq:j_i_large_limit}
\end{align}
When the induced perturbations are on superhorizon scales ($x \ll 1$), $|\mathcal J_\text{h}|$ and $|\mathcal J_0|$ are much smaller than $|\mathcal J_\text{i}|$ and therefore the leading contribution comes only from $\mathcal J_\text{i}$. 
On the other hand, when the induced perturbations are on deep subhorizon scales ($x \gg 1$), $\mathcal J_\text{h}$ asymptotes to $\mathcal J_\text{h,late}$, given as 
\begin{align}
	\mathcal J_\text{h,late}(u,v) \simeq \frac{-9(u^6+v^6) + 5 u^2 v^2 (u^2 + v^2)}{8u^3 v^3} + \frac{3(u^2 -v^2)^2}{4uv} \qquad (u \gg 1, v \gg 1).
	\label{eq:j_h_large_limit}	
\end{align}
Given $|u-v| < 1$ by definition, we can easily see that the cancellation between $\mathcal J_\text{i}$ and $\mathcal J_\text{h,late}$ occurs as $\mathcal J_{\text{i}} + \mathcal J_\text{h,late} \sim \mathcal O(1/u)$ or $\mathcal O(1/v)$.\footnote{In the large-scale limit ($u\gg 1$ and $v \gg 1$), $\mathcal J_0 \sim \mathcal O(1/u^2)$ or $\mathcal O(1/v^2)$ regardless of $x$ and therefore $\mathcal J_0$ is subdominant in the limit. }
In this sense, the contribution from $\mathcal J_\text{h}$, which was neglected in the previous work~\cite{Bartolo:2006fj}, is very important for the late-time ($x \gg 1$) evolution of the perturbations especially on the scales much larger than the source perturbations ($u \gg 1$ and $v \gg 1$).

Similarly to $\Psi^\so$, from Eq.~(\ref{eq:delta_so_exp}), we obtain the analytic solution of $\delta^\so$ as 
\begin{align}
	\delta^\so (\bm k,\eta) =& \int \frac{\dd^3 \tilde k}{(2\pi)^3} uv I_{\delta,\rr}(u,v,x) \left( \frac{2}{3} \right)^2 \zeta(\tilde {\bm k}) \zeta(\bm k - \tilde {\bm k}),
	\label{eq:i_delta_rr_def}
\end{align}
where 
\begin{align}
	I_{\delta,\rr}(u,v,x) \equiv& -2 \left(1+ \frac{x^2}{3} \right) I_{\Psi,\rr}(u,v,x) - 2 x \frac{\dd I_{\Psi,\rr}(u,v,x)}{\dd x} + I_{\delta,\rr,\sss}(u,v,x), \\
	I_{\delta,\rr,\sss}(u,v,x) \equiv& \frac{9 -4 x^2 + 6(u^2 + v^2)(3 - 2x^2) - 27 (u^2 - v^2)^2 }{6 uv} T_{\Phi,\rr}(ux) T_{\Phi,\rr}(vx) \nonumber \\
    &+ \frac{3 + 2x^2 + 2(3 - x^2) (u^2 + v^2) -9 (u^2 - v^2)^2}{6 uv}  \nonumber \\
     & \qquad \qquad \times \left( ux \frac{\dd T_{\Phi,\rr}(ux)}{\dd(ux)} T_{\Phi,\rr}(vx) + vx T_{\Phi,\rr}(ux) \frac{\dd T_{\Phi,\rr}(vx)}{\dd (vx)} \right) \nonumber \\
    & + \frac{x^2 \left[15 + 2x^2 + 2(3 -x^2)(u^2 + v^2) -9 (u^2 - v^2)^2 \right]}{6}  \frac{\dd T_{\Phi,\rr}(ux)}{\dd (ux)} \frac{\dd T_{\Phi,\rr}(vx)}{\dd (vx)}.
\end{align}

For the second-order curvature perturbation, Eq.~(\ref{eq:zeta_so_def}) leads to 
\begin{align}
	\zeta^\so (\bm k,\eta) =& \int \frac{\dd^3 \tilde k}{(2\pi)^3} uv I_{\zeta,\rr}(u,v,x) \left( \frac{2}{3} \right)^2 \zeta(\tilde {\bm k}) \zeta(\bm k - \tilde {\bm k}),
	\label{eq:i_zeta_rr_def}
\end{align}
where 
\begin{align}
	I_{\zeta,\rr}(u,v,x) \equiv& - I_{\Psi,\rr} (u,v,x) + \frac{1}{4} I_{\delta,\rr}(u,v,x)  + I_{\zeta,\rr,\sss}(u,v,x), \\
	I_{\zeta,\rr,\sss}(u,v,x) \equiv& -\frac{1}{uv} \left\{ \frac{1}{8} T_{\delta,\rr}(vx) T_{\delta,\rr}(ux) - \frac{1}{16} \left(vx \frac{\dd T_{\delta,\rr}(v x)}{\dd (v x)} T_{\delta,\rr}(ux) + ux \frac{\dd T_{\delta,\rr}(u x)}{\dd (u x)}T_{\delta,\rr}(vx)   \right) \right. \nonumber\\
	&\qquad  \ \  \ \ 
	+ \frac{1}{4} \left[ T_{\delta,\rr}(vx) \left( ux \frac{\dd T_{\Phi,\rr}(ux)}{\dd (ux)} + 2 T_{\Phi,\rr}(ux) \right) + T_{\delta,\rr}(ux) \left( vx \frac{\dd T_{\Phi,\rr}(vx)}{\dd (vx)} + 2 T_{\Phi,\rr}(vx) \right) \right]  \nonumber \\
	& \qquad \ \ \ \ 
	\left. - \frac{x^2}{32} \left[ \frac{(u^2 - v^2)^2 - 2(u^2 + v^2) + 1}{4} \right]  T_{\delta,\rr}(vx) T_{\delta,\rr}(ux) \right\}.
\end{align}
Note that $I_{\Psi,\rr}$, $I_{\delta,\rr}$, and $I_{\zeta,\rr}$ are symmetric under the exchange $u \leftrightarrow v$.\footnote{
Strictly speaking, we can define $I_{\Psi,\rr}$, $I_{\delta,\rr}$, and $I_{\zeta,\rr}$ so that they are not symmetric under the exchange $u \leftrightarrow v$. 
However, after the integration over $\tilde k$, only the symmetric part contributes to the second-order perturbations.
}

\begin{figure}[htbp]
  \begin{center}
    \begin{tabular}{c}

      \begin{minipage}{0.45\textwidth}
        \begin{center}
          \includegraphics[width=\hsize]{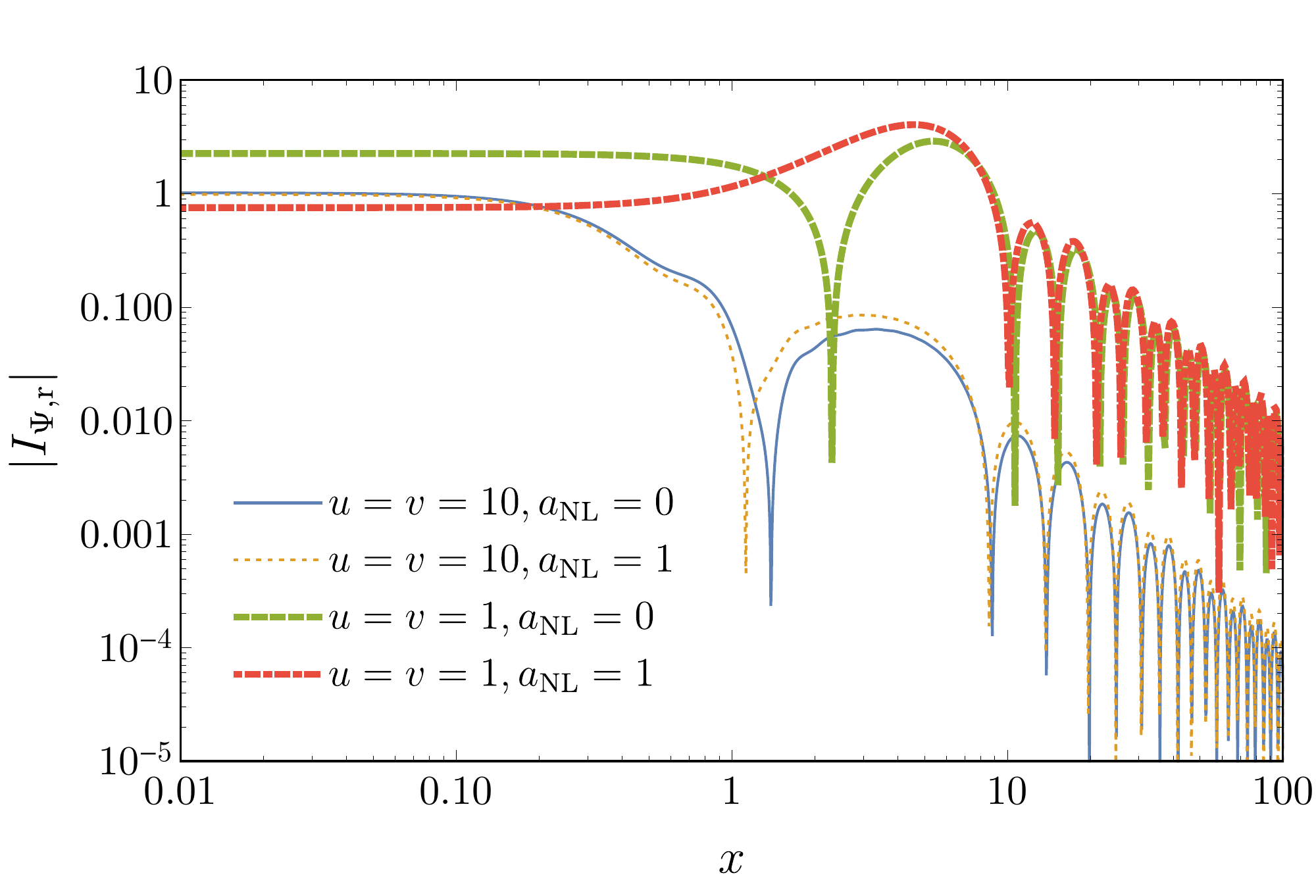}
        \end{center}
      \end{minipage}
	\hspace{0.5cm} 
      \begin{minipage}{0.45\textwidth}
        \begin{center}
          \includegraphics[width=\hsize]{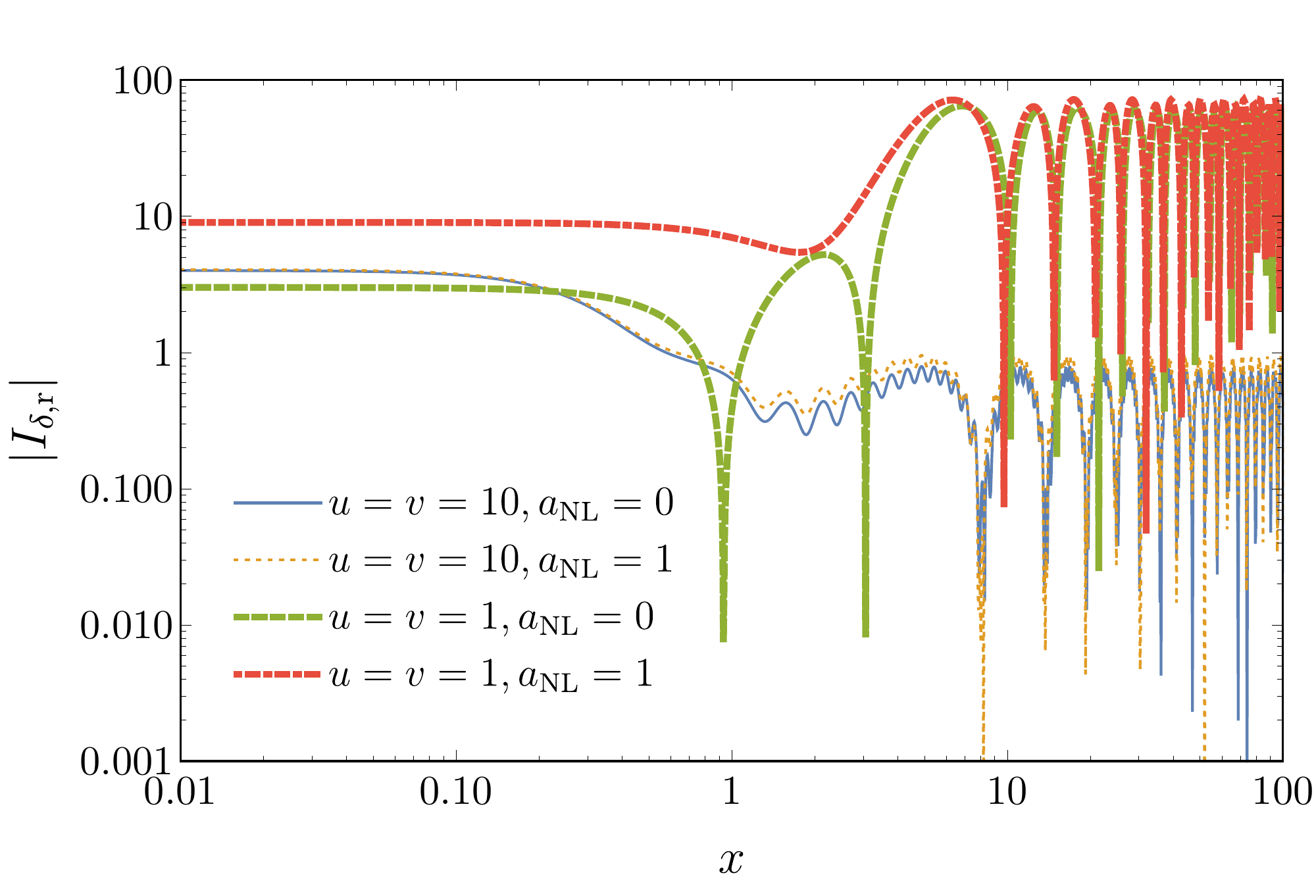}
        \end{center}
      \end{minipage}\\

      \begin{minipage}{0.45\textwidth}
      \vspace{0.6cm}
        \begin{center}
          \includegraphics[width=\hsize]{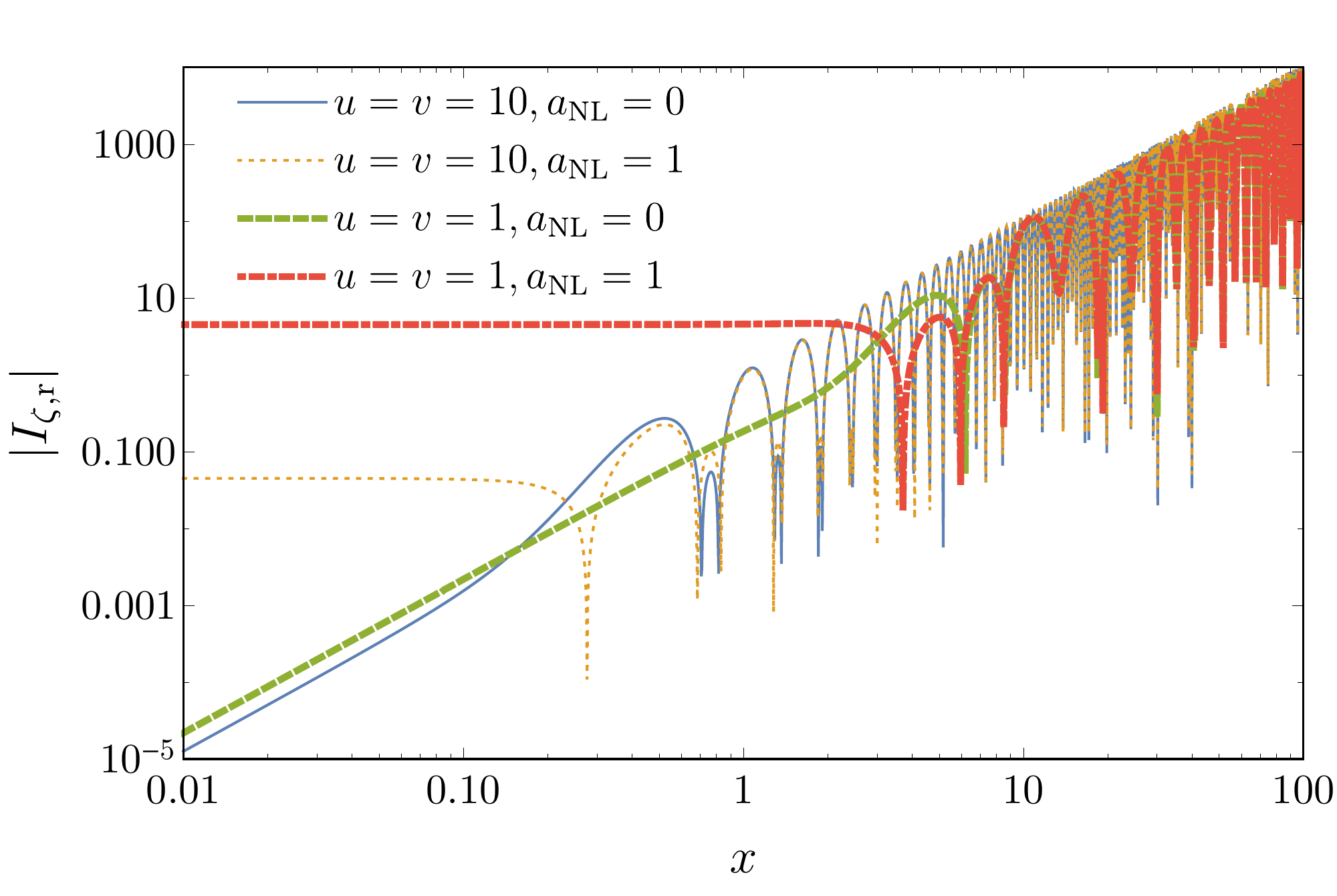}
        \end{center}
      \end{minipage}

    \end{tabular}
    \caption{\small	
    Evolution of $|I_{\Psi,\rr}|$ (top left), $|I_{\delta,\rr}|$ (top right), and $|I_{\zeta,\rr}|$ (bottom), which are defined in Eqs.~(\ref{eq:i_psi_rr_def}), (\ref{eq:i_delta_rr_def}), and (\ref{eq:i_zeta_rr_def}), respectively.
	}
    \label{fig:i_functions_rr}
  \end{center}
\end{figure}

Figure~\ref{fig:i_functions_rr} shows the evolution of the transfer functions, $I_{\Psi,\rr}$, $I_{\delta,\rr}$, and $I_{\zeta,\rr}$, with different $a_\text{NL}$, $u$, and $v$.
The lines for $u=v=1$ show the contributions from the perturbations whose wavenumbers are of the same order as those of the induced scalar perturbations.
On the other hand, the lines for $u=v=10$ show the contributions from the perturbations whose wavenumbers are larger than those of the induced scalar perturbations.
The time $x \sim 1/u$ and $1/v$ corresponds to the horizon entry of the source (or first-order) perturbations and $x \sim 1$ corresponds to that of the induced (or second-order) perturbations.
For $a_\text{NL}$, we take $0$ and $1$ as fiducial examples, which correspond to the perturbations with and without the non-Gaussianities, respectively.

From this figure, we can see that, before the source perturbations enter the horizon ($x \ll 1/u$ and $1/v$), the induced perturbations remain constant except for $\zeta^\so$ with $a_\text{NL} = 0$, which grows even before that.
Note that this behavior of $\zeta^\so$ for $a_\text{NL} = 0$ does not contradict Ref.~\cite{Malik:2003mv} because the reference neglects the spatial derivative of the first-order perturbations.
This behavior just shows the fact that, even if we take the initial condition $\zeta^\so=0$ in the superhorizon limit, $\zeta^\so$ can evolve through the terms suppressed by the gradient factors such as $u k \eta$ and $v k \eta$ ($\ll 1$).

$\Psi^\so$ and $\delta^\so$ start to decay around $x \sim 1/u$ and $1/v$ and oscillate around $x \sim 1$.
On the other hand, $\zeta^\so$ starts to oscillate around $x \sim 1/u$ and $1/v$.
In the late-time ($x \gg 1$), all the second-order perturbations in Fig.~\ref{fig:i_functions_rr} oscillate and their oscillation amplitudes are related to the scale factor as $I_{\Psi,\rr} \propto a^{-2}$, $I_{\delta,\rr} \propto a^0$, and $I_{\zeta,\rr} \propto a^2$.
For $I_{\Psi,\rr}$ and $I_{\delta,\rr}$, their scale factor dependences in the late-time limit are the same as those of the first-order perturbations, $\Psi^\fo \propto a^{-2}$ and $\delta^\fo \propto a^{0}$ up to their oscillations.
On the other hand, $I_{\zeta,\rr}$ evolves differently from the first-order curvature perturbation in the late-time limit (c.f. $\zeta^\fo \propto a^{0}$ up to its oscillation).

\subsection{Auto-power spectra}

In this subsection, we calculate the auto-power spectra of the second-order induced perturbations to see the evolution of the second-order perturbations.
Note that, to compare with observations, we also need to take into account the cross-power spectra between the first-order and the third-order perturbations, which is of the same order of magnitude as the auto-power spectra.
Since the main focus of this paper is the analytic solution of the second-order perturbations, we leave the study of the cross-power spectra for future work.
For this reason, we should keep in mind that the observable power spectra might be different from the auto-power spectra.
The main purpose of the auto-power spectrum in this paper is to show the evolution of the second-order perturbations.

The expectation values of the second-order perturbations can be written as 
\begin{align}
	\vev{\mathcal M^\so ({\bm k}_1,\eta) \mathcal M^\so({\bm k}_2,\eta) } = \int \frac{\dd^3 \tilde k_1}{(2\pi)^3} \int \frac{\dd^3 \tilde k_2}{(2\pi)^3} 
	&(u_1 v_1) (u_2 v_2)  I_{\mathcal M,\rr}(u_1,v_1,x_1) I_{\mathcal M,\rr}(u_2,v_2,x_2) \nonumber \\
	&\times \left( \frac{2}{3} \right)^4 \vev{ \zeta(\tilde {\bm k}_1) \zeta(\bm k_1 - \tilde {\bm k}_1) \zeta(\tilde {\bm k}_2) \zeta(\bm k_2 - \tilde {\bm k}_2) },
	\label{eq:yy_vev}
\end{align}
where $\mathcal M$ indicates $\Psi$, $\delta$, or $\zeta$. The arguments are defined as $v_1 \equiv \tilde{k}_1/k_1$, $u_1 \equiv |\bm k_1 - \tilde{\bm k}_1|/k_1$, $x_1 \equiv k_1\eta$, and similarly for the arguments with the subscript $2$.
Here, we assume that the non-Gaussianity of the primordial curvature perturbation is negligibly small for simplicity. 
Then, we obtain
\begin{align}
\vev{\zeta (\tilde{\bm k}_1)  \zeta ( \bm k_1- \tilde{\bm k}_1) \zeta (\tilde{\bm k}_2) \zeta ( {\bm k}_2- \tilde{\bm k}_2)} 
&= \vev{ \zeta (\tilde{\bm k}_1)  \zeta (\tilde{\bm k}_2)} \vev{   \zeta ( \bm k_1- \tilde{\bm k}_1) \zeta ( {\bm k}_2 - \tilde{\bm k}_2)} \nonumber \\
& \quad + \vev{ \zeta (\tilde{\bm k}_1)  \zeta ( {\bm k}_2 - \tilde{\bm k}_2)} \vev{   \zeta ( \bm k_1- \tilde{\bm k}_1 ) \zeta (\tilde{\bm k}_2)} \nonumber \\
&= (2\pi)^3\delta(\tilde{\bm k}_1 + \tilde{\bm k}_2) (2\pi)^3\delta(\bm k_1 - \tilde{\bm k}_1 + {\bm k}_2 - \tilde{\bm k}_2) \nonumber \\
& \qquad \qquad \times \frac{2\pi^2}{\tilde{k}^3_1} \mathcal P_{\zeta^\fo}(\tilde k_1) \frac{2\pi^2}{|{\bm k}_1 - \tilde{\bm k}_1|^3} \mathcal P_{\zeta^\fo}(|{\bm k}_1 - \tilde{\bm k}_1|)\nonumber \\
& \quad + (2\pi)^3\delta(\tilde{\bm k}_1 +  {\bm k}_2 - \tilde{\bm k}_2) (2\pi)^3\delta(\bm k_1 - \tilde{\bm k}_1 + \tilde{\bm k}_2) \nonumber \\
& \qquad \qquad \times \frac{2\pi^2}{\tilde{k}^3_1} \mathcal P_{\zeta^\fo}(\tilde k_1) \frac{2\pi^2}{|{\bm k}_1 - \tilde{\bm k}_1|^3} \mathcal P_{\zeta^\fo}(|{\bm k}_1 - \tilde{\bm k}_1|),
\label{eq:phi4_phi22}
\end{align}
where we have restored the superscript of the curvature perturbation in the power spectrum, given as
\begin{align}
	\vev{\zeta (\bm k_1) \zeta (\bm k_2)} = (2\pi)^3 \delta(\bm k_1 + \bm k_2) \frac{2 \pi^2}{k^3_1} \mathcal P_{\zeta^\fo}(k_1).
\end{align}
Then, we can rewrite Eq.~(\ref{eq:yy_vev}) as 
\begin{align}
	\label{eq:yy_vev_uv}
	&\vev{\mathcal M^\so ({\bm k}_1,\eta) \mathcal M^\so({\bm k}_2,\eta) } = (2\pi)^3 \delta(\bm k_1 + \bm k_2) \frac{2 \pi^2}{k^3_1} \mathcal P_{\mathcal M^\so}(k_1,\eta), \\
	\label{eq:p_mathcal_m}
	&\mathcal P_{{\mathcal M}^\so}(k,\eta) \equiv \int_0^\infty \dd v \, \int_{|v-1|}^{v+1} \dd u \, I_{\mathcal M,\rr}^2(u,v,x) \left( \frac{2}{3} \right)^4 \mathcal P_{\zeta^\fo}( k v) \mathcal P_{\zeta^\fo}(k u).
\end{align}

\begin{figure}[htbp]
  \begin{center}
    \begin{tabular}{c}

      \begin{minipage}{0.45\textwidth}
        \begin{center}
          \includegraphics[width=\hsize]{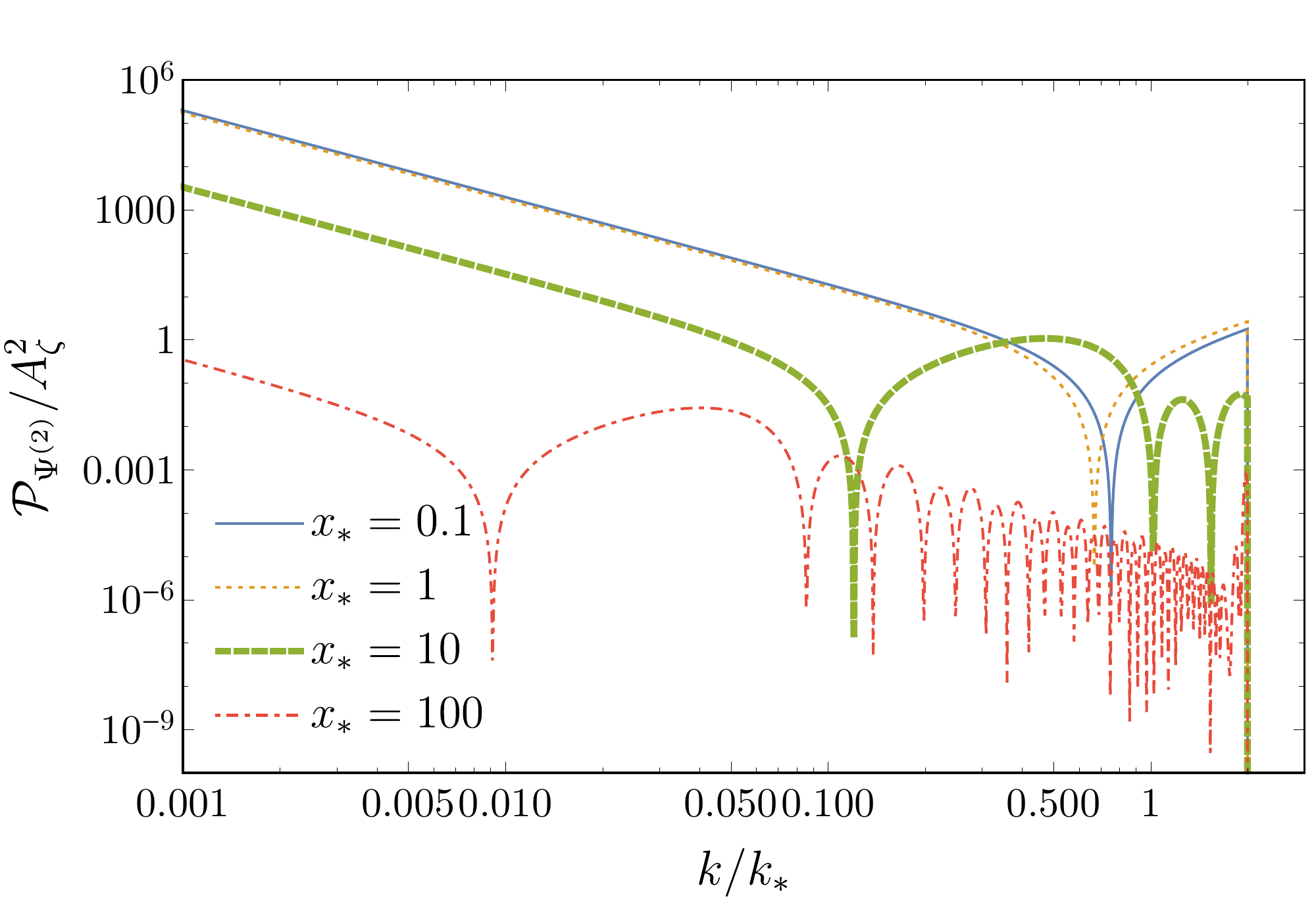}
        \end{center}
      \end{minipage}
	\hspace{0.5cm} 
      \begin{minipage}{0.45\textwidth}
        \begin{center}
          \includegraphics[width=\hsize]{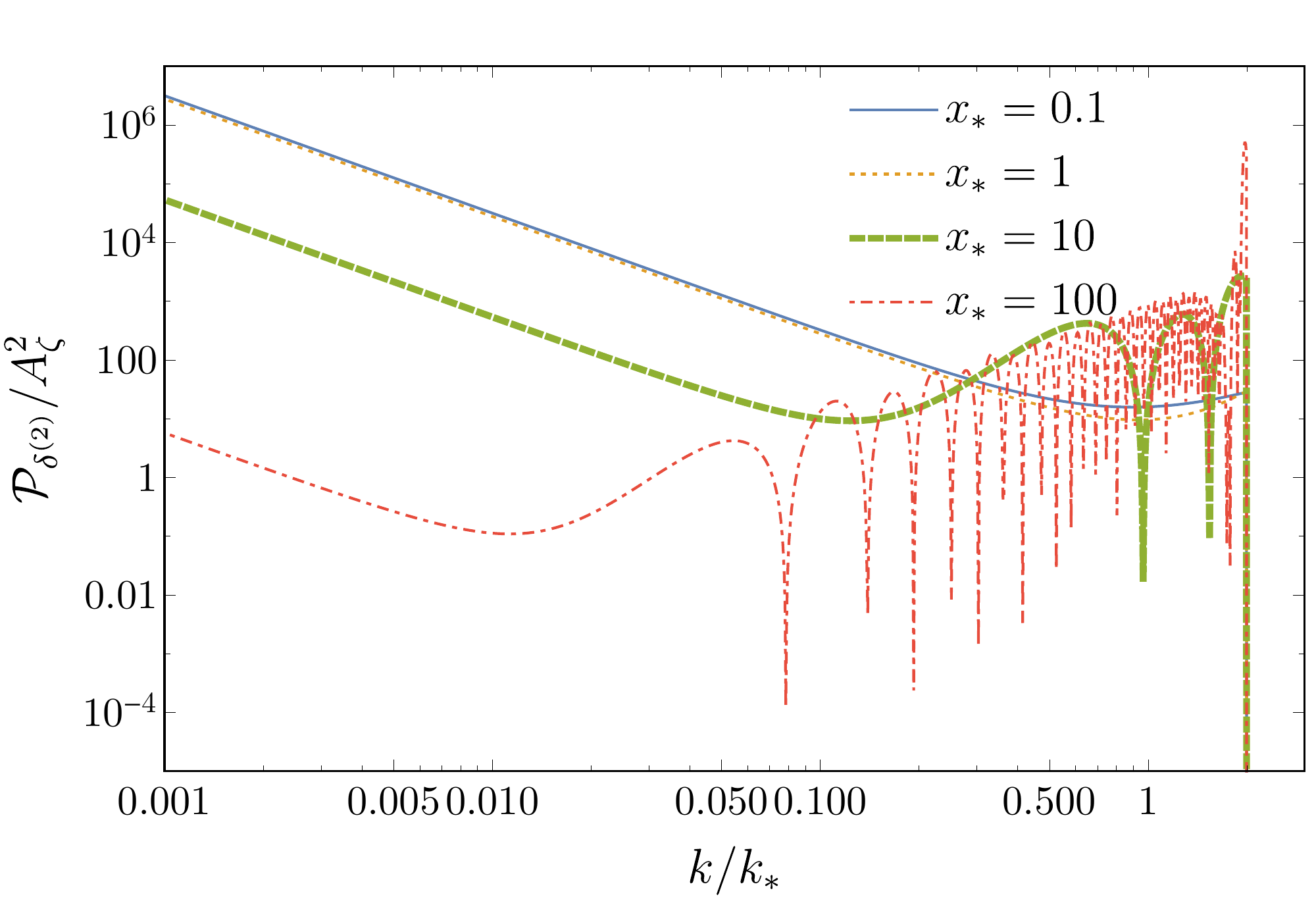}
        \end{center}
      \end{minipage}\\

      \begin{minipage}{0.57\textwidth}
      \vspace{0.6cm}
        \begin{center}
          \includegraphics[width=\hsize]{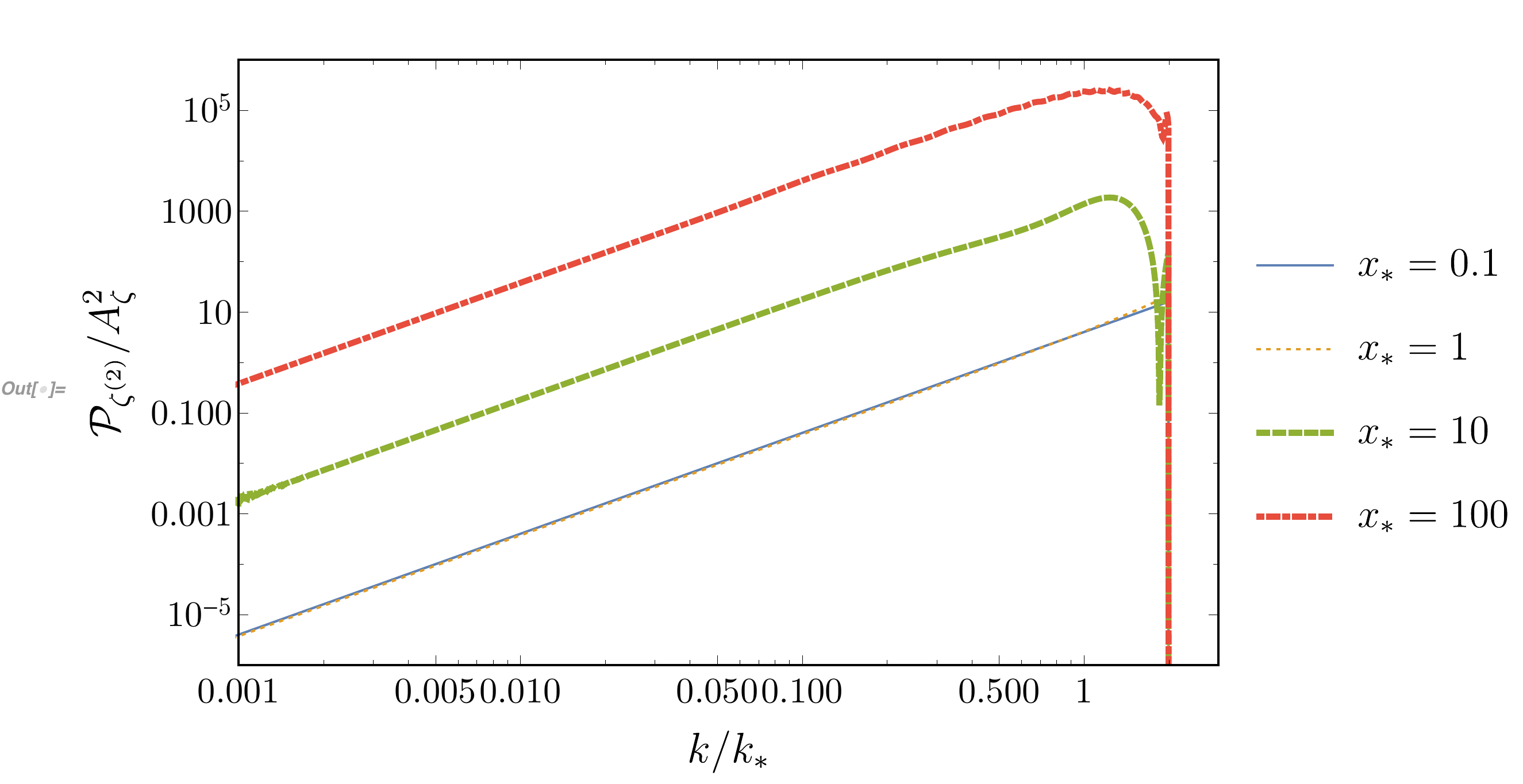}
        \end{center}
      \end{minipage}

    \end{tabular}
    \caption{\small	
    Evolution of $\mathcal P_{\Psi^\so}$ (top left), $\mathcal P_{\delta^\so}$ (top right), and $\mathcal P_{\zeta^\so}$ (bottom) in a RD era with the delta-function power spectrum of the first-order curvature perturbation, defined by Eq.~(\ref{eq:delta_func_pzeta}).
    All of the power spectra are normalized by $A_\zeta^2$.
    We use $x_* \equiv k_* \eta$ as a time variable and take $a_\text{NL} = 1$.
    The relation between the power spectra and the function $I$ is given by Eq.~(\ref{eq:rel_p_so_i_func}).
	}
    \label{fig:p_delta_rr}
  \end{center}
\end{figure}

As a simple example, we here consider the delta-function power spectrum of curvature perturbation
\begin{align} 
	\mathcal P_{\zeta^\fo}(k) = A_\zeta \delta(\text{log}(k/k_*)).
	\label{eq:delta_func_pzeta}
\end{align}
In this case, we can express Eq.~(\ref{eq:p_mathcal_m}) as 
\begin{align}
	\mathcal P_{{\mathcal M}^\so}(k,\eta) =  A_\zeta^2 \left( \frac{2}{3} \right)^4 \left( \frac{k_*}{k} \right)^2 I^2_{\mathcal M,\rr}(k_*/k,k_*/k,x) \Theta(2- k/k_*).
	\label{eq:rel_p_so_i_func}
\end{align}
Figure~\ref{fig:p_delta_rr} shows the evolution of Eq.~(\ref{eq:rel_p_so_i_func}) for $\Psi^\so$, $\delta^\so$, and $\zeta^\so$.
We assume the Gaussian distribution of the curvature perturbations taking $a_\text{NL} = 1$.
The power spectra of the three perturbations remain constant when the source perturbations at $k_*$ are on superhorizon scale ($x_* (\equiv k_* \eta) \lesssim 1$).
Once the source perturbations enter the horizon, the power spectra start to evolve even on superhorizon scales at that time through the non-linear interaction between the perturbations with different wavenumbers.
We can also interpret the evolution on superhorizon scale as the cancellation between $\mathcal J_\text{i}$ and $\mathcal J_\text{h}$, mentioned below Eq.~(\ref{eq:j_h_large_limit}).
See also Appendix~\ref{app:power_rd} for the auto-power spectra in a somewhat more realistic situation, which is often considered in the PBH scenarios.

\section{Induced scalar perturbations in a matter-dominated era}
\label{sec:solution_for_md}

In this section, we discuss the induced scalar perturbations in a MD era ($c_s^2 = w = 0$).

\subsection{Analytic solutions}

First, we derive the analytic solutions.
The equations in a MD era are somewhat more simple than those in a RD era because ${\Phi^\fo}' = 0$ during the era.
From Eq.~(\ref{eq:psi_so_eom_re}), we have
\begin{align}
	{\Psi^\so}''(\bm k,\eta) + \frac{6}{\eta} {\Psi^\so}'(\bm k,\eta) = S^\so_{\text{m}}(\bm k,\eta),
	\label{eq:psi_eom_md}
\end{align}
where we have substituted $\mathcal H = 2/\eta$ and the source term is given as 
\begin{align}
	S^\so_\mm (\bm k,\eta) 
	=&   \int \frac{\dd^3 \tilde k}{(2\pi)^3}  \left[ \tilde {\bm k} \cdot (\bm k - \tilde {\bm k}) 
	- \frac{10}{3k^2}   ( \bm k \cdot \tilde {\bm k} ) ( \bm k \cdot (\bm k - \tilde {\bm k} ) ) \right] \Phi(\tilde {\bm k}) \Phi(\bm k - \tilde {\bm k}) \nonumber \\
	=&
	\int \frac{\dd^3 \tilde k}{(2\pi)^3} k^2 uv f_\mm(u,v) \left( \frac{3}{5} \right)^2 \zeta(\tilde {\bm k}) \zeta(\bm k - \tilde {\bm k}).
	 \label{eq:s_k_md_fourier} 
\end{align}
The source function $f_\mm$ is defined as 
\begin{align}
	f_\mm(u,v) \equiv& - \frac{2 + 3 (u^2 +v^2) - 5 (u^2 -v^2)^2}{6\, u v}.
\end{align}
We define the Green function, which is the solution of the following equation:
\begin{align}
	G''_\mm + \frac{6}{\eta} G_\mm' = \delta(\eta - \bar \eta).
\end{align}
Concretely, we can write down the Green function as
\begin{align}
	&G_\mm = \Theta(\eta - \bar \eta) \frac{{\bar\eta}^6}{5} \left[ \frac{1}{\bar \eta^5} -\frac{1}{\eta^5} \right] \nonumber \\
	\Rightarrow \ 
	&k G_\mm =  \Theta(\eta - \bar \eta) \frac{{\bar x}^6}{5} \left[ \frac{1}{{\bar x}^5} -\frac{1}{x^5} \right].
\end{align}
Using this Green function, we can solve Eq.~(\ref{eq:psi_eom_md}) as
\begin{align}
	\Psi^\so ({\bm k},\eta) =& \Psi^\so(\bm k, 0) + \int^\eta_0 \dd \bar \eta \, G_\mm(k,\eta;\bar \eta) S^\so_\mm ({\bm k},\eta) \nonumber \\
	=&  \Psi^\so(\bm k, 0) + \int \frac{\dd^3 \tilde k}{(2\pi)^3} uv I_{\Psi,\mm,\sss}(u,v,x) \left( \frac{3}{5} \right)^2 \zeta(\tilde {\bm k}) \zeta(\bm k - \tilde {\bm k}), \label{eq:psi_so_sol_md}
\end{align}
where $I_{\Psi,\mm,\sss}$ is defined as 
\begin{align}
	I_{\Psi,\mm,\sss}(u,v,x) \equiv & \int^x_0 \dd \bar x\,  kG_\mm (k,\eta; \bar \eta) f_\mm(u,v) \nonumber \\
	=& \frac{x^2}{14} f_\mm(u,v).
\end{align}
Note that this solution of $\Psi^\so$ was already obtained in Ref.~\cite{Bartolo:2006fj}.
From Eq.~(\ref{eq:psi_so_ini_con}), we can obtain the initial condition for $\Psi^\so$ as
\begin{align}
	\Psi^\so(\bm k, 0) = \int \frac{\dd^3 \tilde k}{(2\pi)^3} uv I_{\Psi,\mm,\text{i}}(u,v) \left( \frac{3}{5} \right)^2 \zeta(\tilde {\bm k}) \zeta(\bm k - \tilde {\bm k}),
\end{align}
where 
\begin{align}
	I_{\Psi,\text{m,i}}(u,v) &\equiv \frac{2(u^2+v^2) - 3(u^2 - v^2)^2 +5 - 10 a_\text{NL}}{3uv}.
\end{align}
Finally, we can rewrite Eq.~(\ref{eq:psi_so_sol_md}) as
\begin{align}
	\Psi^\so ({\bm k},\eta) =&  \int \frac{\dd^3 \tilde k}{(2\pi)^3} uv I_{\Psi,\mm}(u,v,x) \left( \frac{3}{5} \right)^2 \zeta(\tilde {\bm k}) \zeta(\bm k - \tilde {\bm k}), 
	\label{eq:psi_so_final_exp_mm}
\end{align}
where $I_{\Psi,\mm} \equiv I_{\Psi,\mm,\text{i}} + I_{\Psi,\mm,\sss}$.

For the energy density perturbation, Eq.~(\ref{eq:delta_so_exp}) leads to the analytic solution given as
\begin{align}
	\delta^\so (\bm k,\eta) =& \int \frac{\dd^3 \tilde k}{(2\pi)^3} uv I_{\delta,\mm}(u,v,x) \left( \frac{3}{5} \right)^2 \zeta(\tilde {\bm k}) \zeta(\bm k - \tilde {\bm k}), \label{eq:delta_so_final_exp_mm} 
\end{align}
where 
\begin{align}
	I_{\delta,\mm}(u,v,x) \equiv& -\left(2+ \frac{x^2}{6} \right) I_{\Psi,\mm}(u,v,x) - x \frac{\dd I_{\Psi,\mm}(u,v,x)}{\dd x} + I_{\delta,\mm,\sss}(u,v,x), \\
	I_{\delta,\mm,\sss}(u,v,x) \equiv& \frac{1}{uv} \left[ - \frac{2 x^2}{3} \left( u^2 + v^2 \right) 
	 - \frac{5x^2}{18} \frac{1-u^2-v^2}{2} 
	 + \frac{20}{3} \left( \frac{2(u^2+v^2) - 3(u^2 - v^2)^2 +1}{4} \right)  \right].
\end{align}
In the late-time limit, we obtain $I_{\delta,\mm}(u,v,x) \simeq -\frac{x^4}{84} f_\mm (u,v)$, which is consistent with Refs.~\cite{Suto:1990wf,Makino:1991rp}.

Similarly, from Eq.(\ref{eq:zeta_so_def}), we obtain the analytic solution of the curvature perturbations as 
\begin{align}
	\zeta^\so (\bm k,\eta) =& \int \frac{\dd^3 \tilde k}{(2\pi)^3} uv I_{\zeta,\mm}(u,v,x) \left( \frac{3}{5} \right)^2 \zeta(\tilde {\bm k}) \zeta(\bm k - \tilde {\bm k}),\label{eq:zeta_so_final_exp_mm}
\end{align}
where 
\begin{align}
	I_{\zeta,\mm}(u,v,x) \equiv& - I_{\Phi,\mm} (u,v,x) + \frac{1}{3} I_{\delta,\mm}(u,v,x)  + I_{\zeta,\mm,\sss}(u,v,x), \\
	I_{\zeta,\mm,\sss}(u,v,x) \equiv& -\frac{1}{uv}  \left\{ \frac{1}{9}  T_{\delta_\mm}(vx) T_{\delta_\mm}(ux)  - \frac{1}{18} \left(vx \frac{\dd T_{\delta_\mm}(v x)}{\dd (v x)} T_{\delta_\mm}(ux) + ux \frac{\dd T_{\delta_\mm}(u x)}{\dd (u x)}T_{\delta_\mm}(vx)   \right) \right. \nonumber\\
	&\qquad  \ \ 
	+ \frac{2}{3} \left( T_{\delta_\mm}(vx) + T_{\delta_\mm}(ux) \right) \nonumber \\
	& \qquad \ \ 
	\left. - \frac{x^2}{72} \left[ \frac{(u^2 - v^2)^2 - 2(u^2 + v^2) + 1}{4} \right]  T_{\delta_\mm}(vx) T_{\delta_\mm}(ux) \right\}.
\end{align}
Note that we have substituted $T_{\Psi_\mm}(x)=1$.
After the source perturbations enter the horizon ($ux \gg 1$ and $vx \gg 1$), $I_{\zeta,\mm}$ can be approximated as 
\begin{align}
	I_{\zeta,\mm}(u,v,x) \simeq \frac{x^6 uv}{10368} \left( (u^2 - v^2)^2 - 2(u^2 + v^2) + 1 \right).
\end{align}

\begin{figure}[htbp]
  \begin{center}
    \begin{tabular}{c}

      \begin{minipage}{0.45\textwidth}
        \begin{center}
          \includegraphics[width=\hsize]{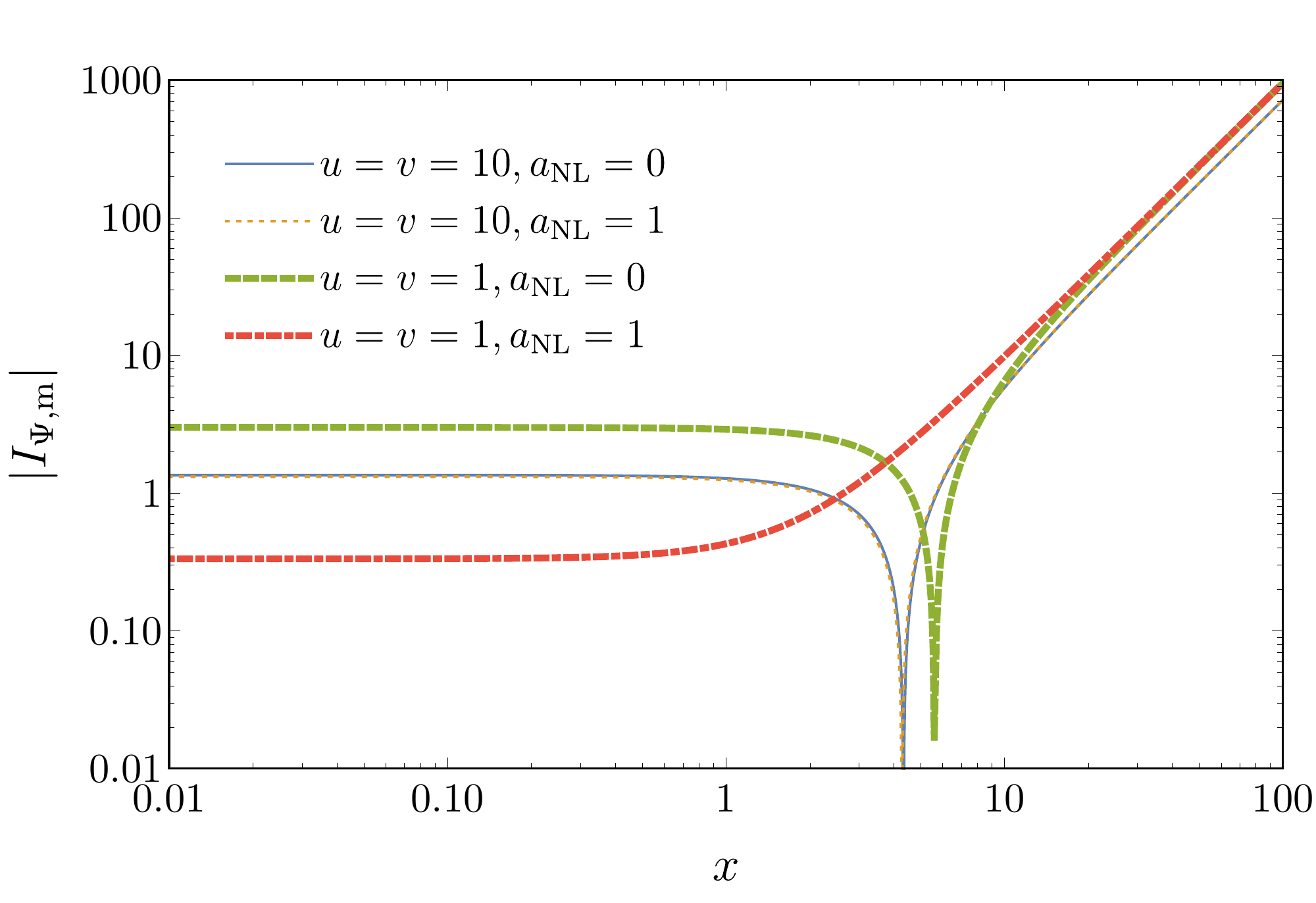}
        \end{center}
      \end{minipage}
	\hspace{0.5cm} 
      \begin{minipage}{0.45\textwidth}
        \begin{center}
          \includegraphics[width=\hsize]{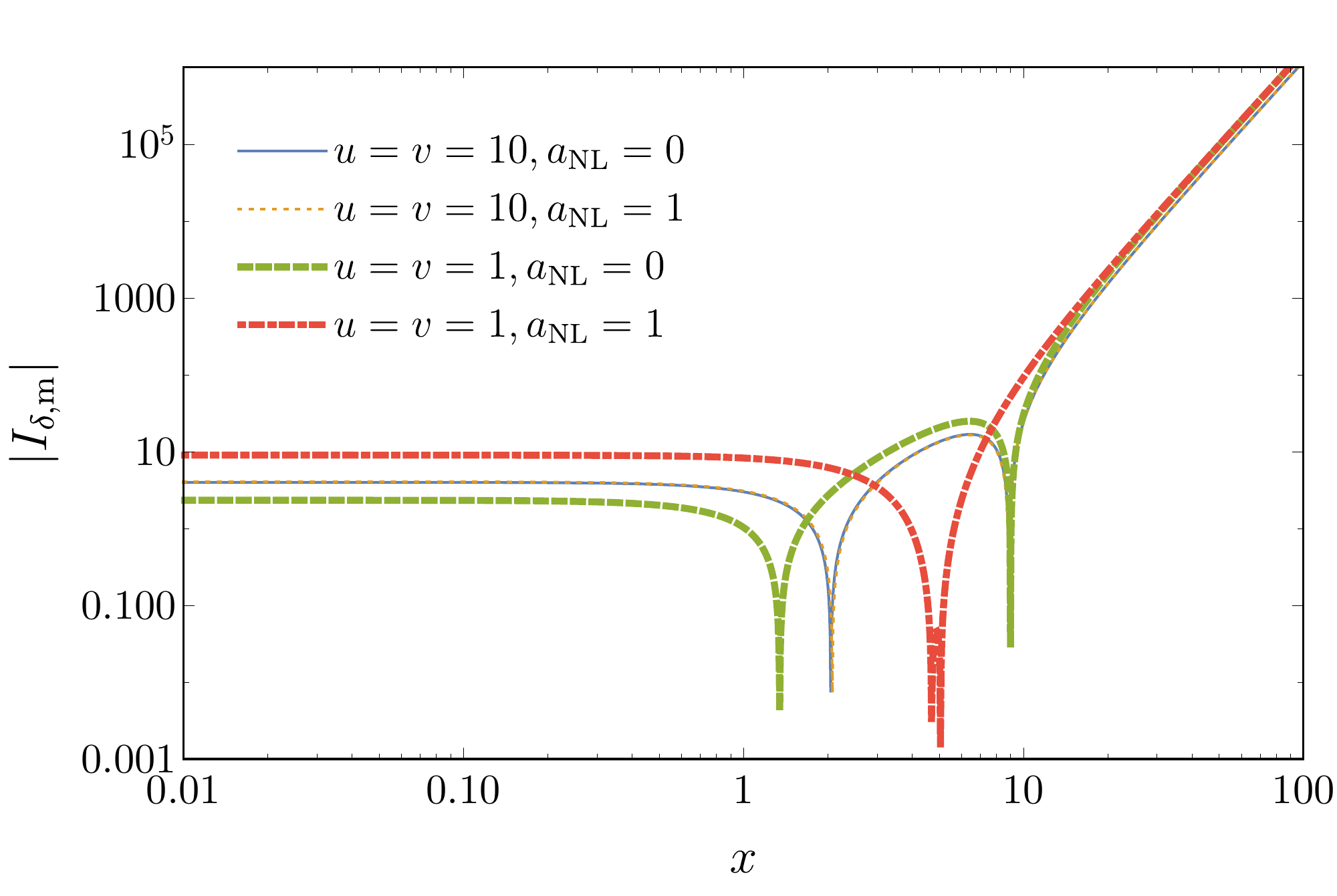}
        \end{center}
      \end{minipage}\\

      \begin{minipage}{0.45\textwidth}
      \vspace{0.6cm}
        \begin{center}
          \includegraphics[width=\hsize]{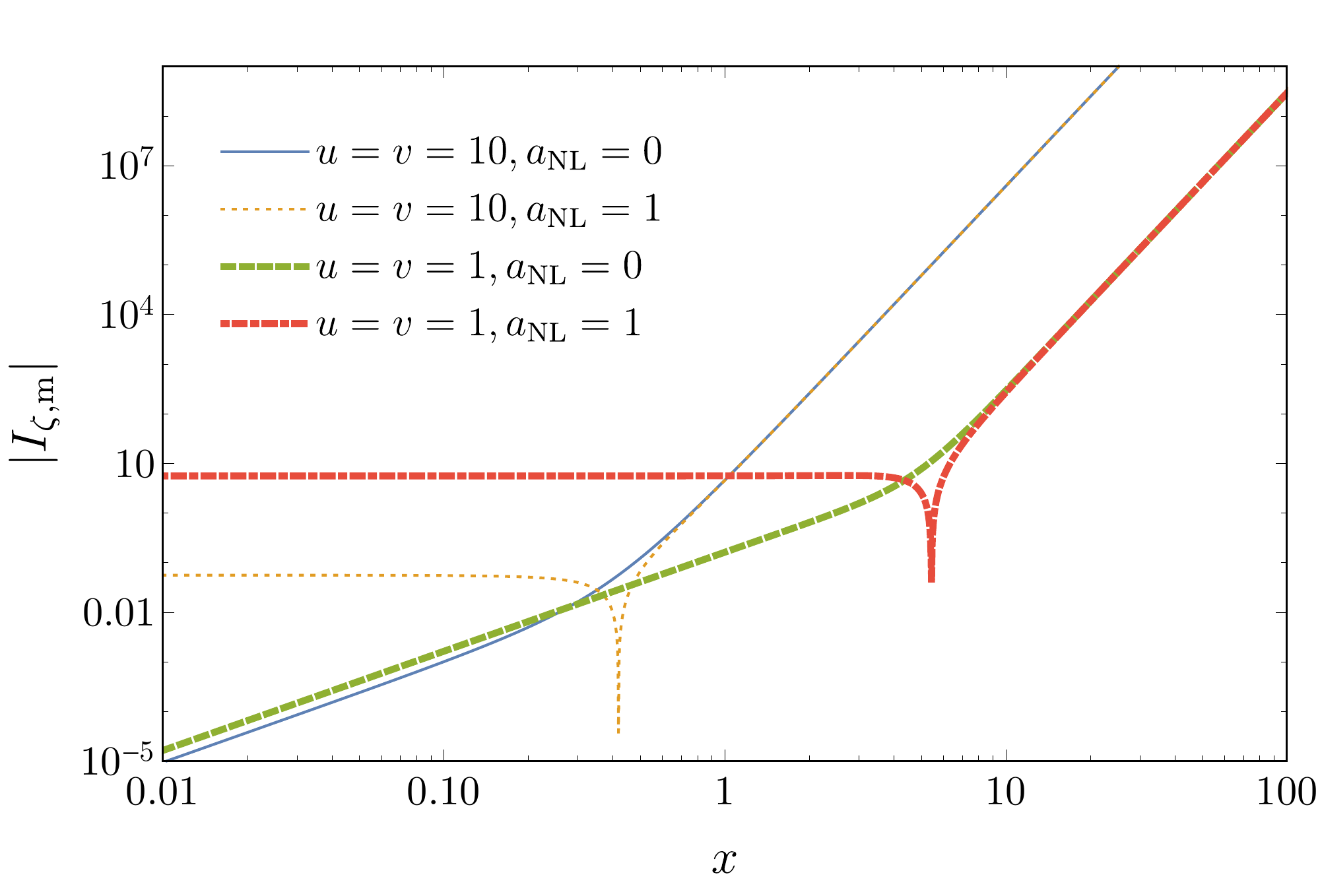}
        \end{center}
      \end{minipage}

    \end{tabular}
    \caption{\small	
        Evolution of $|I_{\Psi,\mm}|$ (top left), $|I_{\delta,\mm}|$ (top right), and $|I_{\zeta,\mm}|$ (bottom), which are defined in Eqs.~(\ref{eq:psi_so_final_exp_mm}), (\ref{eq:delta_so_final_exp_mm}), and (\ref{eq:zeta_so_final_exp_mm}), respectively.
	}
    \label{fig:i_functions_mm}
  \end{center}
\end{figure}

Figure~\ref{fig:i_functions_mm} shows the evolution of $I_{\Psi,\mm}$, $I_{\delta,\mm}$, and $I_{\zeta,\mm}$.
We can see that $I_{\Psi,\mm}$ and $I_{\delta,\mm}$ are constant on superhorizon scales ($x \ll 1$) and start to grow when the induced perturbations enter the horizon ($x \simeq 1$). 
In the late time ($x \gg 1$), $I_{\Psi,\mm}$ and $I_{\delta,\mm}$ are proportional to $x^2 (\propto a)$ and $x^4 (\propto a^2)$, respectively.
We can also see that the values in the late time do not depend on $u$, $v$, and $a_\text{NL}$ so much.
As for $I_{\zeta,\mm}$, it is constant for the case of $a_\text{NL} = 1$ and on the other hand grows proportionally to $x^2$ for $a_\text{NL} = 0$ before the source perturbations enter the horizon ($x \ll 1/u$ and $1/v$).
These behaviors are similar to those in a RD era, discussed in the previous section.
After the source perturbations enter the horizon ($x \gtrsim 1/u$ and $1/v$),  $I_{\zeta,\mm}$ grows proportionally to $x^6 (\propto a^3)$ regardless of the value of $a_\text{NL}$.
From these observations, we conclude that the behaviors of the second-order perturbations, $\Psi^\so$, $\delta^\so$, and $\zeta^\so$, in the late time ($x \gg 1$) are different from those of the first-order perturbations, $\Psi^\fo \propto a^0$, $\delta^\fo \propto a$, and $\zeta^\fo \propto a$.

\subsection{Auto-power spectra}

\begin{figure}[htbp]
  \begin{center}
    \begin{tabular}{c}

      \begin{minipage}{0.45\textwidth}
        \begin{center}
          \includegraphics[width=\hsize]{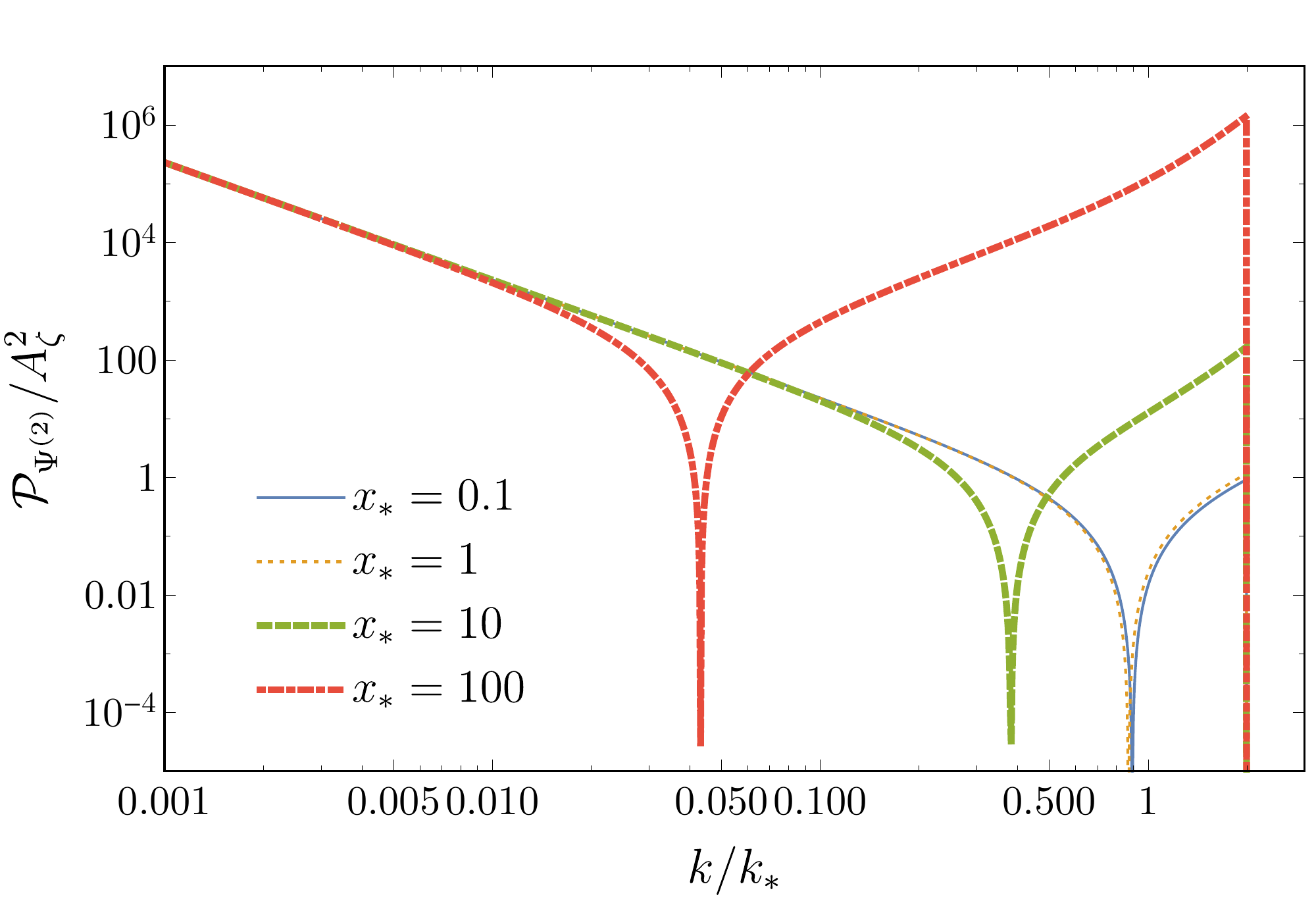}
        \end{center}
      \end{minipage}
	\hspace{0.5cm} 
      \begin{minipage}{0.45\textwidth}
        \begin{center}
          \includegraphics[width=\hsize]{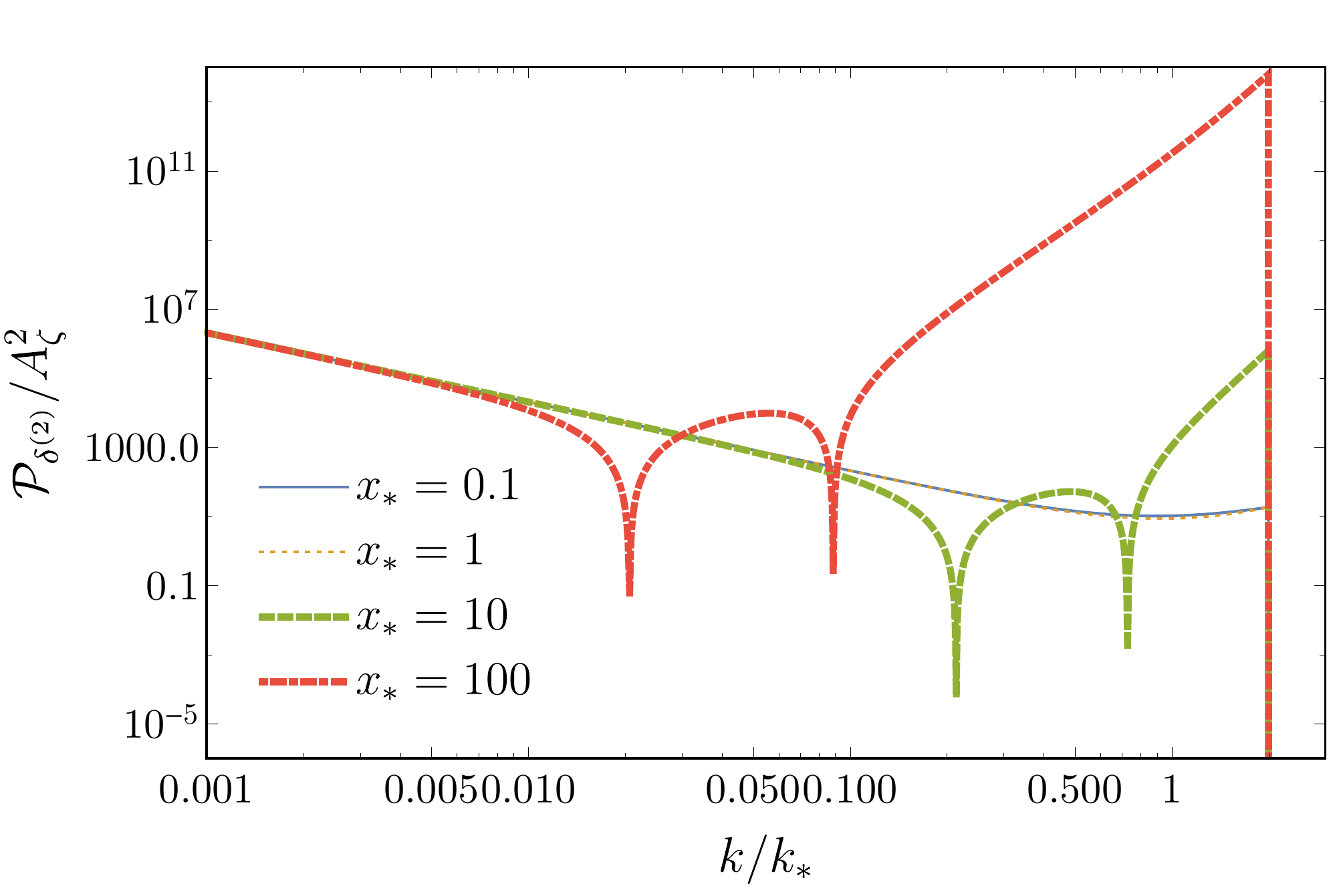}
        \end{center}
      \end{minipage}\\

      \begin{minipage}{0.57\textwidth}
      \vspace{0.6cm}
        \begin{center}
          \includegraphics[width=\hsize]{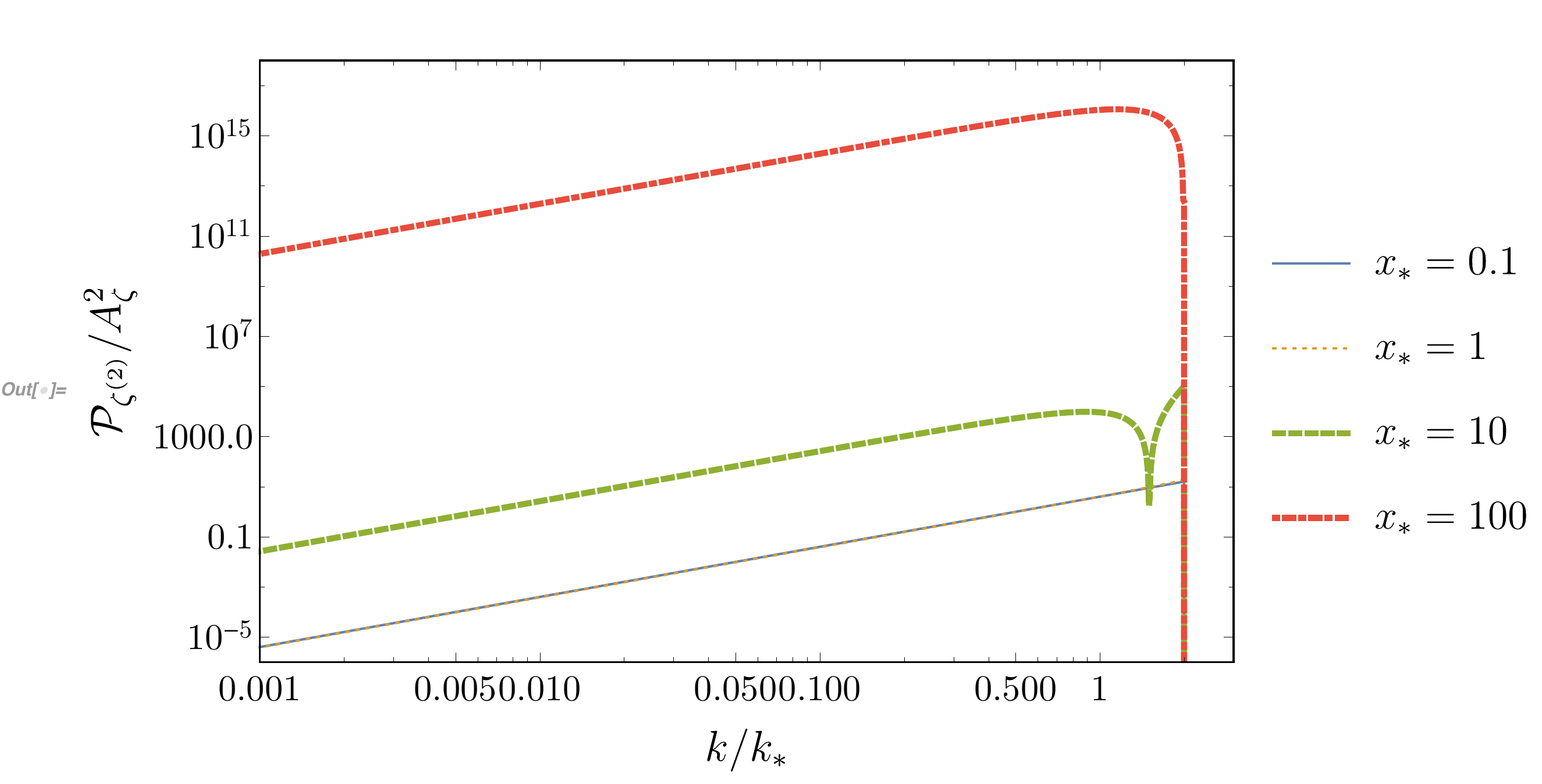}
        \end{center}
      \end{minipage}

    \end{tabular}
    \caption{\small	
   Evolution of $\mathcal P_{\Psi^\so}$ (top left), $\mathcal P_{\delta^\so}$ (top right), and $\mathcal P_{\zeta^\so}$ (bottom) with the delta-function power spectrum of the first-order curvature perturbation defined by Eq.~(\ref{eq:delta_func_pzeta}) in a MD era.
    Similarly to Fig.~\ref{fig:p_math_jh_yh}, we normalize the power spectra by $A_\zeta^2$, use $x_* \equiv k_* \eta$ as a time variable, and take $a_\text{NL} = 1$.
    The relation between the power spectra of the second-order perturbations and the function $I$ is given by Eq.~(\ref{eq:ps_i_rel_mm}).    
	}
    \label{fig:p_delta_mm}
  \end{center}
\end{figure}

Here, similarly to the case of a RD era, we calculate the auto-power spectra of the induced scalar perturbations.
Note again that we just use the power spectra to show the evolution of the second-order perturbations.
To compare with the observation, we must pay attention to the cross-power spectra between the first-order and the third-order perturbations.\footnote{Ref.~\cite{Makino:1991rp} showed that the leading contributions from the auto- and the cross-power spectra are canceled out in the IR limit, where wavenumbers of the two perturbations contributing to the power spectrum are significantly smaller than those of the other two perturbations in the power spectrum.}
The power spectra are given similarly to Eq.~(\ref{eq:p_mathcal_m}) as 
\begin{align}
	\mathcal P_{{\mathcal M}^\so}(k,\eta) \equiv \int_0^\infty \dd v \, \int_{|v-1|}^{v+1} \dd u \, I_{\mathcal M,\mm}^2(u,v,x) \left( \frac{3}{5} \right)^4 \mathcal P_{\zeta^\fo}( k v) \mathcal P_{\zeta^\fo}(k u),
	\label{eq:p_so_md}
\end{align}
where $\mathcal M$ denotes $\Psi$, $\delta$, and $\zeta$. 
Note that the normalization factor $(3/5)^4$ is different from the one in a RD era, given in Eq.~(\ref{eq:p_mathcal_m}).
In the case of the delta-function power spectrum given by Eq.~(\ref{eq:delta_func_pzeta}), Eq.~(\ref{eq:p_so_md}) becomes 
\begin{align}
	\mathcal P_{{\mathcal M}^\so}(k,\eta) =  A_\zeta^2 \left( \frac{3}{5} \right)^4 \left( \frac{k_*}{k} \right)^2 I^2_{\mathcal M,\mm}(k_*/k,k_*/k,x) \Theta(2- k/k_*).
	\label{eq:ps_i_rel_mm}
\end{align}
Figure~\ref{fig:p_delta_mm} shows the evolution of Eq.~(\ref{eq:ps_i_rel_mm}) for $\Psi^\so$, $\delta^\so$, and $\zeta^\so$ with $a_\text{NL} = 1$.
From this figure, we can see that $\mathcal P_{\Psi^\so}$ and $\mathcal P_{\delta^\so}$ are constant on superhorizon scales and start to grow once they enter the horizon ($x \gtrsim 1$).
On the other hand, $\mathcal P_{\zeta^\so}$ grows even on superhorizon scales once the source perturbations enter the horizon, similarly to that in a RD era. 
See also Appendix~\ref{app:power_md} for the auto-power spectra with the top-hat power spectrum of the curvature perturbations.

\section{Conclusion}
\label{sec:conclusion}

In this paper, we have discussed the second-order scalar perturbations induced by the first-order scalar perturbations.
Specifically, we have considered the two eras, a RD and a MD era, and obtained the analytic solutions for the gravitational potential $\Psi^\so$ and the energy density perturbation $\delta^\so$ in the conformal Newtonian gauge, and the curvature perturbation $\zeta^\so$ in the uniform density gauge.
We have also shown the evolution of the second-order perturbations from outside to inside horizon.
Here, we briefly summarize the updates from the previous analyses.

For the case of a RD era, we have taken into account the contributions of $\mathcal J_\text{h}$ and $\mathcal Y_\text{h}$, which is given in Eq.~(\ref{eq:mathcal_j_h_def}) and were neglected in Ref.~\cite{Bartolo:2006fj}.
In particular, we have found that the leading contribution in the scales much larger than the source perturbations ($u \gg 1$ and $v \gg 1$), coming from $\mathcal J_{\text{i}}$, is canceled by $\mathcal J_\text{h}$ in the late-time (subhorizon) limit.

For the case of a MD era, while the analyses on $\delta^\so$ in Refs.~\cite{Suto:1990wf,Makino:1991rp} were based on the Newtonian approximation, we have derived the analytic solutions of the induced perturbations without the approximation.
Our solutions are valid on all scales, including superhorizon scales, and are consistent with the previous results in the subhorizon limit, where the approximation is valid.
Apart from the references, Ref.~\cite{Bartolo:2006fj} also derived the analytic solution of $\Psi^\so$ in a MD era and our solution is consistent with the reference. The updates from the reference for the MD era case is the derivation of the analytic solutions of $\delta^\so$ and $\zeta^\so$.

Finally, we mention some issues for future work.
First, future studies on the effect of the anisotropic stress of the fluid need to be performed.
Although we have assumed a perfect fluid and neglected the anisotropic stress for simplicity, the fluid in a realistic situation can have the anisotropic stress originating from the free-streaming particles, such as photons and neutrinos.
The anisotropic stress causes the diffusion damping of the scalar perturbations on subhorizon scales in a RD era~\cite{Silk:1967kq} and makes a difference between $\Phi^\fo$ and $\Psi^\fo$ in both the eras, which could modify the analytic solutions derived in this paper.
Second, the effects of the transition between a RD era and a MD era need to be discussed.
In reality, the Universe has experienced the transition between the two eras.
Since the second-order scalar perturbations depend on the subhorizon evolution of the first-order perturbations, the perturbations that enter the horizon in the first era and experience the transition at subhorizon scales could make a non-negligible contribution to the second-order perturbations in the subsequent era.
In particular, this transition effect is closely related to the Meszaros effect, which is already discussed in the context of the second-order perturbations in Ref.~\cite{Bartolo:2006fj}.
Third, future work on the third-order perturbations needs to be performed.
As we mentioned in Sec.~\ref{sec:solution_for_rd} and \ref{sec:solution_for_md}, to compare with observation, we need to take into account the cross-power spectra between the first-order and the third-order perturbations, which are of the same order of magnitude as the auto-power spectrum of the second-order perturbations.
In particular, although the third-order perturbations in a MD era are already discussed in the context of the large scale structure~\cite{1981MNRAS.197..931J,1983MNRAS.203..345V,1984MNRAS.209..139J,Suto:1990wf,Makino:1991rp} and CMB anisotropies~\cite{DAmico:2007ngk}, those in a RD era have never been discussed and need to be studied in future work.

\acknowledgments
We thank Takahiro Terada for explaining how to perform the integrals in Ref.~\cite{Kohri:2018awv} and Wayne Hu for useful comments on the manuscript.
We also acknowledge the anonymous referee for her/his helpful comments.
We used the Mathematica package xPand~\cite{Pitrou:2013hga} during our work.
This work was supported by JSPS KAKENHI Grant Numbers, 15H02082 and 20H05248.

\appendix

\section{Derivation of equations for the perturbations}
\label{app:deriv_eq_for_pertb}

In this appendix, we derive the equations given in Sec.~\ref{sec:eom_so_scalar} from the Einstein equations.
Note that we assume a perfect fluid and follow the notation of the perturbations in Sec.~\ref{sec:eom_so_scalar} unless otherwise noted.

\subsection{First-order perturbations}

First, we discuss the perturbations at first order.
The Einstein and the energy-momentum tensors at first order are given as 
\begin{align}
  \delta {G^0_{\ 0}}^\fo =& \frac{1}{a^2} \left( 6 \mathcal H^2 \Phi^\fo + 6 \mathcal H {\Psi^\fo}' 
  - 2 \Psi^{\fo,i}_{\quad \ ,i} 
  \right), \\
  \label{eq:einstein_prtb_0i}  
  \delta {G^0_{\ i}}^\fo =& \frac{1}{a^2}\left( -2 \mathcal H \Phi^\fo_{,i} - 2 {\Psi^\fo}'_{, i} \right), \\
  \label{eq:einstein_prtb_ij}    
  \delta {G^i_{\ j}}^\fo =& \frac{1}{a^2} \left[\left( 2 \mathcal H^2 \Phi^\fo + 4 \mathcal H' \Phi^\fo + 2 \mathcal H {\Phi^\fo}' + 4 \mathcal H {\Psi^\fo}' + 2 {\Psi^\fo}'' + \Phi^{\fo ,k}_{\quad \  ,k} - \Psi^{\fo ,k}_{\quad \  ,k}  \right) \delta^i_{\ j} \right. \nonumber \\
   & \qquad \left. -  \Phi^{\fo ,i}_{\quad \  ,j} + \Psi^{\fo ,i}_{\quad \  ,j} \right], \\
  \delta {T^0_{\ 0}}^\fo =& - \delta \rho^\fo, \\
  \delta {T^0_{\ i}}^\fo =& (\rho^\zo + P^\zo) \delta v^\fo_{\quad ,i}, \\
  \delta {T^i_{\ j}}^\fo =& \delta P^\fo \delta^i_{\ j}.
\end{align}
From the perturbed Einstein equations ${G^\mu_{\ \nu}}^\fo = {T^\mu_{\ \nu}}^\fo/M_\Pl^2$, we can derive some equations.
The $(i,j)$ entry with $i\neq j$ leads to $\Phi^\fo = \Psi^\fo$ and the $(0,i)$ entry leads to
\begin{align}
	\label{eq:v_i_phi}
	\delta v_i = \frac{1}{3 (1 + w) \mathcal H^2} \left( -2 \mathcal H \Phi^\fo_{,i} - 2 {\Phi^\fo}'_{, i} \right),
\end{align}
where we have used the relation $\Phi^\fo = \Psi^\fo$.
Using the $(0,0)$ entry and the $(i,j)$ entry with $i=j$, we can derive the equations for the gravitational potential and the energy density perturbation (Eqs.~(\ref{eq:phi_fo_eom_re}) and (\ref{eq:delta_fo_eom_re})), 
\begin{align}
	\label{eq:phi_fo_eom}
	&{\Phi^\fo}'' + 3 (1+c_\sss^2) \mathcal H {\Phi^\fo}' - c_\sss^2 \Phi^{\fo ,i}_{\quad \ ,i} + (2 \mathcal H' + (1 + 3 c_\sss^2) \mathcal H^2) \Phi^\fo = 0, \\
		&\delta^\fo \equiv \frac{\delta \rho^\fo}{\rho^\zo} = - \left( 2 \Phi^\fo + \frac{2}{\mathcal H} {\Phi^\fo}' - \frac{2}{3\mathcal H^2} \Phi^{\fo\, ,i}_{\qquad ,i} \right).
\end{align}
Solving these equations, we obtain Eqs.~(\ref{eq:phi_zeta_rel})-(\ref{eq:trans_delta_mm}).

\subsection{Second-order perturbations}

Next, we discuss the perturbations at second order.
The Einstein tensor at second order is given as 
\begin{align}
	\delta {G^0_{\ 0}}^{\so} = &\frac{1}{a^2} \left[ 3 \mathcal H^2 \Phi^\so + 3 \mathcal H {\Psi^\so}' - \Psi^{\so,i}_{\quad \  ,i} -12 \mathcal H^2 \left(\Phi^\fo \right)^2 -12 \mathcal H \Phi^\fo {\Psi^\fo}'  +12 \mathcal H \Psi^\fo {\Psi^\fo}' \right. \nonumber \\
	&\left. \qquad 
	- 3 \left({\Psi^\fo}' \right)^2 - 8 \Psi^\fo \Psi^{\fo ,i}_{\quad \ ,i} - 3 \Psi^{\fo ,i} \Psi^\fo_{\quad ,i} \right], \label{eq:einstein_00} \\
	\delta{G^i_{\ j}}^\so = & \frac{1}{a^2} \left\{ \left( \mathcal H^2 \Phi^\so + 2 \mathcal H' \Phi^\so + \mathcal H {\Phi^\so}' + 2 \mathcal H {\Psi^\so}' + {\Psi^\so}'' + \frac{1}{2} \Phi^{\so ,k}_{\quad \  ,k} - \frac{1}{2} \Psi^{\so ,k}_{\quad \  ,k}  \right) \delta^i_{\ j} \right. \nonumber \\
	& \qquad \ \left.  - \frac{1}{2} \Phi^{\so ,i}_{\quad \  ,j} + \frac{1}{2} \Psi^{\so ,i}_{\quad \  ,j} \right. \nonumber \\
	&  \qquad 
	+ \left[ -4 \mathcal H^2 \left( \Phi^\fo \right)^2 - 8 \mathcal H' \left( \Phi^\fo \right)^2 - 8 \mathcal H \Phi^\fo {\Phi^\fo}' - 8 \mathcal H \Phi^\fo {\Psi^\fo}' - 2 {\Phi^\fo}' {\Psi^\fo}'  \right. \nonumber \\ 
	& \qquad \qquad 
	+ 8 \mathcal H \Psi^\fo {\Psi^\fo}' + \left( {\Psi^\fo}' \right)^2 -4 \Phi^\fo {\Psi^\fo}'' + 4 \Psi^\fo {\Psi^\fo}'' - 2 \Phi^\fo \Phi^{\fo ,k}_{\quad \ ,k}  \nonumber \\
	& \ 
	\left. \phantom{\left( \Phi^\fo \right)^2} + 2 \Psi^\fo \Phi^{\fo ,k}_{\quad \ ,k} - 4 \Psi^\fo \Psi^{\fo ,k}_{\quad \ ,k} - \Phi^{\fo ,k} \Phi^\fo_{\quad ,k} - 2 \Psi^{\fo ,k} \Psi^\fo_{\quad ,k} \right] \delta^i_{\ j} \nonumber \\
	& \qquad 
	+ \Phi^{\fo ,i} \Phi^\fo_{\quad \ ,j} - \Psi^{\fo ,i} \Phi^\fo_{\quad \ ,j} - \Phi^{\fo ,i} \Psi^\fo_{\quad \ ,j} + 3 \Psi^{\fo ,i} \Psi^\fo_{\quad \ ,j} \nonumber \\
	& \left. \quad \phantom{\frac{1}{2}}
	+ 2 \Phi^\fo \Phi^{\fo ,i}_{\quad \ ,j} - 2 \Psi^\fo \Phi^{\fo ,i}_{\quad \ ,j} + 4 \Psi^\fo \Psi^{\fo ,i}_{\quad \ ,j} \right\} + \left( \text{terms with } V_i^\so \text{ or } h^\so_{ij}  \right). \label{eq:einstein_ij}
\end{align}
Substituting $\Phi^\fo = \Psi^\fo$ into these equations, we obtain
\begin{align}
	\delta {G^0_{\ 0}}^{\so} 
	= &
	\frac{1}{a^2} \left[ 3 \mathcal H^2 \Phi^\so + 3 \mathcal H {\Psi^\so}' - \Psi^{\so,i}_{\quad \  ,i} -12 \mathcal H^2 \left(\Phi^\fo \right)^2 
	- 3 \left({\Phi^\fo}' \right)^2 - 8 \Phi^\fo \Phi^{\fo ,i}_{\quad \ ,i} - 3 \Phi^{\fo ,i} \Phi^\fo_{\quad ,i}  \right], \label{eq:einstein_00_2} \\
	\delta{G^i_{\ j}}^\so 
	=
	& \frac{1}{a^2} \left\{ \left( \mathcal H^2 \Phi^\so + 2 \mathcal H' \Phi^\so + \mathcal H {\Phi^\so}' + 2 \mathcal H {\Psi^\so}' + {\Psi^\so}'' + \frac{1}{3} \Phi^{\so ,k}_{\quad \  ,k} - \frac{1}{3} \Psi^{\so ,k}_{\quad \  ,k}  \right) \delta^i_{\ j} \right. \nonumber \\
	& \qquad 
	- \frac{1}{2}  \Lambda^{i\ \, k}_{\ j\ \, l} \Phi^{\so ,l}_{\quad \  ,k} +  \frac{1}{2}  \Lambda^{i\ \, k}_{\ j\ \, l} \Psi^{\so ,l}_{\quad \  ,k} \nonumber \\
	&  \qquad 
	+ \left[ -4 \mathcal H^2 \left( \Phi^\fo \right)^2 - 8 \mathcal H' \left( \Phi^\fo \right)^2 - 8 \mathcal H \Phi^\fo {\Phi^\fo}' - \left({\Phi^\fo}'\right)^2  \right. \nonumber \\
	& \qquad  \qquad  
	\left. - \frac{8}{3} \Phi^\fo \Phi^{\fo ,k}_{\quad \ ,k} - \frac{7}{3} \Phi^{\fo ,k} \Phi^\fo_{\quad ,k}  \right] \delta^i_{\ j} \nonumber \\ 
	& \quad \left. \phantom{\frac{1}{2}}
	+   2  \Lambda^{i\ \, k}_{\ j\ \, l} \Phi^{\fo ,l} \Phi^\fo_{\quad \ ,k} + 4  \Lambda^{i\ \, k}_{\ j\ \, l} \Phi^\fo \Phi^{\fo ,l}_{\quad \ ,k} \right\} + \left( \text{vector and tensor modes}  \right), \label{eq:einstein_ij_2}
\end{align}
where $ \Lambda^{i\ \, k}_{\ j\ \, l}$ is the projection operator onto the traceless longitudinal mode and explicitly expressed as\footnote{Note that $\frac{\partial_i \partial_j}{\nabla^2}$ is the projection operator onto longitudinal modes.}
\begin{align}
	\Lambda^{i\ \, k}_{\ j\ \, l} A^l_{\ k} (\bm x) = \frac{3}{2} \left( \frac{\partial^i \partial_j}{\nabla^2}- \frac{1}{3} \delta^i_{\ j} \right) \left( \frac{\partial^k \partial_l}{\nabla^2} - \frac{1}{3} \delta^k_{\ l} \right) A^l_{\ k} (\bm x).
\end{align}
With the notation of the perturbations in Eqs.~(\ref{eq:rho_pertb_def})-(\ref{eq:u_i_pertb_def}), we obtain the second-order energy-momentum tensor, 
\begin{align}
	\label{eq:t_0_0_so}
	\delta {T^0_{\ 0}}^\so &= - \left( \frac{1}{2} \delta \rho^\so + (\rho^\zo + P^\zo) \delta v^{\fo ,i} \delta v^\fo_{\quad ,i} \right), \\
	\label{eq:t_i_j_so}	
	\delta {T^i_{\ j}}^\so &=  \frac{1}{2} \delta P^\so \delta^i_{\ j} + (\rho^\zo + P^\zo) \delta v^{\fo ,i} \delta v^\fo_{\quad ,j}.
\end{align}
From the traceless longitudinal mode of the perturbed Einstein equation $\delta {G^\mu_{\ \nu}}^\so = {\delta T^\mu_{\ \nu}}^\so/M_\Pl^2$, we obtain the following relation:
\begin{align}
	\frac{1}{a^2} N^j_{\ i} \left( -\frac{1}{2} \Phi^{\so,i}_{\quad \ ,j} + \frac{1}{2} \Psi^{\so,i}_{\quad \ ,j} +  2 \Phi^{\fo ,i} \Phi^\fo_{\quad ,j} + 4 \Phi^\fo \Phi^{\fo ,i}_{\quad \ ,j}  \right) 
	 = \frac{\rho^\zo + P^\zo}{M_\Pl^2} N^j_{\ i} (\delta v^{\fo ,i} \delta v^\fo_{\quad ,j}).
\end{align}
Note again $N^j_{\ i}$ is defined in Eq.~(\ref{eq:n_ji_def}).
Substituting Eq.~(\ref{eq:v_i_phi}) into this equation, we obtain
\begin{align}
  &N^j_{\ i} \left[-\frac{1}{2} \Phi^{\so,i}_{\quad \ ,j} + \frac{1}{2} \Psi^{\so,i}_{\quad \ ,j} + \frac{2(1+3w)}{3(1+w)} \Phi^{\fo ,i} \Phi^\fo_{\quad ,j} - \frac{4}{3(1+w) \mathcal H} \left(\Phi^{\fo ,i} {\Phi^\fo_{\quad ,j}}' + {\Phi^{\fo ,i}}' \Phi^\fo_{\quad ,j} \right) \right. \nonumber \\
  & \qquad \qquad \qquad \qquad \qquad \qquad \qquad \qquad 
  \left. - \frac{4}{3(1+w) \mathcal H^2} {\Phi^{\fo ,i}}' {\Phi^\fo_{\quad ,j}}' + 4 \Phi^\fo \Phi^{\fo ,i}_{\quad \ ,j} \right] = 0 \nonumber \\
\Rightarrow \quad &  
\Phi^{\so} = \Psi^{\so} + 4 \left(\Phi^\fo \right)^2 - N^j_{\ i} {B^i_{\ j}}^\so, 
\end{align}
where ${B^i_{\ j}}^\so$ is defined as 
\begin{align}
 {B^i_{\ j}}^\so \equiv \left[\frac{4(5+3w)}{3(1+w)} \Phi^{\fo ,i} \Phi^\fo_{\quad ,j} 
+ \frac{8}{3(1+w) \mathcal H} \left(\Phi^{\fo ,i} {\Phi^\fo_{\quad ,j}}\right)' 
 + \frac{8}{3(1+w) \mathcal H^2} {\Phi^{\fo ,i}}' {\Phi^\fo_{\quad ,j}}'\right].
\end{align}
Here, we use the relation $\delta P^\so = c_\sss^2 \delta \rho^\so$, which is valid for the adiabatic perturbations with a constant sound speed. 
Then, from Eqs.~(\ref{eq:einstein_00}) and (\ref{eq:t_0_0_so}) and the diagonal part of Eqs.~(\ref{eq:einstein_ij}) and (\ref{eq:t_i_j_so}), we obtain (Eqs.(\ref{eq:psi_so_eom_re}) and (\ref{eq:psi_so_source_re}))
\begin{align}
&\frac{c_\sss^2}{a^2} \left[ 3 \mathcal H^2 \Phi^\so + 3 \mathcal H {\Psi^\so}' - \Psi^{\so,i}_{\quad \  ,i} -12 \mathcal H^2 \left(\Phi^\fo \right)^2 
	- 3 \left({\Phi^\fo}' \right)^2 - 8 \Phi^\fo \Phi^{\fo ,i}_{\quad \ ,i} - 3 \Phi^{\fo ,i} \Phi^\fo_{\quad ,i}  \right] \nonumber \\
	& + \frac{1}{a^2} \left\{ \mathcal H^2 \Phi^\so + 2 \mathcal H' \Phi^\so + \mathcal H {\Phi^\so}' + 2 \mathcal H {\Psi^\so}' + {\Psi^\so}'' + \frac{1}{3} \Phi^{\so ,i}_{\quad \  ,i} - \frac{1}{3} \Psi^{\so ,i}_{\quad \  ,i}  \right. \nonumber \\
	&  \qquad \left.
	- \left[ 4 \mathcal H^2 \left( \Phi^\fo \right)^2 + 8 \mathcal H' \left( \Phi^\fo \right)^2 + 8 \mathcal H \Phi^\fo {\Phi^\fo}' + \left({\Phi^\fo}'\right)^2  + \frac{8}{3} \Phi^\fo \Phi^{\fo ,i}_{\quad \ ,i} + \frac{7}{3} \Phi^{\fo ,i} \Phi^\fo_{\quad ,i}  \right] \right\} \nonumber \\
	&
	= \frac{\rho^\zo + P^\zo}{M_\Pl^2} \left(\frac{1}{3} - c_\sss^2 \right) \delta v^{\fo ,i} \delta v^\fo_{\quad ,i} \nonumber \\
	\Rightarrow \quad  & 
	{\Psi^\so}'' + 3 (1 + c_\sss^2) \mathcal H {\Psi^\so}' + \left[2\mathcal H' + (3c_\sss^2 + 1) \mathcal H^2 \right] \Psi^\so - c_\sss^2 \Psi^{\so,i}_{\quad \  ,i} = S^\so, 
	\label{eq:psi_so_eom}
	\end{align}
where $S^\so$ is the source term, defined as 
\begin{align}
	S^\so \equiv& \left(3c_\sss^2 -\frac{1}{3} \right) \Phi^{\fo ,i} \Phi^\fo_{\quad ,i} + 8c_\sss^2 \Phi^\fo \Phi^{\fo ,i}_{\quad \ ,i} + (3c_\sss^2 + 1) \left({\Phi^\fo}' \right)^2 + \left[ (3c_\sss^2 + 1)\mathcal H^2 + 2 \mathcal H' \right]  N^j_{\ i} {B^i_{\ j}}^\so \nonumber \\
	& \ 
	 + \mathcal H  N^j_{\ i} \left({B^i_{\ j}}^\so \right)' + \frac{1}{3}  N^j_{\ i} \left({B^i_{\ j}}^\so \right)^{,k}_{\ ,k} + \left(\frac{1}{3} - c_\sss^2 \right) \frac{4}{3(1+w) \mathcal H^2}  \left( \mathcal H \Phi^{\fo,i} +  {\Phi^{\fo,i}}' \right)  \left(  \mathcal H \Phi^\fo_{,i} + {\Phi^\fo}'_{, i} \right).
\end{align}

From the (0,0) entry of the perturbed Einstein equation, $\delta {G^0_{\ 0}}^\so = {\delta T^0_{\ 0}}^\so/M_\Pl^2$, we can obtain the equation for the second-order energy density perturbation, 
\begin{align}
	\delta \rho^\so =& -\frac{2M_\Pl^2}{a^2} \left[ 3 \mathcal H^2 \Phi^\so + 3 \mathcal H {\Psi^\so}' - \Psi^{\so,i}_{\quad \  ,i} -12 \mathcal H^2 \left(\Phi^\fo \right)^2 
	- 3 \left({\Phi^\fo}' \right)^2 - 8 \Phi^\fo \Phi^{\fo ,i}_{\quad \ ,i} - 3 \Phi^{\fo ,i} \Phi^\fo_{\quad ,i}  \right] \nonumber \\
	&-2(\rho^\zo + P^\zo) \delta v^{\fo ,i} \delta v^\fo_{\quad ,i}.
	\end{align}
Substituting $\Phi^\fo = \Psi^\fo$ and Eq.~(\ref{eq:v_i_phi}), we obtain (Eq.~(\ref{eq:delta_so_exp}))
\begin{align}
	\delta^\so \equiv& \frac{\delta \rho^\so}{\rho^\zo} \nonumber \\
	=& -2 \Psi^\so +2 N^j_{\ i} B^i_{\ j} -\frac{2}{\mathcal H} {\Psi^\so}' + \frac{2}{3\mathcal H^2} \Psi^{\so,i}_{\quad \  ,i} 
	+ \frac{2}{\mathcal H^2} \left({\Phi^\fo}' \right)^2 + \frac{16}{3\mathcal H^2} \Phi^\fo \Phi^{\fo ,i}_{\quad \ ,i}  \nonumber \\
	& + \frac{1}{\mathcal H^2} \left( 2 - \frac{8}{9(1+w)} \right) \Phi^{\fo ,i} \Phi^\fo_{\quad ,i} - \frac{8}{9(1+w) \mathcal H^3} (\Phi^{\fo ,i} \Phi^\fo_{\quad ,i} )'  - \frac{8}{9(1+w) \mathcal H^4} {\Phi^\fo}'_{, i}  {\Phi^{\fo,i}}'.
\end{align}

Finally, we summarize the concrete expression of the second-order curvature perturbation in Ref.~\cite{Malik:2008im}.\footnote{We have fixed some typos in Ref.~\cite{Malik:2008im}.}
To show the expression in a general gauge, we restore some metric perturbations that are killed in the conformal Newtonian gauge as
\begin{align} 
 \dd s^2 &= g_{\mu\nu} \dd x^\mu \dd x^\nu \nonumber \\
& = a^2 \left\{- (1 + 2 \Phi^{(1)} + \Phi^{(2)})\dd \eta^2  + (2{B_{,i}}^\fo + B_{,i}^\so) \dd \eta \dd x^i \right. \nonumber \\
& \qquad  \qquad 
\left.+ \left[ (1-2 \Psi^{(1)} - \Psi^{(2)}) \delta_{ij} + 2E^\fo_{,ij} + E^\so_{,ij} \right] \dd x^i \dd x^j \right\},
\label{eq:def_metric_pertb_app}
\end{align}
where we have neglected vector and tensor perturbations. 
One of the often-used gauge-invariant quantities is the curvature perturbation in the uniform density hypersurfaces ($\delta \rho = 0$).
Although the condition of uniform density is enough to define the gauge-invariant curvature perturbation at first order in perturbations, it is not enough to define the second-order curvature perturbation.
In this paper, following Ref.~\cite{Malik:2008im}, we impose the additional condition $E^\fo =0$ to define the second-order curvature perturbation.
Then, we can define the curvature perturbation as~\cite{Malik:2008im} 
\begin{align}
\zeta^\so = -\Psi^\so + \mathcal H \alpha^\so + \frac{1}{4} \mathcal X^k_{\ k} - \frac{1}{4} \nabla^{-2} \mathcal X_{ij}^{\ \ ,ij}, \label{eq:psi_trans_so_2}
\end{align}
where the quantities in a general gauge are given as
\begin{align}
		\alpha^\so \equiv & - \frac{\delta \rho^\so}{{\rho^\zo}'} + \frac{\delta \rho^\fo \delta {\rho^\fo}'}{({\rho^\zo}')^2} + \frac{\delta \rho^\fo_{,k}}{{\rho^\zo}'}E^{\fo \, ,k}, \\
	 \mathcal X_{ij} \equiv&   2 \left[ \left(2\mathcal H^2 + \mathcal H' - \mathcal H \frac{{\rho^\zo}''}{{\rho^\zo}'} \right) \left( \frac{\delta \rho^\fo}{{\rho^\zo}'} \right)^2 + \mathcal H \frac{ \delta \rho^\fo \delta {\rho^\fo}'}{ ({\rho^\zo}')^2} + \mathcal H  \frac{\delta \rho^\fo_{,k}}{{\rho^\zo}'} E^{\fo\, ,k} \right] \delta_{ij} \nonumber \\
&  + 4 \left[\frac{\delta \rho^\fo}{{\rho^\zo}'} ({\Psi^{\fo}}' + 2 \mathcal H \Psi^\fo) + \Psi^\fo_{,k} E^{\fo\, ,k} \right] \delta_{ij} \nonumber \\
 & + 8 \Psi^\fo E^{\fo}_{,ij} - 2  \frac{\delta \rho^\fo}{{\rho^\zo}'} {E^{\fo}_{,ij}}' -4 E^\fo_{,ik} E^{\fo \, ,k}_{\qquad ,j} - 2 \left(B^\fo_{,i}  \frac{\delta \rho^\fo_{,j}}{{\rho^\zo}'} + B^\fo_{,j} \frac{\delta \rho^\fo_{,i}}{{\rho^\zo}'}\right)  -2 \frac{\delta \rho^\fo_{,i }\delta \rho^\fo_{,j}}{ ({\rho^\zo}')^2}  \nonumber \\
 & - 2 E^\fo_{,ijk} E^{\fo \, ,k}+  \left( {E^\fo_{,i}}' \frac{\delta \rho^\fo_{,j}}{{\rho^\zo}'} + {E^\fo_{,j}}' \frac{\delta \rho^\fo_{,i}}{{\rho^\zo}'} \right). \label{eq:math_x_ij_ij_u_den_re} 
\end{align}
Then, we obtain
\begin{align}
	\mathcal X^k_{\ k} 
	=&  6 \left[ \left(2\mathcal H^2 + \mathcal H' - \mathcal H \frac{{\rho^\zo}''}{{\rho^\zo}'} \right) \left( \frac{\delta \rho^\fo}{{\rho^\zo}'} \right)^2 + \mathcal H \frac{ \delta \rho^\fo \delta {\rho^\fo}'}{ ({\rho^\zo}')^2} + \mathcal H  \frac{\delta \rho^\fo_{,k}}{{\rho^\zo}'} E^{\fo\, ,k}\right] \nonumber \\
	& + 12 \left[ \Psi^\fo_{,k} E^{\fo\, ,k} + \frac{\delta \rho^\fo}{{\rho^\zo}'} \left( {\Psi^\fo}' +2 \mathcal H \Psi^\fo\right) \right]    \nonumber \\
	&
	- 2 E^{\fo \, ,k} \nabla^2 E^\fo_{\,,k}  -2  \frac{\delta \rho^\fo}{{\rho^\zo}'} \nabla^2 {E^\fo}'  +  8 \Psi^\fo \nabla^2 E^{\fo} -4 E^{\fo\, ,kl} E^\fo_{,kl} \nonumber \\
	& - 2\frac{\delta \rho^\fo_{\,k}}{{\rho^\zo}'} \left( 2B^{\fo\, ,k} - {E^{\fo \, ,k}}' \right)  - 2 \frac{\delta \rho^{\fo\, ,k} \delta \rho^\fo_{,k}}{({\rho^\zo}')^2},
\end{align}	
\begin{align}
	\nabla^{-2} \mathcal X_{ij}^{\ \ , ij} 
 =&  2 \left[ \left(2\mathcal H^2 + \mathcal H' - \mathcal H \frac{{\rho^\zo}''}{{\rho^\zo}'} \right)  \left( \frac{\delta \rho^\fo}{{\rho^\zo}'} \right)^2 + \mathcal H \frac{ \delta \rho^\fo \delta {\rho^\fo}'}{ ({\rho^\zo}')^2} + \mathcal H  \frac{\delta \rho^\fo_{,k}}{{\rho^\zo}'} E^{\fo\, ,k} \right]  \nonumber \\
& + 4  \left[\frac{\delta \rho^\fo}{{\rho^\zo}'} ({\Psi^{\fo}}' + 2 \mathcal H \Psi^\fo) + \Psi^\fo_{,k} E^{\fo\, ,k} \right]  \nonumber \\ 
 &
 + \nabla^{-2} \left[ 8  \Psi^\fo E^{\fo}_{,ij} - 2 \frac{\delta \rho^\fo}{{\rho^\zo}'} {E^{\fo}_{,ij}}'  -4 E^\fo_{,ik} E^{\fo \, ,k}_{\qquad ,j}  - 4 B^\fo_{,i}  \frac{\delta \rho^\fo_{,j}}{{\rho^\zo}'}  -2 \frac{\delta \rho^\fo_{,i }\delta \rho^\fo_{,j}}{ ({\rho^\zo}')^2} \right. \nonumber \\
 & \left. \qquad \qquad 
 - 2 E^\fo_{,ijk} E^{\fo \, ,k} + 2 {E^\fo_{,i}}' \frac{\delta \rho^\fo_{,j}}{{\rho^\zo}'} \right]^{,ij}. \label{eq:math_x_ij_ij_u_den_re}
\end{align}
Note that if we take the gauge satisfying $\delta \rho^\fo = \delta \rho^\so = E^\fo = 0$, we have $\zeta^\so = \Psi^\so$ by definition.
Taking the Newtonian gauge ($E = B =0$), we finally obtain Eq.~(\ref{eq:zeta_so_def}).

\section{Complete expressions of $M^a_{nm}$, $\mathcal J_\text{h}$ and $\mathcal Y_\text{h}$}
\label{app:ana_sol_rd}

In this appendix, we summarize the complete expressions of $M^a_{nm}$, given in Eq.~(\ref{eq:i_a_expand}), and $\mathcal J_\text{h}$ and $\mathcal Y_\text{h}$, defined in Eq.~(\ref{eq:mathcal_j_h_def}). 

First, we show the expressions of $M^a_{nm}(u,v)$ ($1 \leq n \leq 8, \, 0 \leq m \leq 7$):
\begin{align}
	&M^j_{1 0}(u,v) = 0, \\
	&M^j_{1 1}(u,v) = -\frac{3 \left(3 u^4-2 u^2 \left(3 v^2+1\right)+3 v^4-2 v^2-1\right) (u (v (u+v+1)-2)-2 v)}{16 u^3 v^3}, \\
	&M^j_{1 2}(u,v) = 0, \\	
	&M^j_{1 3}(u,v) = \frac{1}{8 u^4 v^4} (9 \left(-3 u^7 (v-1)+u^6 (3 (7-3 v) v+3)+u^5 (v (3 v (v+5)+11)-11) \right. \nonumber \\
	& \qquad  \quad\qquad  \quad 
	 +u^4 (v (3 v (v (6 v-13)+1)-23)-2)+u^3 (v (v (v (3 (v-13)
   v-14)+2)-11)+11) \nonumber \\
	& \qquad  \quad \qquad  \quad 
   +u^2 (v (v (v (3 v ((5-3 v) v+1)+2)-5)+17)-1) \nonumber \\
	& \qquad  \quad \qquad  \quad 
   +u (v (v (v (v (v (11-3 (v-7) v)-23)-11)+17)-5)+5) \nonumber \\
	& \qquad  \quad \qquad  \quad \left.         
   +v \left(3
   v^2+1\right) \left(v \left(v \left(v^2+v-4\right)-1\right)+5\right)\right)), \\
	&M^j_{1 4}(u,v) = 0, \\	   
   &M^j_{1 5}(u,v) = \frac{1}{8 u^4 v^4} (27 \left(21 u^6+54 u^5 (v-1)-u^4 (3 v (7 v+18)+23)-36 u^3 (v-1) \left(3 v^2+1\right) \right. \nonumber \\
	& \qquad  \quad \qquad  \quad 
   +u^2 (v (v (3 (36-7 v) v-10)+36)+17)+18
   u (v-1) \left(3 v^4-2 v^2+1\right) \nonumber \\
	& \qquad  \quad  \qquad  \quad \left.
   +v (v (v (v (3 v (7 v-18)-23)+36)+17)-18)+5\right)), \\
	&M^j_{1 6}(u,v) = 0, \\	   
   &M^j_{1 7}(u,v) = -\frac{729 \left(3 u^4-2 u^2 \left(3 v^2+1\right)+3 v^4-2 v^2+1\right)}{4 u^4 v^4}, \\
   &M^j_{2 m}(u,v) = -M^j_{1 m}(-u,v), \\
   &M^j_{3 m}(u,v) = -M^j_{1 m}(u,-v), \\
   &M^j_{4 m}(u,v) = M^j_{1 m}(-u,-v), \\   
   &M^j_{5 0}(u,v) = \frac{\sqrt{3} \left(-3 u^4+u^2 \left(6 v^2+2\right)-3 v^4+2 v^2+1\right)}{16 u^2 v^2}, \\   
   &M^j_{5 1}(u,v) = 0, \\
   &M^j_{5 2}(u,v) = \frac{1}{16 u^4 v^4} (3 \sqrt{3} \left(3 u^7 (v-2) v+3 u^6 (v ((v-6) v-2)+2) \right.  \nonumber \\
   	& \qquad  \quad \qquad  \quad  
   +2 u^5 v (v (-3 (v-1) v-4)+11)+u^4 (2 v (v (v (5-3 (v-6)
   v)+3)+2)-4) \nonumber \\
	& \qquad  \quad  \qquad  \quad 
   +u^3 v (v (v (v (3 v (v+2)+10)-28)+3)-22) \nonumber \\
	& \qquad  \quad  \qquad  \quad 
   +u^2 (v (v (v (v (v (3 (v-6) v-8)+6)+3)-10)+2)-2)  \nonumber \\
	& \qquad  \quad  \qquad  \quad \left. 
   -2 u v \left(3
   v^2+1\right) \left(v \left(v \left(v^2+v-4\right)-1\right)+5\right)+6 v^6-4 v^4-2 v^2\right))), \\   
   &M^j_{5 3}(u,v) = 0, \\   
   &M^j_{5 4}(u,v) = -\frac{1}{8 u^4 v^4} (9 \sqrt{3} \left(3 u^7+21 u^6 (v-1)+u^5 (3 v (5 v-18)-11) \right. \nonumber \\
& \qquad  \quad  \qquad  \quad    
   +u^4 (v (3 (7-13 v) v-23)+23)+u^3 (v (v (3 (36-13 v)
   v+2)+36)+11)\nonumber \\
& \qquad  \quad  \qquad  \quad 
   +u^2 (v (v (v (3 v (5 v+7)+2)+10)+17)-17)\nonumber \\
& \qquad  \quad  \qquad  \quad 
   +u (v (v (v (v (3 v (7 v-18)-23)+36)+17)-18)+5)\nonumber \\
& \qquad  \quad  \qquad  \quad  \left.   
   +v (v (v (v (v (3 (v-7)
   v-11)+23)+11)-17)+5)-5\right)), \\     
   &M^j_{5 5}(u,v) = 0, \\       
   &M^j_{5 6}(u,v) = \frac{243 \sqrt{3} (u+v-1) \left(3 u^4-2 u^2 \left(3 v^2+1\right)+3 v^4-2 v^2+1\right)}{4 u^4 v^4}, \\          
   &M^j_{5 7}(u,v) = 0, \\  
   &M^j_{6 m}(u,v) = -M^j_{5 m}(-u,v), \\
   &M^j_{7 m}(u,v) = -M^j_{5 m}(u,-v), \\   
   &M^j_{8 m}(u,v) = M^j_{5 m}(-u,-v), \\      
   &M^y_{1 m}(u,v) = -M^j_{5 m}(u,v), \\
   &M^y_{2 m}(u,v) = M^j_{5 m}(-u,v), \\   
   &M^y_{3 m}(u,v) = M^j_{5 m}(u,-v), \\      
   &M^y_{4 m}(u,v) = -M^j_{5 m}(-u,-v), \\       
   &M^y_{5 m}(u,v) = M^j_{1 m}(u,v), \\          
   &M^y_{6 m}(u,v) = -M^j_{1 m}(-u,v), \\             
   &M^y_{7 m}(u,v) = -M^j_{1 m}(u,-v), \\                
   &M^y_{8 m}(u,v) = M^j_{1 m}(-u,-v).
\end{align}

Next, we show the expressions of $\mathcal J_\text{h}$ and $\mathcal Y_\text{h}$, defined in Eq.~(\ref{eq:mathcal_j_h_def}).
To perform the integral, we use the following relation:\footnote{This relation is used to calculate the GWs induced by scalar perturbations analytically~\cite{Ananda:2006af,Kohri:2018awv}.}
\begin{align}
		\int^{x_2}_{x_1} \dd \bar x \frac{\sin (\alpha \bar x + \phi)}{\bar{x}^m} =& \left[ \sum^{m-2}_{k=0} \frac{(m-k-2)!}{(m-1)!} \alpha^k \sin\left( \alpha \bar x + \phi + \frac{(k+2)}{2}\pi \right) {\bar x}^{1+k-m} \right]^{x_2}_{x_1} \nonumber \\
	& - \frac{\alpha^{m-1}}{(m-1)!} \int^{x_2}_{x_1} \dd \bar x \frac{1}{\bar x} \sin \left( \alpha \bar x + \phi + \frac{(m+1)}{2}\pi \right).
\end{align}
After tedious calculations, we finally obtain
\begin{align}
	\mathcal J_\text{h}(u,v,x) =& -\frac{3\sqrt{3}}{8 u^4 v^4 x^6} \left[ x \sin\left( \frac{x}{\sqrt{3}} \right) \left( 2u x \cos\left(  \frac{u x}{\sqrt{3}} \right) \left\{ \phantom{\frac{1}{2}}   \right. \right. \right. \nonumber \\
	& v x \left[ 
	 3 (x^2 + 18) u^4 -2 (3 (x^2 + 18)v^2 + 2(x^2 +9)) u^2 \right. \nonumber \\
	& \qquad \left.  -3(x^2 -6) -4 v^2(x^2 +9) + 3v^4(x^2 + 18) \right] \cos\left(  \frac{v x}{\sqrt{3}} \right) \nonumber \\
	 &
	 + \sqrt{3}\left[ 6 x^2 v^6 - (7x^2 + 54) v^4 + 6(x^2 + 6) v^2 + 3(x^2 -6) + u^4( (6v^2 -3)x^2 -54) \right. \nonumber \\
	 & \left. \phantom{\frac{1}{2}} \left. \qquad
	 + 2 u^2(-6 x^2 v^4 + (x^2 + 54)v^2 + 2(x^2 + 9)) \right] \sin\left(  \frac{v x}{\sqrt{3}} \right) \right\} \nonumber \\
	 &+ \sin \left(  \frac{u x}{\sqrt{3}} \right) \biggl\{ 2\sqrt{3} v x \left[ 6 x^2 u^6 - ((12 v^2 +7)x^2 + 54)u^4  \right.  \nonumber \\
	 &\qquad \qquad\qquad\qquad\qquad
	  + 2 (3x^2 v^4 + (x^2 + 54)v^2 + 3 (x^2 + 6))u^2 \nonumber \\
	 &\qquad \qquad\qquad\qquad\qquad
	 \left. + 3(x^2 -6) + 4v^2(x^2 +9)-3 v^4(x^2 + 18) \right] \cos\left(  \frac{v x}{\sqrt{3}} \right) \nonumber \\
	 & + \left[ ((6v^2 -3) x^4 - 36 x^2) u^6 + (-(12 v^4 +v^2 +3)x^4 + 6(6v^2 + 7)x^2 + 324) u^4  \right. \nonumber \\
	 & + (6x^4 v^6 -x^2 (x^2 -36) v^4 + 4(x^4 + 3x^2 - 162)v^2 + 3(x^4 - 12x^2 -72)) u^2 \nonumber\\
	 & -3 (x^2(x^2 + 12) v^6 + (x^4 - 14 x^2 - 108)v^4 \nonumber \\
	 & \left. + (-x^4 + 12 x^2 + 72) v^2 - x^4 + 6 x^2 -36) \right] \sin\left(  \frac{v x}{\sqrt{3}} \right) \biggr\} \nonumber \\
	 &+ \cos \left(  \frac{x}{\sqrt{3}} \right) \biggr( -u x \cos \left( \frac{u x}{\sqrt{3}} \right)\biggl\{ 
	 \left[3x^4 u^6 + ((3-9v^2)x^4 + (54-36v^2)x^2+324) u^4 \right.  \nonumber \\
	 & + 3(3x^2 (x^2+8) v^4 -2(x^4 +14 x^2 +  108)v^2 -x^4 - 16 x^2 - 72)u^2 - 3v^6 x^2(x^2 + 12) \nonumber \\
	 &\left. -3(x^4 + 2 x^2 - 36) + 3 v^4 (x^4 + 26x^2 + 108) - v^2(5x^4 + 60x^2 + 216) \right] \sin \left(  \frac{v x}{\sqrt{3}} \right) \nonumber \\
	 &-2\sqrt{3} v x \left[ 9(x^2 + 6) u^4 -2(9(x^2 + 6)v^2 + 4 x^2 + 18) u^2 - x^2 \right. \nonumber \\
	 & \left. + 9 v^4(x^2 + 6) -4v^2(2x^2+9) + 18  \right] \cos\left(  \frac{v x}{\sqrt{3}} \right) \biggr\} \nonumber \\
	 & - \sin \left(  \frac{u x}{\sqrt{3}} \right) \biggl\{ v x  \left[ -3 x^2 (x^2 + 12)u^6 + 3((3v^2 + 1) x^4 + (24 v^2 + 26) x^2 + 108) u^4  \right. \nonumber \\ 
	 & - (9 x^2(x^2 + 4)v^4 + 6(x^4 + 14 x^2 + 108)v^2 + 5x^4 + 60x^2 + 216 ) u^2 \nonumber  \\
	 & \left. + 3(x^4 v^6 + (x^4 + 18 x^2 + 108) v^4 - (x^4 + 16 x^2 + 72) v^2 -x^4 - 2x^2 + 36) \right] \cos \left(  \frac{v x}{\sqrt{3}} \right) \nonumber \\
	 & + \sqrt{3} \left[ 3x^2 ((1-2v^2) x^2 + 12) u^6 + ((12 v^4 + v^2 -3) x^4 -6(6v^2 + 13)x^2 -324) u^4  \right. \nonumber \\
	 & + (-6x^4 v^6 + x^2 (x^2 -36)v^4 + (8x^4 + 60 x^2 + 648) v^2 + 5x^4 + 60 x^2 + 216) u^2 \nonumber \\
	 & \left. +  3v^6 x^2 (x^2 + 12) + 3(x^4 + 2 x^2 -36) - 3v^4 (x^4 + 26 x^2 + 108) \right. \nonumber \\
	 & \left. + v^2 (5 x^4 + 60 x^2 + 216) \right] \sin \left(  \frac{v x}{\sqrt{3}} \right) \biggr\} \biggr) \biggr] \nonumber \\
	 &+ \frac{3}{32 u^4 v^4} \left( 3u^4(u^4 - 2) + 3v^4(v^4 -2) + 3 - 8u^2v^2(1 + u^2 + v^2) -6 u^4 v^4 \right) \nonumber \\
	& \qquad 
	\times \left( \Ci \left(\frac{|-1+u-v|}{\sqrt{3}} x \right) + \Ci \left(\frac{1+u-v}{\sqrt{3}} x\right) - \Ci\left(\frac{-1+u+v}{\sqrt{3}} x \right) - \Ci \left(\frac{1+u+v}{\sqrt{3}} x \right)  \right) \nonumber \\ 
	&+ \frac{1}{8 u^3 v^3}
	\left[-9(u^6 + u^4 -u^2) -9(v^6 + v^4 - v^2) + 9 + 6 u^2 v^2 (u^2 - v^2)^2  \right. \nonumber \\
	& \left. \qquad \qquad \ 
	+ 5 u^2 v^2(u^2 + v^2) + 8 u^2 v^2 \right] \nonumber \\
	&-\frac{3}{32 u^4 v^4} \left( 3u^4(u^4 - 2) + 3v^4(v^4 -2) + 3 - 8u^2v^2(1 + u^2 + v^2) -6 u^4 v^4 \right) \text{log}\left( \frac{1-(u-v)^2 }{(u+v)^2 -1} \right),
\end{align}
\begin{align}
	\mathcal Y_\text{h} (u,v,x) = & \frac{3\sqrt{3}}{8 u^4 v^4 x^6} \left[ x \cos\left( \frac{x}{\sqrt{3}} \right) \left( 2u x \cos\left(  \frac{u x}{\sqrt{3}} \right) \left\{ \phantom{\frac{1}{2}}   \right. \right. \right. \nonumber \\
	& v x \left[ 
	 3 (x^2 + 18) u^4 -2 (3 (x^2 + 18)v^2 + 2(x^2 +9)) u^2 \right. \nonumber \\
	& \left. \qquad 
	 -3(x^2 -6) -4 v^2(x^2 +9) + 3v^4(x^2 + 18) \right] \cos\left(  \frac{v x}{\sqrt{3}} \right) \nonumber \\
	 &
	 + \sqrt{3}\left[ 6 x^2 v^6 - (7x^2 + 54) v^4 + 6(x^2 + 6) v^2 + 3(x^2 -6) + u^4( (6v^2 -3)x^2 -54) \right. \nonumber \\
	 & \left. \phantom{\frac{1}{2}} \left. \qquad
	 + 2 u^2(-6 x^2 v^4 + (x^2 + 54)v^2 + 2(x^2 + 9)) \right] \sin\left(  \frac{v x}{\sqrt{3}} \right) \right\} \nonumber \\
	 &+ \sin \left(  \frac{u x}{\sqrt{3}} \right) \biggl\{ 2\sqrt{3} v x \left[ 6 x^2 u^6 - ((12 v^2 +7)x^2 + 54)u^4  \right.  \nonumber \\
	 & \qquad \qquad\qquad\qquad\qquad
	 + 2 (3x^2 v^4 + (x^2 + 54)v^2 + 3 (x^2 + 6))u^2 \nonumber \\
	 &\qquad \qquad\qquad\qquad\qquad
	 \left. + 3(x^2 -6) + 4v^2(x^2 +9)-3 v^4(x^2 + 18) \right] \cos\left(  \frac{v x}{\sqrt{3}} \right) \nonumber \\
	 & + \left[ ((6v^2 -3) x^4 - 36 x^2) u^6 + (-(12 v^4 +v^2 +3)x^4 + 6(6v^2 + 7)x^2 + 324) u^4  \right. \nonumber \\
	 & + (6x^4 v^6 -x^2 (x^2 -36) v^4 + 4(x^4 + 3x^2 - 162)v^2 + 3(x^4 - 12x^2 -72)) u^2 \nonumber\\
	 & \left. -3 (x^2(x^2 + 12) v^6 + (x^4 - 14 x^2 - 108)v^4 \right. \nonumber \\
	 & \left. + (-x^4 + 12 x^2 + 72) v^2 - x^4 + 6 x^2 -36) \right] \sin\left(  \frac{v x}{\sqrt{3}} \right) \biggr\} \nonumber \\
	 &+ \sin \left(  \frac{x}{\sqrt{3}} \right) \biggr( u x \cos \left( \frac{u x}{\sqrt{3}} \right)\biggl\{ 
	 \left[3x^4 u^6 + ((3-9v^2)x^4 + (54-36v^2)x^2+324) u^4 \right.  \nonumber \\
	 & + 3(3x^2 (x^2+8) v^4 -2(x^4 +14 x^2 +  108)v^2 -x^4 - 16 x^2 - 72)u^2 - 3v^6 x^2(x^2 + 12) \nonumber \\
	 &\left. -3(x^4 + 2 x^2 - 36) + 3 v^4 (x^4 + 26x^2 + 108) - v^2(5x^4 + 60x^2 + 216) \right] \sin \left(  \frac{v x}{\sqrt{3}} \right) \nonumber \\
	 &-2\sqrt{3} v x \left[ 9(x^2 + 6) u^4 -2(9(x^2 + 6)v^2 + 4 x^2 + 18) u^2 - x^2  \right. \nonumber \\
	 & \left. + 9 v^4(x^2 + 6) -4v^2(2x^2+9) + 18  \right] \cos\left(  \frac{v x}{\sqrt{3}} \right) \biggr\} \nonumber \\
	 & + \sin \left(  \frac{u x}{\sqrt{3}} \right) \biggl\{ v x  \left[ -3 x^2 (x^2 + 12)u^6 + 3((3v^2 + 1) x^4 + (24 v^2 + 26) x^2 + 108) u^4  \right. \nonumber \\ 
	 & - (9 x^2(x^2 + 4)v^4 + 6(x^4 + 14 x^2 + 108)v^2 + 5x^4 + 60x^2 + 216 ) u^2 \nonumber  \\
	 & \left. + 3(x^4 v^6 + (x^4 + 18 x^2 + 108) v^4 - (x^4 + 16 x^2 + 72) v^2 -x^4 - 2x^2 + 36) \right] \cos \left(  \frac{v x}{\sqrt{3}} \right) \nonumber \\
	 & + \sqrt{3} \left[ 3x^2 ((1-2v^2) x^2 + 12) u^6 + ((12 v^4 + v^2 -3) x^4 -6(6v^2 + 13)x^2 -324) u^4  \right. \nonumber \\
	 & + (-6x^4 v^6 + x^2 (x^2 -36)v^4 + (8x^4 + 60 x^2 + 648) v^2 + 5x^4 + 60 x^2 + 216) u^2 \nonumber \\
	 & \left. +  3v^6 x^2 (x^2 + 12) + 3(x^4 + 2 x^2 -36) - 3v^4 (x^4 + 26 x^2 + 108) \right. \nonumber \\
	 & \left. + v^2 (5 x^4 + 60 x^2 + 216) \right] \sin \left(  \frac{v x}{\sqrt{3}} \right) \biggr\} \biggr) \biggr] \nonumber \\
	& -\frac{3}{32 u^4 v^4} \left( 3u^4(u^4 - 2) + 3v^4(v^4 -2) + 3 - 8u^2v^2(1 + u^2 + v^2) -6 u^4 v^4 \right)  \nonumber \\
& \qquad \times \left( \Si \left(\frac{-1+u-v}{\sqrt{3}} x \right) - \Si \left(\frac{1+u-v}{\sqrt{3}} x\right) - \Si\left(\frac{-1+u+v}{\sqrt{3}} x \right) + \Si \left(\frac{1+u+v}{\sqrt{3}} x \right)  \right),
\end{align}
where Si and Ci denote the sine and cosine integrals, defined as 
\begin{align}
	\text{Si}(x) &\equiv \int^x_0 \dd t \frac{\sin t}{t}, \\
	\text{Ci}(x) &\equiv -\int^{\infty}_x  \dd t \frac{\cos t}{t}.
\end{align}
In the late-time limit ($x\gg 1$), $\mathcal J_\text{h}$ and $\mathcal Y_\text{h}$ become 
\begin{align}
	\label{eq:mathcal_jh_limit}
	\mathcal J_\text{h}(u,v,x (\gg 1)) \simeq& \frac{1}{8 u^3 v^3}
\left[-9(u^6 + u^4 -u^2) -9(v^6 + v^4 - v^2) + 9 \right. \nonumber \\
& \left. \qquad \qquad  + 6 u^2 v^2 (u^2 - v^2)^2 + 5 u^2 v^2(u^2 + v^2) + 8 u^2 v^2 \right] \nonumber \\
	&-\frac{3}{32 u^4 v^4} \left( 3u^4(u^4 - 2) + 3v^4(v^4 -2) + 3 - 8u^2v^2(1 + u^2 + v^2) -6 u^4 v^4 \right) \nonumber \\
& \qquad \qquad
 \times \text{log}\left( \frac{1-(u-v)^2 }{(u+v)^2 -1} \right), \\
	\label{eq:mathcal_yh_limit}
	\mathcal Y_\text{h}(u,v,x (\gg 1)) \simeq & \frac{3 \pi }{32 u^4 v^4} \left( 3u^4(u^4 - 2) + 3v^4(v^4 -2) + 3 - 8u^2v^2(1 + u^2 + v^2) -6 u^4 v^4 \right).
\end{align}
These expressions correspond to $\mathcal J_\text{h,late}$ and $\mathcal Y_\text{h,late}$, defined in Eqs.~(\ref{eq:j_h_late_def}) and (\ref{eq:y_h_late_def}).

\section{Power spectra in other situations}
\label{app:power_realistic}

In this appendix, we discuss the auto-power spectra of the second-order scalar perturbations in situations not considered in the main body of this paper.

\subsection{Radiation-dominated era}
\label{app:power_rd}

First, we discuss the power spectra in a RD era.
Throughout this paper, we assume a perfect fluid, in which there is no anisotropic stress.
However, in realistic situations, the fluid deviates from a perfect fluid and the scalar perturbations can be diffused on deeply subhorizon scales in a RD era.
This phenomenon is called the diffusion (or Silk) damping~\cite{Silk:1967kq}.
The CMB distortion, which can constrain the small-scale perturbations, is caused by the energy injection to CMB through the diffusion damping.
Because of this, we need to be careful when we compare the second-order perturbations with the observational constraints.

For the first-order perturbations, the CMB distortion is caused only by the perturbation on $k \lesssim 10^{4}$Mpc$^{-1}$. 
This is because the photon-number changing process such as the double Compton scattering, which prevents the CMB distortion, is effective at the moment when the smaller-scale ($k \gtrsim  10^{4}$Mpc$^{-1}$) perturbations are erased through the diffusion damping (see e.g. Ref.~\cite{Chluba:2012gq} for the detail of the CMB distortion).

For the second-order perturbations, the situation is a little bit different.
The contributions of $\mathcal J_0$ and $\mathcal Y_0$ continue to depend on the evolution of the source perturbations even on deep subhorizon scales.
$\mathcal J_0$ and $\mathcal Y_0$ are modified (might be erased) on all scales at the moment when the {\it source} perturbation, whose wavenumber is $k \gg 10^4\,\text{Mpc}^{-1}$, is erased by the diffusion damping.
Note that, even if the contributions of $\mathcal J_0$ and $\mathcal Y_0$ on $k \lesssim 10^{4}$Mpc$^{-1}$ are also erased at that moment, the distortion is not produced because the photon-number changing process is effective at that time.
On the other hand, the other contributions from $\mathcal J_\text{h}$, $\mathcal J_\text{i}$, and $\mathcal Y_\text{h}$ do not depend on the source perturbations when the induced perturbations are on deep subhorizon scales\footnote{Conversely, $\mathcal J_\text{h}$ and $\mathcal Y_\text{h}$ do not converge until the induced perturbations enter the horizon ($x \sim 1$), which means that if the source perturbations are diffused at $\eta_\text{d}$, the convergence values are different from $\mathcal J_\text{h,late}$ and $\mathcal Y_\text{h,late}$ on the scales $k \lesssim 1/\eta_\text{d}$. Here, we assume that the source perturbations are diffused well after entering the horizon and the convergence values are given by $\mathcal J_\text{h,late}$ and $\mathcal Y_\text{h,late}$ on all scales for simplicity.} and therefore they are not affected by the diffusion damping of the {\it source} perturbations at least on the subhorizon scales.
Similarly to the first-order perturbations, the contributions of $\mathcal J_\text{h}$, $\mathcal J_\text{i}$, and $\mathcal Y_\text{h}$ are erased at the moment when their scales are smaller than the diffusion scale and, if their scales are $k \lesssim 10^{4}$Mpc$^{-1}$, they produce the distortion, which is constrained by the CMB observation.
Since the analysis of the diffusion damping effect on $\mathcal J_0$ and $\mathcal Y_0$ is beyond the scope of this paper, we just focus on the other contributions from $\mathcal J_\text{h}$, $\mathcal J_\text{i}$, and $\mathcal Y_\text{h}$ hereafter.
Note that, even if we neglect $\mathcal J_0$ and $\mathcal Y_0$, $\Psi^{(2)}$ still depends on time because $\mathcal J$ and $\mathcal Y$ are the coefficients in front of the first-order solutions, which are proportional to $j_1(x/\sqrt{3})/x$ and $y_1(x/\sqrt{3})/x$ (see Eq.~(\ref{eq:i_psi_r_s_re}))\footnote{The solution of $y_1(x/\sqrt{3})/x$ does not appear in the analytic solution of $\Phi^\fo$ (Eq.~(\ref{eq:phi_zeta_rel})) because of its initial condition on superhorizon scales. }.

To compare the induced perturbations with the primordial first-order perturbations, we define the following power spectra:
\begin{align}
	\mathcal P_{\mathcal J_\text{const}}(k) \equiv& \int_0^\infty \dd v \, \int_{|v-1|}^{v+1} \dd u \, \left(\mathcal J_\text{h,late}(u,v) + \mathcal J_\text{i} (u,v) \right)^2 \left( \frac{2}{3} \right)^4 \mathcal P_{\zeta^\fo}( k v) \mathcal P_{\zeta^\fo}(k u), 
	\label{eq:math_j_h_ps_def}  \\
	\mathcal P_{\mathcal Y_\text{const}}(k) \equiv& \int_0^\infty \dd v \, \int_{|v-1|}^{v+1} \dd u \, \mathcal Y_\text{h,late}^2(u,v) \left( \frac{2}{3} \right)^4 \mathcal P_{\zeta^\fo}( k v) \mathcal P_{\zeta^\fo}(k u).
	\label{eq:math_y_h_ps_def} 
\end{align}
Note that these power spectra do not depend on time because they only include the contribution independent of the source perturbations on deep subhorizon scale.
At first order in perturbations, the coefficient $\mathcal J$ corresponds to the amplitude of the curvature perturbations on superhorizon scale (see Eq.~(\ref{eq:phi_zeta_rel})).
For this reason, we compare the Eqs.~(\ref{eq:math_j_h_ps_def}) and (\ref{eq:math_y_h_ps_def}) with $\mathcal P_{\zeta^\fo}$ with the appropriate normalization, explained below. 
At first order, the power spectra of $\Psi^\fo$ and $\zeta^\fo$ are related as $\mathcal P_{\zeta^\fo} \simeq (9/4)\mathcal P_{\Psi^\fo}$ on superhorizon scales in a RD era. 
From this observation, we compare $(9/16) \mathcal P_{\mathcal J_\text{const}}$ and $(9/16)\mathcal P_{\mathcal Y_\text{const}}$ with $\mathcal P_{\zeta^\fo}$, where we have additionally multiplied them by $1/4$ because the normalization of $\Psi^\so$ is different from $\Psi^\fo$ by $1/2$ (see Eq.~(\ref{eq:def_metric_pertb})).
For $\mathcal P_{\zeta^\fo}$, in addition to the delta-function power spectrum, we also consider the top-hat power spectrum, given as 
\begin{align}
	\mathcal P_{\zeta^\fo}(k) = A_\zeta \Theta(k_* - k) \Theta(k - k_{*,l}),
	\label{eq:tophat_ps_zeta}
\end{align}
where $\Theta(x)$ denotes the Heaviside step function.

\begin{figure}[htbp]
  \begin{center}
    \begin{tabular}{c}

      \begin{minipage}{0.45\textwidth}
        \begin{center}
          \includegraphics[width=\hsize]{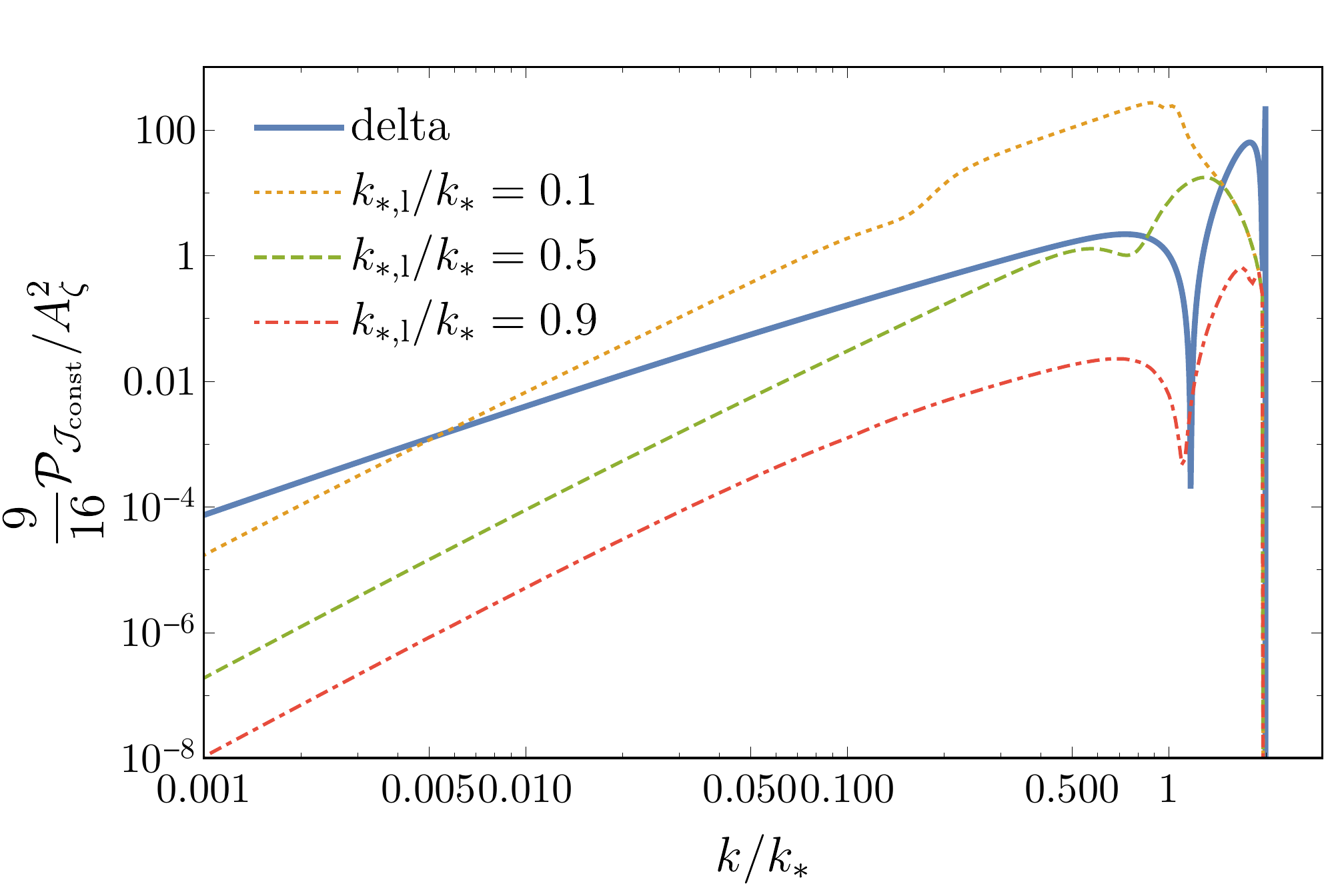}
        \end{center}
      \end{minipage}
	\hspace{0.5cm} 
      \begin{minipage}{0.45\textwidth}
        \begin{center}
          \includegraphics[width=\hsize]{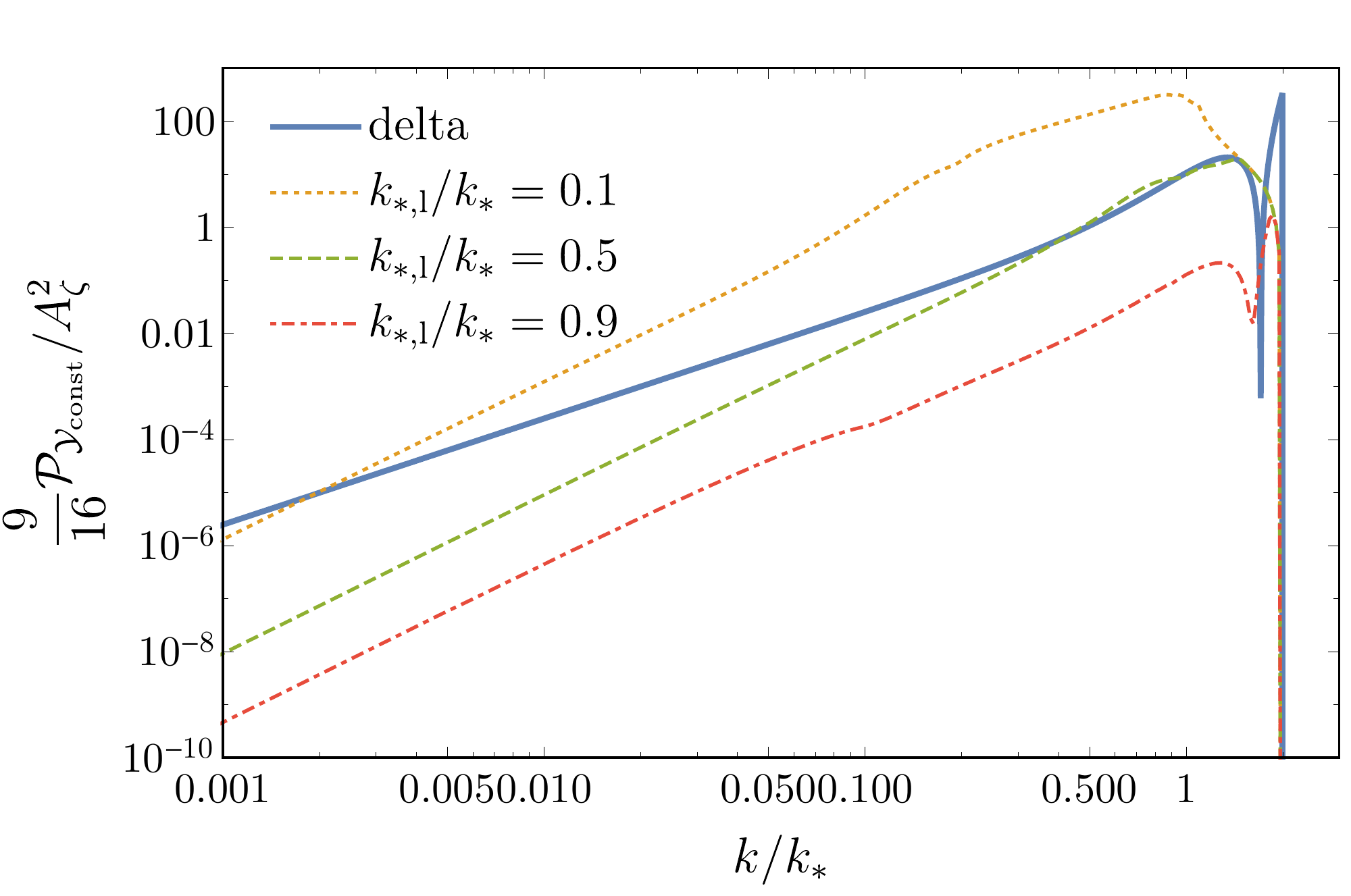} 
        \end{center}
      \end{minipage}\\

    \end{tabular}
    \caption{\small	
	The power spectra $\mathcal P_{\mathcal J_\text{const}}$ (left) and $\mathcal P_{\mathcal Y_\text{const}}$ (right), defined in Eqs.~(\ref{eq:math_j_h_ps_def}) and (\ref{eq:math_y_h_ps_def}) and multiplied by $9/(16 A_\zeta^2)$. 
	The power spectra of the first-order curvature perturbations are given by the delta-function one (Eq.~(\ref{eq:delta_func_pzeta})) for the blue solid lines and by the top-hat one (Eq.~(\ref{eq:tophat_ps_zeta})) for the other lines.
	We take $a_\text{NL} = 1$ for all the lines.}
    \label{fig:p_math_jh_yh}
  \end{center}
\end{figure}

Figure~\ref{fig:p_math_jh_yh} shows the normalized spectra, $(9/16) \mathcal P_{\mathcal J_\text{const}}/A_\zeta^2$ and $(9/16)\mathcal P_{\mathcal Y_\text{const}}/A_\zeta^2$, with the delta-function and the top-hat power spectra. 
From this figure, we can see that the peaks of the power spectra become higher for the broader $\mathcal P_{\zeta^\fo}$ and the wavenumber dependences of $\mathcal P_{\mathcal J_\text{const}}$ and $\mathcal P_{\mathcal Y_\text{const}}$ on large scales are given by $\sim k^3$ for the top-hat power spectrum and $\sim k^2$ for the delta-function one up to the logarithmic factor. 
These wavenumber dependences are the same as those of the GWs induced by scalar perturbations~\cite{Pi:2020otn}.

\begin{figure}[ht] 
        \centering \includegraphics[width=0.9\columnwidth]{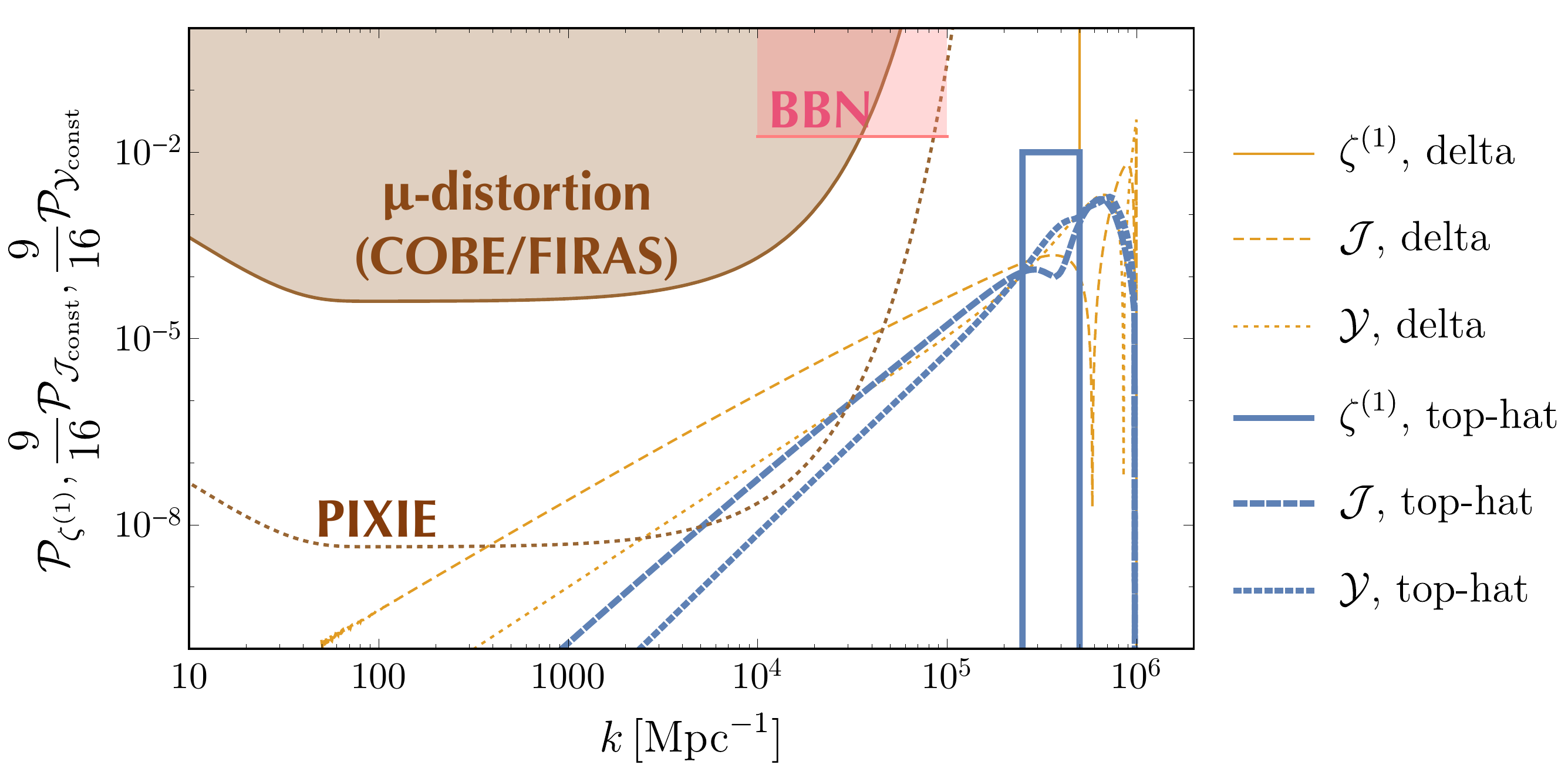}
        \caption{
        The power spectra $\mathcal P_{\mathcal J_\text{const}}$ and $\mathcal P_{\mathcal Y_\text{const}}$ in the case the scalar perturbations produce the sizable amount of PBHs, which could explain BHs detected by LIGO-Virgo collaborations. 
        The thin orange and the thick blue lines show the spectra for the delta-function and the top-hat power spectra of the first-order curvature perturbations, respectively.
        The solid, dashed, and dotted lines show $\mathcal P_{\zeta^\fo}$, $(9/16)\, \mathcal P_{\mathcal J_\text{const}}$, and $(9/16)\, \mathcal P_{\mathcal Y_\text{const}}$ respectively.
        We take $a_\text{NL} = 1$, $A_\zeta = 0.01$, $k_* = 5 \times 10^5\,$Mpc$^{-1}$, and $k_{*,\text{l}} = 2.5 \times 10^{5}\,$Mpc$^{-1}$ as fiducial values.
        For comparison, we also show the constraints from the big bang nucleosynthesis (BBN)~\cite{Inomata:2016uip} (see also Refs.~\cite{Jeong:2014gna,Nakama:2014vla}) and the $\mu$-distortion of CMB spectrum~\cite{Chluba:2013dna} with COBE/FIRAS result ($|\mu| < 9 \times 10^{-5}$~\cite{Fixsen:1996nj}) and the future sensitivity of PIXIE ($|\mu| < 10^{-8}$~\cite{Kogut:2011xw,kogut2016primordial,Kogut:2019vqh}), where the shaded regions are already excluded by current observations.
        }
        \label{fig:induced_scalar_real}
\end{figure}

Since the small-scale perturbations are often discussed in the context of PBHs, we here consider the power spectrum, which predicts a sizable amount of PBHs that can explain BHs detected by LIGO-Virgo collaborations.
Specifically, we take $A_\zeta = 0.01$ and $k_* = 5 \times 10^5\,$Mpc$^{-1}$ as fiducial values~\cite{Inomata:2017vxo}.
Figure~\ref{fig:induced_scalar_real} shows the spectra of the induced scalar perturbations with the fiducial values.
Although the power spectra are consistent with the null detection of the CMB $\mu$-distortion and the light element abundance determined by the big bang nucleosynthesis so far, they could possibly be investigated in the future by a PIXIE-like spectrometer~\cite{Kogut:2011xw,kogut2016primordial,Kogut:2019vqh}. 
Note again that we have neglected the contributions from $\mathcal J_0$ and $\mathcal Y_0$ and those from the cross-power spectrum between the first- and the third-order perturbations for simplicity.
For these reasons, we should keep in mind that the observable spectra of the induced perturbations might be different from those in Fig.~\ref{fig:induced_scalar_real}.

\subsection{Matter-dominated era}
\label{app:power_md}

\begin{figure}[htbp]
  \begin{center}
    \begin{tabular}{c}

      \begin{minipage}{0.45\textwidth}
        \begin{center}
          \includegraphics[width=\hsize]{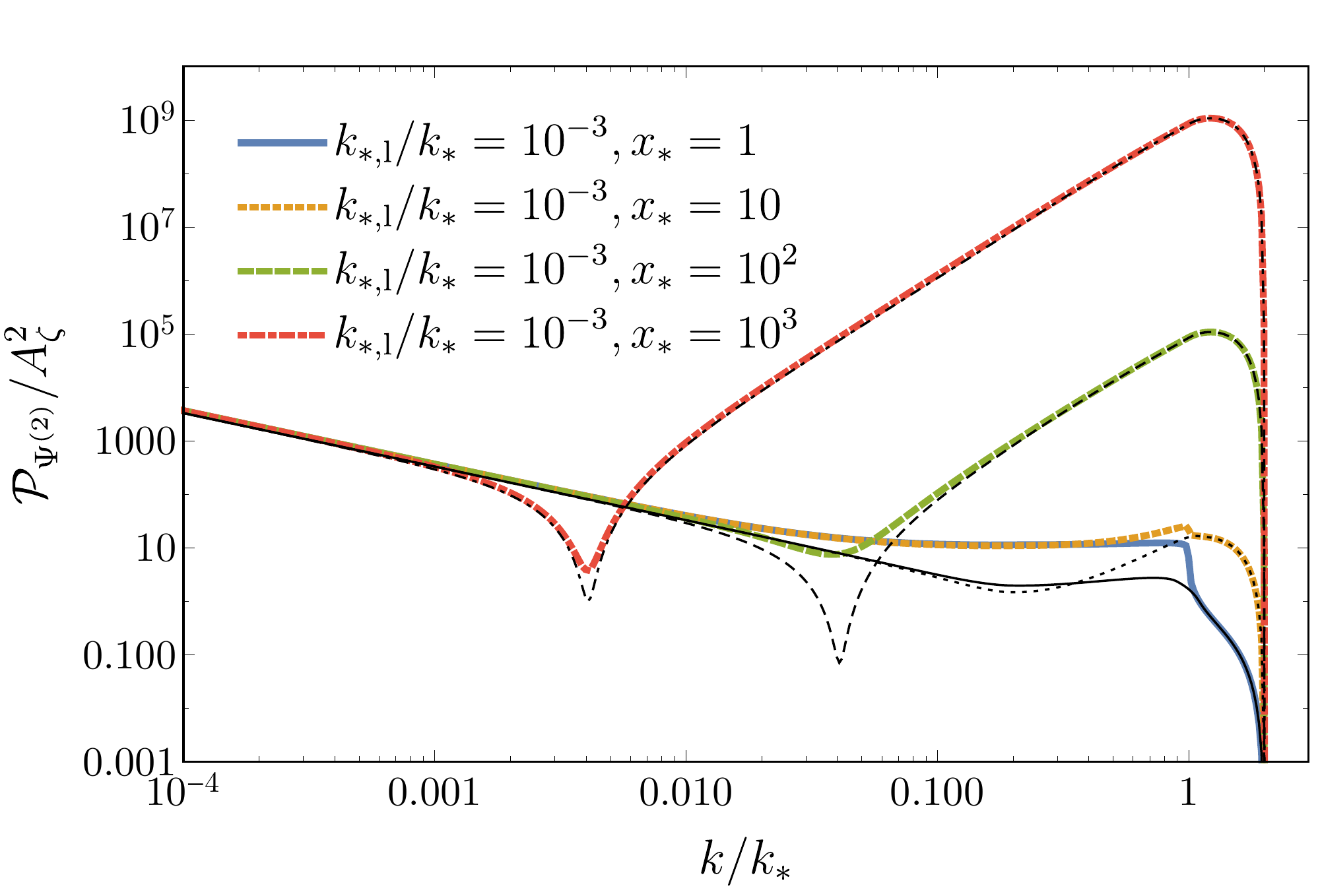} 
        \end{center}
      \end{minipage}
	\hspace{0.5cm} 
      \begin{minipage}{0.45\textwidth}
        \begin{center}
          \includegraphics[width=\hsize]{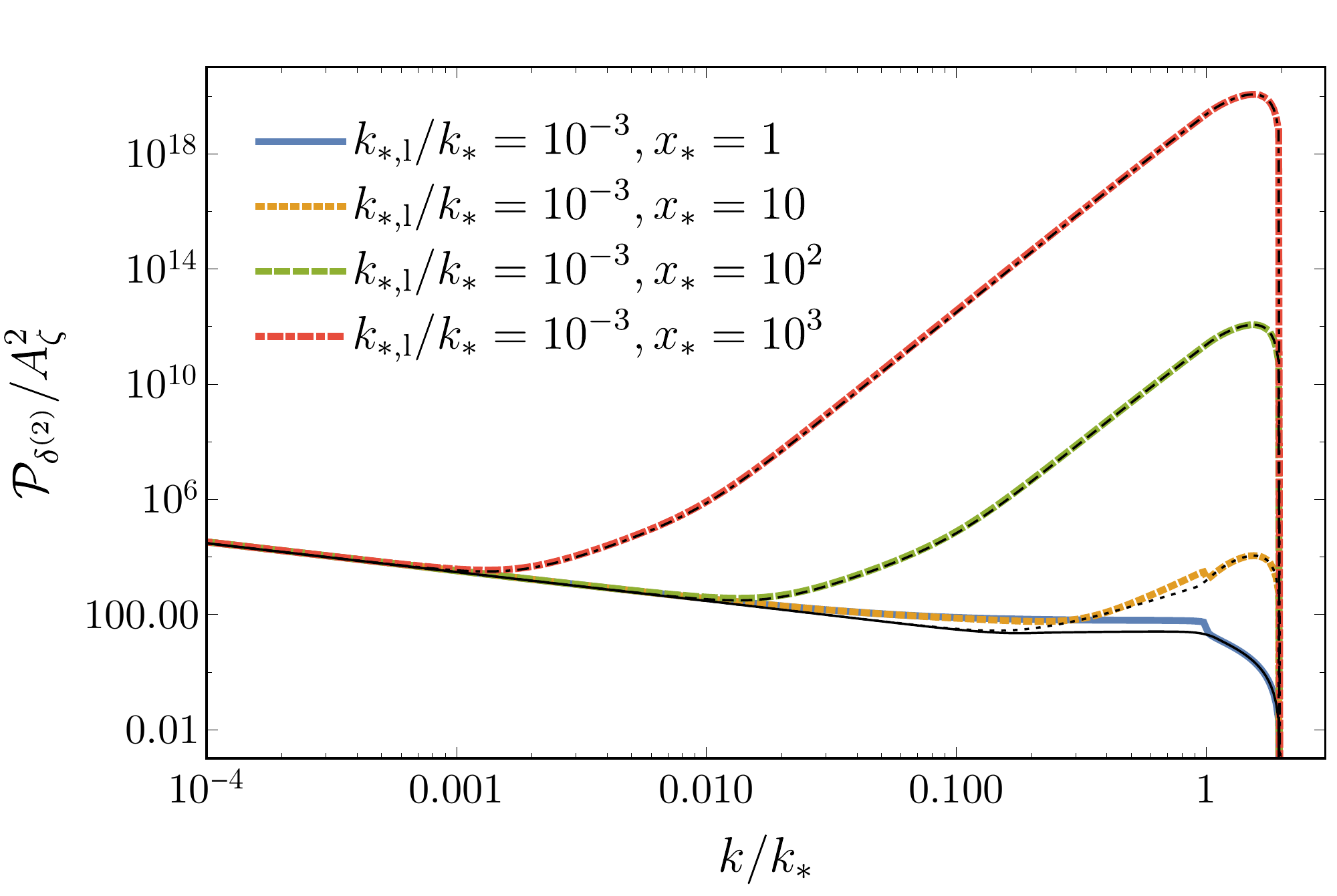}        
        \end{center}
      \end{minipage}\\

      \begin{minipage}{0.67\textwidth}
      \vspace{0.6cm}
        \begin{center}
          \includegraphics[width=\hsize]{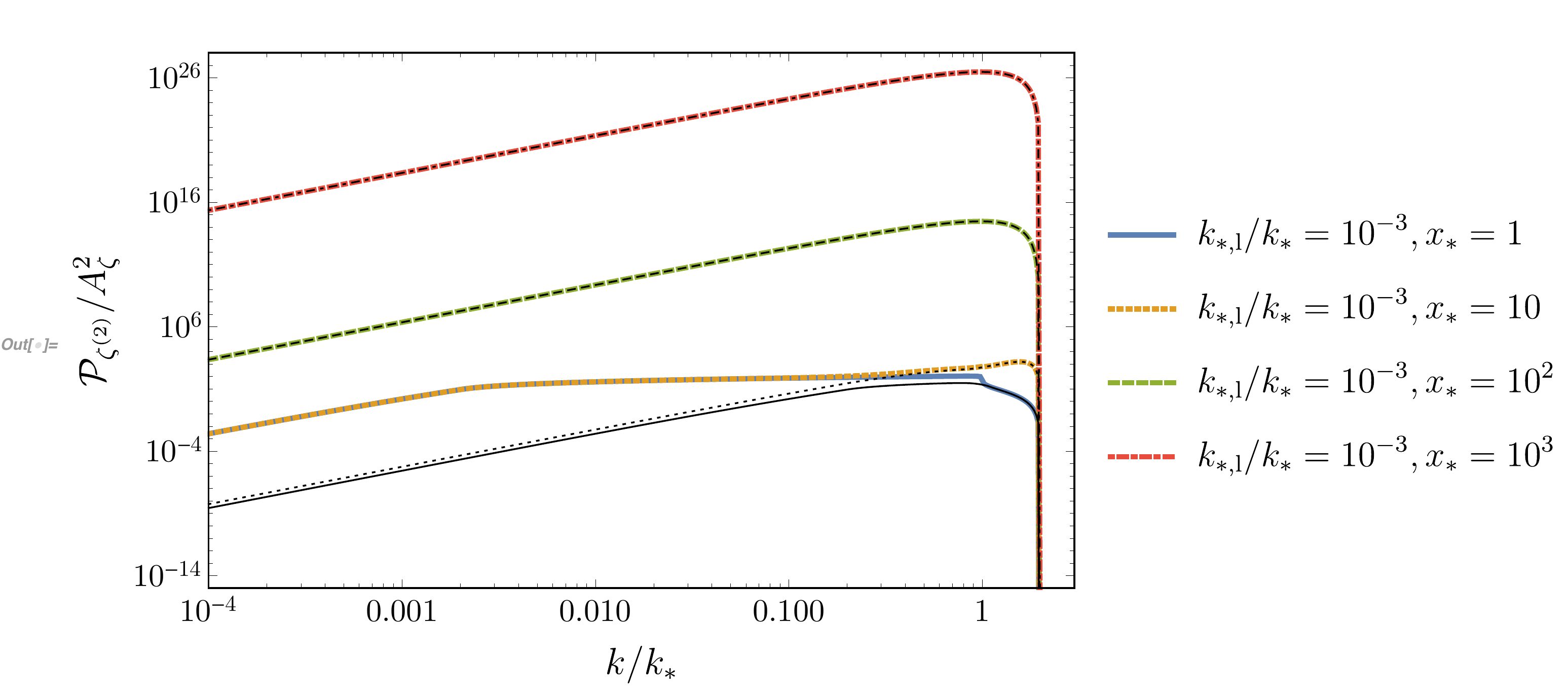}
        \end{center}
      \end{minipage}

    \end{tabular}
    \caption{\small	
   Evolution of $\mathcal P_{\Psi^\so}$ (top left), $\mathcal P_{\delta^\so}$ (top right), and $\mathcal P_{\zeta^\so}$ (bottom) with the top-hat power spectrum defined by Eq.~(\ref{eq:tophat_ps_zeta}) in a MD era.
   For colored lines, we take $k_{*,l}/k_* = 10^{-3}$ and for black lines, $k_{*,l}/k_* = 10^{-1}$.
   The same line styles correspond to the same values of $x_*$, e.g. the solid lines indicate $x_* =1$ regardless of their colors.
    We take $a_\text{NL} = 1$ for all the lines.
	}
    \label{fig:p_md_realistic}
  \end{center}
\end{figure}

Next, we discuss the power spectra in a MD era.
Similarly to the previous subsection, we consider the top-hat power spectrum, given in Eq.~(\ref{eq:tophat_ps_zeta}).
Since the observations of CMB and the large scale structure have revealed that the Universe has experienced the MD era starting around $z \simeq 3400$ and the $\mathcal P_{\zeta^\fo}$ is almost scale-invariant on the large scales ($k \lesssim 1$\,Mpc$^{-1}$), we especially focus on the case of a broad top-hat power spectrum with $k_{*,l} \ll k_*$ as a fiducial example.
Figure~\ref{fig:p_md_realistic} shows the evolution of the power spectra with $k_{*,l}/k_* = 10^{-3}$.
For comparison, we also show the spectra with $k_{*,l}/k_* = 10^{-1}$ (black lines), from which we can see the spectra do not depend on $k_{*,l}$ so much especially for $x_* = 10^2$ and $10^3$.
From this observation, we conclude that the main contribution comes from the perturbations on the smallest scales ($\sim k_{*}$).
Similarly to the spectra in Fig.~\ref{fig:p_delta_mm}, $\mathcal P_{\Psi^\so}$ and $\mathcal P_{\delta^\so}$ remain constant on superhorizon scales and start to evolve once they enter the horizon.
Finally, $\mathcal P_{\Psi^\so}$ and $\mathcal P_{\delta^\so}$ on subhorizon scales grow proportionally to $\eta^4 \propto a^{2}$ and $\eta^8 \propto a^{4}$, respectively.
On the other hand, $\mathcal P_{\zeta^\so}$ finally grows proportionally to $\eta^{12} \propto a^6$ on all scales, including superhorizon scales, well after the perturbations around the smallest scales enter the horizon.
In the late time ($x_* \gg 1$), the wavenumber dependences of the spectra can be roughly summarized as follows:
\begin{align}
	&\mathcal P_{\Psi^\so} \propto \begin{cases}
	k^{-1} \quad &(x \ll 1) \\
	k^{3} \quad &(1 \ll x \lesssim x_*)
	\end{cases}, \\
	&\mathcal P_{\delta^\so} \propto \begin{cases}
	k^{-1} \quad &(x \ll 1) \\
	k^{7} \quad &(1 \ll x \lesssim x_*)
	\end{cases}, \\
	&\mathcal P_{\zeta^\so} \propto
	k^{3} \quad (x \lesssim x_*).
\end{align}

\small
\bibliographystyle{apsrev4-1}
\bibliography{draft_secondary_scalar}{}

\end{document}